\documentclass[11pt]{article}
\usepackage[T1]{fontenc}
\usepackage[utf8]{inputenc}
\usepackage{color}
\usepackage{array}
\usepackage{booktabs}
\usepackage{url}
\usepackage{multirow}
\usepackage{amsmath}
\usepackage{amssymb}
\usepackage{graphicx}
\usepackage[authoryear]{natbib}
\usepackage[unicode=true,
 bookmarks=false,
 breaklinks=false,pdfborder={0 0 1},% backref=section,
 colorlinks=false]
 {hyperref}
\usepackage{siunitx}
\usepackage{url}
\usepackage{tabularx}
\usepackage{array}
\usepackage{mathtools}
\usepackage{booktabs}
\usepackage{subfig}
\usepackage{bm}
\usepackage{tikz-cd}
\graphicspath{{./art/}}
\allowdisplaybreaks
\usepackage[plain,noend]{algorithm2e}
\usepackage{color}

\makeatletter

\usepackage{latexsym}
\usepackage{multirow}\usepackage{bm}%\usepackage[latin9]{inputenc}
\usepackage{url}% not crucial - just used below for the URL 
\usepackage{tabularx}
\usepackage{booktabs}
\usepackage{setspace}

\usepackage{enumerate}

\def\T{{ \mathrm{\scriptscriptstyle T} }}

\newcommand{\spacingset}[1]{\renewcommand{\baselinestretch}%
{#1}\small\normalsize}
\newcommand{\op}{\mathcal{O}}

\newcommand{\E}{\mathrm{E}}
\newcommand{\V}{\mathrm{var}}
\newcommand{\pr}{\mathrm{P}}

\newcommand{\cov}{\textnormal{cov}}

\newcommand{\de}{\mathrm{d}}

\newcommand{\eff}{\textnormal{eff}}

\newcommand{\iif}{\textnormal{IF}}

\newcommand{\rct}{\textnormal{rct}}

\newcommand{\rwe}{\textnormal{rwd}}
\newcommand{\meta}{\textnormal{meta}}
\newcommand{\adj}{\textnormal{adj}}

\newcommand{\R}{\mathcal{R}}
\newcommand{\bbP}{\mathrm{P}}

\newcommand*{\indep}{%
{\mbox{$\perp\!\!\!\perp$}}
}
\newcommand*{\nindep}{%
	\mathbin{%                   % The final symbol is a binary math operator
		\mathpalette{\@indep}{\not}% \mathpalette helps for the adaptation
	}%
}

\usepackage{amsmath}
\usepackage{amssymb}
\usepackage{amsfonts}
\usepackage{multirow}
\usepackage{amsthm}

\newtheorem{theorem}{Theorem}
\newtheorem{lemma}{Lemma}

\newtheorem{Assumption}{Assumption}
\newtheorem{Proposition}{Proposition}
\newtheorem{condition}{Condition}
\theoremstyle{definition}
\newtheorem{definition}{Definition}

\newtheorem{remark}{Remark}

\begin{document}


\title{\textbf{Data fusion methods for the heterogeneity of treatment effect and confounding function}}

%\iffalse 
\author{Shu Yang$^1$, Siyi Liu$^1$, Donglin Zeng$^2$, Xiaofei Wang$^{3}$} 

\date{\vspace{-5ex}}

\maketitle
\begin{center}
$^{1}$Department of Statistics, North Carolina State University, Raleigh, NC, USA \\
$^{2}$Department of Biostatistics, University of North Carolina, Chapel Hill, NC, USA \\
$^{3}$Department of Biostatistics and Bioinformatics, Duke University, Durham, NC, USA
\end{center}
\spacingset{1.5} 
\begin{abstract}
The heterogeneity of treatment effect (HTE) lies at the heart of precision medicine. Randomized controlled trials are gold-standard for treatment effect estimation but are typically underpowered for heterogeneous effects. In contrast, large observational studies have high predictive power but are often confounded due to the lack of randomization of treatment. We show that an observational study, even subject to hidden confounding, may be used to empower trials in estimating the HTE using the notion of confounding function. The confounding function summarizes the impact of unmeasured confounders on the difference between the observed treatment effect and the causal treatment effect, given the observed covariates, which is unidentifiable based only on the observational study. Coupling the trial and observational study, we show that the HTE and confounding function are identifiable. We then derive the semiparametric efficient scores and the integrative estimators of the HTE and confounding function. We clarify the conditions under which the integrative estimator of the HTE is strictly more efficient than the trial estimator. Finally, we illustrate the integrative estimators via simulation and an application.

\noindent \textbf{keywords:} Estimating equation; goodness of fit; over-identification test; semiparametric efficiency; structural model.
\end{abstract}
\newpage{}

%%%%%%%%%%%%%%%%%%%%%%%%%%%%%%%%%%%%%%%%%%%%%%
%%%% Main text entry area:
\section{Introduction}

Randomized controlled trials are the cornerstone of evidence-based
medicine for treatment effect evaluation because randomization of
treatment ensures that treatment groups are comparable and biases
are minimized. Recently, considerable interest has been in understanding
the heterogeneity of treatment effects, a critical path toward personalized
medicine \citep{collins2015new}. However, due to eligibility criteria
for recruiting patients, the trial sample is often limited in the
patient diversity, which renders the trial underpowered to estimate
the heterogeneity of treatment effect. On the other hand, large observational
studies are increasingly available for research purposes, such as
electronic health records, claims databases, and disease registries,
with much broader demographic and diversity than trial cohorts. However,
they also present challenges such as confounding due to the lack of
randomization.

Existing approaches to harmonize evidence from trial and observational
studies include meta-analysis \citep{verde2015combining} and joint
analysis of the pooled data \citep{prentice2008estrogen}. As related,
\citet{chen2008semiparametric} proposed an efficient generalized-method-of-moments
estimator combining primary and auxiliary samples under missingness
at random, i.e., no unmeasured confounding in our context. However,
these approaches assume no hidden confounders, which is unlikely to
be true in practice. The no unmeasured confounding assumption requires
researchers to measure all relevant predictors of treatment and outcome.
However, it is always possible that certain important confounders
are unavailable in uncontrolled, real-world settings. For example,
doctors use patients' symptoms not captured in the medical charts
to assign treatments. Alternatively, certain prognostic factors are
measured with errors due to technological limitations. Unmeasured
confounding presents a major threat to causal inference from observational
studies. Classical approaches to mitigating bias due to unmeasured
confounding include instrumental variables \citep{angrist1996identification},
negative controls \citep{kuroki2014measurement}, and sensitivity
analysis \citep{robins2000sensitivity}. In particular, sensitivity
analysis is often recommended to assess the robustness of the study
conclusion to no unmeasured confounding. Many authors have implemented
sensitivity analysis using the so-called confounding function \citep{robins2000sensitivity,yang2017sensitivity,kasza2017assessing};
namely, the difference of the potential outcome means between the
treatment groups given the measured covariates due to the unmeasured
confounders. Because the observational studies carry no information
about confounding biases due to unmeasured confounders, the confounding
function is not identifiable based solely on the observational studies.

In this paper, we leverage observational studies to improve trial
analysis of the heterogeneity of treatment effect (HTE) with a vector
of known effect modifiers. We focus on the setting where the transportability
of the heterogeneous treatment effect holds from the trial to the
observational study but the observational study may be subject to
hidden confounding. Transportability is a minimal requirement for
data integration and has been considered in a vast literature \citep[e.g.,][]{stuart2011use,tipton2013improving,buchanan2018generalizing,dahabreh2019generalizing}.
It holds if the sample is randomly selected from the population or
treatment effect modifiers are fully captured. We also introduce a
new and natural confounding function to capture the impact of unmeasured
confounding in observational studies on the difference between the
observed treatment effect and the causal treatment effect given measured
covariates. Under structural model assumptions, we show that the trial
can be leveraged to identify the HTE and confounding function, in
contrast to sensitivity analysis.

The identification results motivate a broad class of consistent estimators
of the model parameters. However, naive choices lead to inefficient
estimators. We derive the semiparametric efficient score combining
the two data sources to guide constructing efficient estimators and
accelerate the full potential of trial and observational studies.
The theoretical task is challenging because of restrictions on the
parameters of interest induced from the identification assumptions,
such that the existing semiparametric efficiency theory for data integration
\citep[e.g.,][]{chen2008semiparametric} cannot apply. To overcome
the challenges, we translate the restrictions into the likelihood
function by re-parameterization and follow the geometric approach
\citep[e.g., ][]{bickel1993efficient,tsiatis2007semiparametric} to
derive the efficient score. Built upon the efficient score, we propose
an integrative estimator, which enables a fast root-$N$ rate of convergence
under weaker conditions on nuisance function approximation, e.g.,
using consistent but flexible semiparametric and nonparametric methods.
We clarify the conditions under which the gain of efficiency is strictly
positive by data integration over the trial-only estimator.{{}
The improvement in efficiency arises when certain predictors in the
heterogeneous treatment effect function are absent in the confounding
function. The formulation of the heterogeneous treatment effect and
confounding functions can be grounded in domain expertise or determined
through variable selection based on training data. Additionally, we
propose goodness-of-fit tests for assessing the structural assumptions
based on over-identification tests.} A simulation study shows that
the integrative estimator outperforms the trial-only estimator in
two settings with and without unmeasured confounding in the observational
study. In addition, we apply the proposed method to estimate the heterogeneous
treatment effect of chemotherapy for non-small cell lung cancer. {The
proofs of the semiparametric efficient score and its asymptotic properties
are presented in the last section of the main paper, with more technical
proofs given in the supplementary material \citep{supplementary_material}.}

% \subsection{{Related works}}

\textcolor{black}{Our work is motivated by related works and addresses challenges in the areas of data fusion, projection, and nonparametric structural models, each of which is elaborated as follows.}

\textit{{Data Fusion}}. \citet{yang2020combining}
developed integrative causal analyses of the average treatment effect
by calibrating auxiliary information from the validation sample to
the big main sample with unmeasured confounders for efficiency gain.
However, their approach requires the validation and main samples to
be comparable in providing consistent estimators of auxiliary parameters.
This requirement may be stringent because randomized controlled trials
often have strict inclusion and exclusion criteria lending their patient
compositions different from the observational population \citep{stuart2015assessing}.
\citet{yang2023elastic} pretested the comparability between trial
and observational studies and customized the subsequent analysis based
the pretest result. Another line of research for combining trial and
observational studies is to generalize the average treatment effect
from trials to the target population (\citealp{buchanan2018generalizing,lee2021improving};
\citealp{lee2022doubly}), where the observational sample provides
a representative covariate distribution of the target population.
\citet{yang2022rwd} and \citet{colnet2024causal} provided comprehensive
reviews. As a by-product of the proposed framework, we derive an efficient
plug-in sample estimator of the population average treatment effect
in the data integration context. Most existing methods rely on the
overlap assumption of the covariate distribution between the trial
and observational samples. Our method does not require the overlap
assumption but utilizes parametric structural assumptions on the HTE,
thus offering an alternative means for causal generalization. Nonetheless,
we add a caveat that the lack of overlap renders the structural assumptions
fragile, and one relies on model extrapolation. In practice, we still
advocate checking the overlap assumption for generalizing the treatment
effect from trial to a target population.

\textit{{Projection Parameters}}{.
In clinical applications, parametric models are preferable for their
straightforward interpretability. We operate under the assumption
that the structural models are parametric and are correctly specified.
However, in cases of misspecification, they can still be interpreted
as projection parameters---the projection of nonparametric structural
models onto a constrained model space. Projection-based interpretation
has gained popularity and facilitates deriving nonparametric efficient
scores for the projection parameters \citep[e.g.,][]{neugebauer2005prefer,chernozhukov2018generic,kennedy2019robust,kennedy2021semiparametric}.
Unlike these approaches, our semiparametric efficiency scores take
into account the constraints on the structural parameters imposed
by the identification assumptions. Semiparametric efficiency for models
subject to constraints can be of independent interest.}

\textit{{Nonparametric Structural Models}}. Aside
from the interpretability, a technical reason to consider parametric
models is that the nonparametric HTE and confounding functions are
often local parameters that are not pathwise differentiable, so their
efficient scores with finite variances do not exist \citep{bickel1993efficient}.
{Exceptions include the cases of fully discrete data
and functions valued in general Hilbert space \citep{luedtke2023one}.
}With the nonparametric models, one can alternatively study the bounds
on the asymptotic minimax risk indicating the best possible performance
of any estimator in the worst case scenarios (\citealp{kennedy2022minimax}
and references therein). For example, \citet{kennedy2022minimax}
derived a lower bound on the minimax rate of HTE estimation when the
HTE and nuisance functions are Holder-smooth in studies without hidden
confounding.{{} When considering nonparametric rates
of HTE estimation in the data fusion context, one strategy involves
initially approximating the HTE by projecting it onto a finite-dimensional
approximating space and then balancing estimation accuracy and approximation
error. However, in such analyses, the objective differs from what
is presented in this paper; specifically, there is no need to derive
a precise asymptotic distribution of the estimation for the finite-dimensional
projection. Instead, the focus is on controlling the mean-squared
error as the dimension of the projection increases. This research
topic will be pursued in the future.}

\section{Basic setup and identification\label{sec:Basic-setup}}

\subsection{Notation, causal effects, and two data sources}

Let $A$ be the binary treatment, $X$ be a vector of pre-treatment
covariates with the first component being $1$, and $Y$ be the outcome
of interest. The target population consists of all patients with certain
diseases where the new treatment is intended to be given. We use the
potential outcomes to define causal effects. Let $Y(a)$ be the potential
outcome had the subject been given treatment $a$, for $a=0,1$. Based
on the potential outcomes, the individual treatment effect becomes
$Y(1)-Y(0),$ the HTE can be characterized through $\tau(X)=\E[Y(1)-Y(0)\mid X]$,
and the average treatment effect is $\tau_{0}=\E[\tau(X)]$.

We consider two independent data sources: one is a randomized trial
study, and the other is an observational study. Let $S=1$ denote
trial participation, and let $S=0$ denote observational study participation.
Let $\mathcal{A}$ and $\mathcal{B}$ be sample index sets for the
two data sources with sample sizes $|\mathcal{A}|=n$ and {$|B|=m$},
and the total sample size is $N=n+m$. The trial data consist of $\{V_{i}=(A_{i},X_{i},Y_{i},S_{i}):i\in\mathcal{A},S_{i}=1\}$,
where the observations i.i.d. follow $f(X,A,Y\mid S=1)$, and the
observational data consist of $\{V_{i}:i\in\mathcal{B},S_{i}=0\}$,
where the observations i.i.d. follow $f(X,A,Y\mid S=0)$. Figure \ref{fig:demo}
displays the envisioned data structure within the target population.
To link the observed outcome and potential outcomes, we make the typical
causal consistency assumption of $Y=Y(A)$. This assumption rules
out the interference between subjects and treatment version relevance
between samples and population \citep{tipton2013improving}. The implication
is that $Y(a)$ has consistent meaning and value across the trial
and observational studies. This assumption requires the same treatment
or comparison conditions to be given to both studies, and being in
the trial should not affect the values of the potential outcomes.
To simplify the exposition, we define 
\begin{align*}
e(X,S) & =\pr(A=1\mid X,S),\mu_{a}(X,S)=\E[Y\mid A=a,X,S],\\
\sigma_{a}^{2}(X,S) & =\V[Y\mid A=a,X,S],\mu(X,S)=\E[Y\mid X,S],
\end{align*}
where $e(X,S)$ is the propensity score, $\mu_{a}(X,S)$ and $\sigma_{a}^{2}(X,S)$
are the treatment-specific outcome mean and variance functions, for
$a=0,1$, and $\mu(X,S)$ is the outcome mean function marginalized
over treatment. For any $g(V),$ define $\epsilon_{g(V)}=g(V)-\E[g(V)\mid X,S]$,
e.g., $\epsilon_{A}=A-e(X,S)$.
\begin{figure}
{\centering}{
\begin{centering}
\resizebox{0.6\textwidth}{!}{% %
\begin{tikzpicture}[every node/.style=scale=1.1, node distance = 3cm, auto]\tikzstyle{block} = [rectangle, draw, text width=18em, text centered, rounded corners, minimum height=4em]
\tikzstyle{every node} = [font = \footnotesize]
% Place nodes
\node [block] (A) {	\begin{tabular}{c} 	% content
%	\textbf{Latent Data}\\
    \textbf{Target future population} \\
    	(Superpopulation)\\
	$\left\{X_i,Y_i(0),Y_i(1)\right\}_{i=1}^\infty$
\end{tabular}
};
\node at (-4.5,-4) [block] (B) {	\begin{tabular}{l} 	% content
%	\textbf{Latent Data}\\
	$\left\{X_i,Y_i(0),Y_i(1),S_i=1\right\}_{i\in\mathcal{A}}$
\end{tabular}
};
\node at (4.5,-4) [block] (C) {	\begin{tabular}{l} 	% content
%	\textbf{Latent Data}\\
    % \textbf{Finite OS population}
	$\left\{X_i,Y_i(0),Y_i(1),S_i=0\right\}_{i\in\mathcal{B}}$
\end{tabular}
};
\node at (-4.5,-8) [block] (D) {
	\begin{tabular}{l} 	% content
	\textbf{Observed trial sample}\\
    $\left\{ (X_i,A_i,Y_i): S_i=1 \right\}_{i\in\mathcal{A}}$ \\
    $\qquad \stackrel{\text{i.i.d.}}{\sim} f(X,A,Y\mid S=1)$
	\end{tabular}
};
\node at (4.5,-8)  [block] (E) {\begin{tabular}{l} 
%	\textbf{Latent Data}\\
	\textbf{Observed OS sample}\\
    $\left\{ (X_i,A_i,Y_i): S_i=0 \right\}_{i\in\mathcal{B}}$ \\
    $\qquad \stackrel{\text{i.i.d.}}{\sim} f(X,A,Y\mid S=0)$
\end{tabular}
};
\node at (-2.5,-2.2)(p-RCT){Trial sampling};
\node at (2.5,-2.2)(p-RWE){OS sampling};
\node at (-4.5,-5.7)(s-RCT){Trial treatment $A_i \sim$ Randomization};
\node at (4.5,-5.7)(s-RWE){OS treatment $A_i \sim$ Unknown};
% Draw edges
\draw [->,shorten >=4pt] 

(A) edge (p-RCT)
(p-RCT) edge (B) 
(A) edge (p-RWE)
(p-RWE) edge (C)
(B) edge (s-RCT)
(s-RCT) edge (D)
(C) edge (s-RWE)
(s-RWE) edge (E);\end{tikzpicture}}
\par\end{centering}
\caption{Demonstration of the data structure for the trial and observational
study (OS) samples within the target population.\label{fig:demo}}

}
\end{figure}

\subsection{Assumptions, confounding function, and nonparametric identification}

Due to the fundamental problem that the potential outcomes can never
be jointly observed for a particular subject, $\tau(X)$ is not identifiable
in general. We make the following assumptions.

{\begin{Assumption}[Transportability and randomized trial design]\label{Asump:transp}
(i) $\E[Y(1)-Y(0)\mid X,{S=s}]=\tau(X)$, for $s=0,1$,
(ii) $Y(a)\indep A\mid(X,S=1)$ for $a\in\{0,1\}$, and $0<e(X,S)<1$
almost surely.

\end{Assumption}}

{Assumption \ref{Asump:transp}(i) states that the treatment effect
function is transportable from the trial and observational samples
to the target population. This assumption is common in the data integration
literature. Stronger versions of Assumption \ref{Asump:transp}(i)
include the ignorability of study participation \citep[e.g.,][]{buchanan2018generalizing}
and the mean exchangeability \citep[e.g.,][]{dahabreh2019generalizing}.
Assumption \ref{Asump:transp}(i) holds if $X$ captures the heterogeneity
of effect modifiers or the study sample is a random sample from the
target population. To ensure this assumption holds, variables and
samples should be carefully chosen with consultations of subject knowledge;
e.g., collect data on likely effect modifiers that affect study participation.
Under the structural equation model framework, \citet{pearl2011transportability}
provided graphical conditions for transportability. %The graphical
%representation can aid the investigator to assess the plausibility
%of Assumption \ref{Asump:transp}}(i).
Assumption \ref{Asump:transp}(ii) holds by a well-designed trial
with good patient compliance.}

{Unlike trials, treatment randomization is typically unrealistic
for observational studies. To take into account the possible unmeasured
confounders, we define the confounding function 
\[
\lambda(X)=\mu_{1}(X,S=0)-\mu_{0}(X,S=0)-\tau(X),
\]
which measures the difference between the observed treatment effect
and the causal treatment effect given $X$. In the absence of unmeasured
confounders, we have $\lambda(X)=0$. In the presence of unmeasured
$U$ that is related to both $\{Y(0),Y(1)\}$ and $A$ after controlling
for $X$, we have $\lambda(X)\neq0$.}

\begin{Proposition}\label{prop:id}

Under Assumption \ref{Asump:transp}, $\tau(X)$ and $\lambda(X)$
are identifiable by 
\begin{align}
\tau(X) & =\mu_{1}(X,S=1)-\mu_{0}(X,S=1)=\E[\widetilde{Y}\mid X,S=1],\label{eq:id1}\\
\lambda(X) & =\mu_{1}(X,S=0)-\mu_{0}(X,S=0)-\tau(X)=\E[\widetilde{Y}\mid X,S=0]-\tau(X),\label{eq:id2}
\end{align}
where $\widetilde{Y}=Y\{A-e(X,S)\}/[e(X,S)\{1-e(X,S)\}]$.

\end{Proposition} {That is, the transportability and randomized
trial design identify $\tau(X)$. The observed treatment effect from
the observational study is attributable to both $\tau(X)$ and $\lambda(X)$,
but coupling the trial and observational samples identifies $\lambda(X)$.
}{The second equality on (\ref{eq:id1}) and (\ref{eq:id2}) follows
by $\E(\widetilde{Y}\mid X,S)=\mu_{1}(X,S)-\mu_{0}(X,S)$. Proposition
\ref{prop:id} provides two identification strategies relying on different
components of the observed data distribution, one using the outcome
mean functions and the other using the propensity score via $\widetilde{Y}$.
}

\subsection{Parametric structural models and identification results}

In clinical settings, the parametric models of the HTE are desirable
due to their easy interpretation. These models offer a transparent
way of describing how the treatment effect varies across patients'
characteristics and can be used to tailor the treatment to an individual's
characteristics \citep{chakraborty2013statistical}. We make the following
parametric structural assumptions.

\begin{Assumption}[Parametric structural models]\label{asmp:HTE}

The HTE and confounding functions are 
\begin{equation}
\tau(X)=\tau_{\varphi_{0}}(X),\ \ \ \lambda(X)=\lambda_{\phi_{0}}(X),\label{eq:SNMM}
\end{equation}
where $\tau_{\varphi}(X)$ and $\lambda_{\phi}(X)$ are known continuous
functions of $\varphi\in\Theta_{1}$ and $\phi\in\Theta_{2}$, $\varphi_{0}$
and $\phi_{0}$ are unique but unknown values, and $\Theta_{1}$ and
$\Theta_{2}$ are compact sets in $\R^{p_{1}}$ and $\R^{p_{2}},$
respectively. 

\end{Assumption}

{When $X$ is discrete and models are saturate, the
structural models are nonparametric.} The continuity of $\tau_{\varphi}(X)$
and $\lambda_{\phi}(X)$ and compactness of $\Theta_{1}$ and $\Theta_{2}$
are imposed for identification and are standard in the literature.
The treatment effect model $\tau_{\varphi_{0}}(X)$ is a special case
of structural nested mean models \citep{robins1994correcting} with
a single treatment. \citet{lu2014asimplemethod} and \citet{vansteelandt2014structural}
considered a linear treatment effect $\tau_{\varphi_{0}}(X)=X^{\T}{\varphi_{0}}$,
where the first component of $X$ is one, specifying an intercept
term. This model entails that on average, the treatment would increase
the mean of the outcome by $X^{\T}{\varphi_{0}}$, and
the magnitude of the increase depends on $X$. Moreover, each component
of ${\varphi_{0}}$ quantifies the magnitude of the treatment
effect of each $X$. Assume that higher values are indicative of better
outcomes. If $X^{\T}{\varphi_{0}}>0$, it indicates that
the treatment is beneficial for the subject with $X$. Other flexible
models can also be considered, such as single-index models \citep{song2017semiparametric}
and multiple-index models \citep{chen2011single}. Modeling $\lambda_{\phi_{0}}(X)$
follows the large sensitivity analysis literature \citep{robins2000sensitivity},
which typically requires domain knowledge to identify the possible
unmeasured confounders and their relationships with the observed data.

Nonparametric identification, established in (\ref{eq:id1}) and (\ref{eq:id2}),
leads to identification of $\psi_{0}=(\varphi_{0}^{\T},\phi_{0}^{\T})^{\T}$
under Assumption \ref{asmp:HTE}: 
\begin{align}
\varphi_{0} & =\arg\min_{\varphi\in\Theta_{1}}\E[S\{\mu_{1}(X,S=1)-\mu_{0}(X,S=1)-\tau_{\varphi}(X)\}^{2}],\label{eq:id3}\\
\phi_{0} & =\arg\min_{\phi\in\Theta_{2}}\E[(1-S)\{\mu_{1}(X,S=0)-\mu_{0}(X,S=0)-\lambda_{\phi}(X)-\tau_{\varphi_{0}}(X)\}^{2}],\label{eq:id4}
\end{align}
and $\varphi_{0}$ and $\phi_{0}$ are the unique values that satisfy
(\ref{eq:id3}) and (\ref{eq:id4}). With model misspecification,
$\tau_{\varphi_{0}}(X)$ and $\lambda_{\phi_{0}}(X)$ can be interpreted
as the best approximations of $\tau(X)$ and $\lambda(X)$ in the
overlap population \citep{li2016balancing}; see Remark \ref{rmk:proj}.
A substantial body of literature has advocated for the use of parametric
structural models and projection-based interpretations \citep[e.g.,][]{neugebauer2005prefer,chernozhukov2018generic,kennedy2019robust,kennedy2021semiparametric}. 

\subsection{Direct estimators and the need for improved estimators}

Proposition \ref{prop:id} gives two identification strategies and
motivates two direct estimators. To construct the direct estimators,
let the adjusted outcomes be $Y_{i}^{\adj,1}=\widehat{\mu}_{1}(X_{i},S_{i}=1)-\widehat{\mu}_{0}(X_{i};S_{i}=1)\}$
and $Y_{i}^{\adj,2}=Y_{i}\{A_{i}-\widehat{e}(X_{i},S_{i})\}/\allowbreak[\widehat{e}(X_{i},S_{i})\{1-\widehat{e}(X_{i},S_{i})\}]$,
where $\widehat{\mu}_{a}(X,S)$ and $\widehat{e}(X,S)$ are estimators
of $\mu_{a}(X,S)$ and $e(X,S)$ for $a=0,1$. For $k=1$ or $2,$
one can then fit the adjusted outcome $Y_{i}^{\adj,k}$ with mean
$\tau_{\varphi}(X)$$+(1-S)\lambda_{\phi}(X)$ to obtain the direct
estimators. However, the two direct estimators require either the
correct specification of the outcome mean function or the propensity
score. One can use flexible semiparametric or nonparametric models
to estimate the two nuisance functions; however, the corresponding
direct estimators will suffer from a slower rate of convergence due
to the slower-rate of convergence of the nuisance function estimators
\citep{chernozhukov2018double}. This calls for the construction of
more principled estimators that provide more attractive statistical
properties. It is well-known that estimators constructed based on
efficient scores are doubly robust in the sense that they are consistent
if either one of the parametric models for the nuisance functions
is correctly specified \citep{robins1994estimation}. More recently,
many authors have shown that doubly robust estimators possess a ``rate-double
robustness'' property \citep{rotnitzky2019characterization} in the
sense that they retain a root-$N$ convergence rate under weaker conditions
on consistent but otherwise flexible semi-/non-parametric models of
the nuisance functions \citep{chernozhukov2018double}. In the next
section, we derive the semiparametric efficiency score for $\tau_{\varphi}(X)$
and $\lambda_{\phi}(X)$ to motivate a new estimator. 

\section{Semiparametric efficiency theory for $\tau_{\varphi_{0}}(X)$ and
$\lambda_{\phi_{0}}(X)$ \label{sec:Main-Theory}}

\subsection{Semiparametric models with conditional moment restrictions }

{{}Our semiparametric model consists of structural models (\ref{eq:SNMM}),
Assumption \ref{Asump:transp}, and other unspecified components of
the likelihood function. We show that Assumption \ref{Asump:transp}
imposes restrictions on the structural parameters of interest.}{

}

{{}For the trial participants ($S=1$), the transportability assumption
of $\tau(X)$ and trial design lead to 
\begin{equation}
\E[Y\mid A=1,X,S=1]-\mu_{0}(X,S=1)=\tau(X).\label{eq:H1}
\end{equation}
Moreover, for the observational participants ($S=0$), the transportability
assumption of $\tau(X)$ and definition of $\lambda(X)$ lead to 
\begin{equation}
\E[Y\mid A=1,X,S=0]-\mu_{0}(X,S=0)=\tau(X)+(1-S)\lambda(X).\label{eq:H0}
\end{equation}
Combining (\ref{eq:H1}) and (\ref{eq:H0}) results in 
\begin{equation}
\E[Y\mid A=1,X,S]=\tau(X)+(1-S)\lambda(X)+\mu_{0}(X,S).\label{eq:H10}
\end{equation}
Based on (\ref{eq:H10}), the key insight is to introduce 
\begin{equation}
H_{\psi_{0}}=Y-\{\tau_{\varphi_{0}}(X)+(1-S)\lambda_{\phi_{0}}(X)\}A,\label{eq:def of H(k)-1}
\end{equation}
which enjoys a mean exchangeability property of $\E[H_{\psi_{0}}\mid A,X,S]=\mu_{0}(X,S)$
and consequently a conditional moment restriction.}

\begin{Proposition}[Conditional moment restriction]\label{prop:CondIndep}

Under Assumptions \ref{Asump:transp} and \ref{asmp:HTE}, for $H_{\psi_{0}},$
we have 
\begin{equation}
\E[H_{\psi_{0}}\mid A,X,S]=\E[H_{\psi_{0}}\mid X,S]=\mu_{0}(X,S).\label{eq:model-part}
\end{equation}
\end{Proposition}

\subsection{Semiparametric efficient score \label{subsec:semi_effscore}}

The likelihood function based on a single variable $V$ is $\mathcal{L}(\psi_{0},\theta;V)=f(\epsilon_{H,\psi_{0}}\mid A,X,S)f(A\mid X,S)f(X,S),$
where 
\begin{equation}
\epsilon_{H,\psi_{0}}=H_{\psi_{0}}-\E[H_{\psi_{0}}\mid X,S]=Y-\mu(X,S)-\{\tau_{\varphi_{0}}(X)+(1-S)\lambda_{\phi_{0}}(X)\}\{A-e(X,S)\},\label{eq:RoR}
\end{equation}
and $\theta$ is a infinite-dimensional nuisance parameter. The general
geometric approach of \citet{bickel1993efficient} to obtaining the
efficient score requires deriving the nuisance tangent space $\Lambda$
of $\theta$ and the projection of the score function of $\psi_{0}$
onto $\Lambda^{\bot}$ , the orthogonal complement space of $\Lambda.$
This task is non-trivial because Assumption \ref{Asump:transp} imposes
restrictions on $\psi_{0}$ by (\ref{eq:model-part}) or equivalently
$\E[\epsilon_{H,\psi_{0}}\mid A,X,S]=0$. To resolve this challenge,
following \citet{robins1994correcting}, we will translate the restrictions
directly into the observed data likelihood function, leading to an
unconstrained likelihood function of $\psi_{0}$, see {\eqref{eq:full-data lik}},
and finally the efficient score $S_{\psi_{0}}(V)$. {A
detailed roadmap and relevant propositions are provided in }$\mathsection${\ref{subsec:proof-thm1}
to illustrate the derivation. }

\begin{theorem}[Semiparametric efficient score of $\psi_0$]\label{Thm: semipar}

Suppose Assumptions \ref{Asump:transp} and \ref{asmp:HTE} hold.
The efficient score of $\psi_{0}$ is 
\begin{equation}
{S_{\psi_{0}}(V)=\left(\begin{array}{c}
\frac{\partial\tau_{\varphi_{0}}(X)}{\partial\varphi}\\
(1-S)\frac{\partial\lambda_{\phi_{0}}(X)}{\partial\phi}
\end{array}\right)\left(A-\E\left[AW\mid X,S\right]\E[W\mid X,S]^{-1}\right)W\epsilon_{H,\psi_{0}}},\label{eq:effEE}
\end{equation}
where $W=\{\sigma_{A}^{2}(X,S)\}^{-1}$, and $V_{\eff}=(\E[S_{\psi_{0}}(V)\allowbreak S_{\psi_{0}}^{\T}(V)])^{-1}$
is the semiparametric efficiency bound.

\end{theorem}

{{}Theorem \ref{Thm: semipar} provides a benchmark for gauging
the efficiency of estimators of $\psi_{0}$. The efficient score in
(\ref{eq:effEE}) depends on the unknown distribution through the
nuisance functions $\vartheta=(e,\mu,\sigma_{a}^{2})$, indicating
$e(X,S)$, $\mu(X,S)$, and $\sigma_{a}^{2}(X,S)$, respectively.
%\textcolor{black}{{} because
%$\E\left[AW\mid X,S\right]=\{\sigma_{1}^{2}(X,S)\}^{-1}e(X,S)$ and
%$\E[W\mid X,S]=\{\sigma_{1}^{2}(X,S)\}^{-1}e(X,S)+\{\sigma_{0}^{2}(X,S)\}^{-1}\{1-e(X,S)\}$}.
To facilitate the estimation of $\psi_{0}$, we approximate the nuisance
functions by flexible semiparametric or nonparametric models and solve
the estimating equation of $\psi_{0}$ with the approximated nuisance
functions{{} based on the observed data}. The variance
function $\sigma_{a}^{2}(X,S)$ can be estimated by fitting a model
for $\log\{Y_{i}-\widehat{\mu}_{a}(X_{i},S_{i})\}^{2}$ against $X_{i}$,
separately for the treatment group and data source, and transforming
the fitted models to the exponential scale. In the simulation study,
we adopt the super learner \citep{van2007super}, with the candidate
learners including generalized linear models, generalized additive
models, and multivariate adaptive regression splines, which can be
carried out using off-the-shelf software, e.g., the ``SuperLearner''
function with specified algorithms in R. To emphasize the dependence
on the nuisance functions $\vartheta$, we write $S_{\psi}(V)$ in
(\ref{eq:effEE}) as $S_{\psi}(V;\vartheta)$. The proposed estimator
$\widehat{\psi}=(\widehat{\varphi}^{\T},\widehat{\phi}^{\T})^{\T}$
solves $\bbP_{N}S_{\psi}(V;\widehat{\vartheta})=0$ with $e(X,S)$,
$\mu(X,S)$, and $\sigma_{a}^{2}(X,S)$ replaced by their estimators
$\widehat{e}(X,S)$, $\widehat{\mu}(X,S)$, and $\widehat{\sigma}_{a}^{2}(X,S)$,
respectively.}

{The decomposition (\ref{eq:RoR}) has been utilized in different
contexts, including partially linear models \citep{robinson1988root,chernozhukov2018double},
structural nested mean models \citep{robins2004optimal,yang2021semiparametric},
causal random forest \citep{athey2019generalized}, and R-learner
of the HTE \citep{nie2021quasi}. This particular formulation leads
to a desirable statistical property of $S_{\psi_{0}}(V)$ known as
Neyman orthogonality, which in turn results in the rate-double robustness
property of $\widehat{\psi}$ concerning $\widehat{e}(X,S)$ and $\widehat{\mu}(X,S)$.
In the next section, we will delve further into studying this property.
}

\section{Asymptotic property\label{sec:Asymptotic-properties-rate}}

\subsection{Robustness to slower rates for nuisance functions\label{subsec:robustness}}

To protect the estimator from model misspecification, suppose $\widehat{\vartheta}=(\widehat{e},\widehat{\mu},\widehat{\sigma}_{a}^{2})$
include general semi-/non-parametric estimators $\widehat{e}(X,S)$,
$\widehat{\mu}(X,S)$, and $\widehat{\sigma}_{a}^{2}(X,S)$. Let $\vartheta_{0}$
be the probability limit of $\widehat{\vartheta}$. For a vector $v$,
we use $\Vert v\Vert_{2}=(v^{\top}v)^{1/2}$ to denote its Euclidean
norm. For a function $g(V)$, denote $\bbP g(V)=\int g(v)\de\bbP(v)$
and the $L_{2}$-norm $\Vert g(V)\Vert=\left\{ \int g(v)^{2}\de\mathbb{P}(v)\right\} ^{1/2}$.
For simplicity of the exposition, we assume the sample size ratio
$n/m\rightarrow p\in(0,1)$, as $n\rightarrow\infty$, so that the
asymptotic regime is the same for $n\rightarrow\infty,$ $m\rightarrow\infty$
or $N=n+m\rightarrow\infty$. We discuss the case with $n/m\rightarrow0$
in Remark \ref{rmk:mHuge}. Suppose that $\lVert\widehat{e}(X,S)-e(X,S)\rVert=o_{\bbP}(N^{-\alpha_{e}})$
and $\lVert\widehat{\mu}(X,S)-\mu(X,S)\rVert=o_{\bbP}(N^{-\alpha_{\mu}})$.
The following theorem summarizes the regularity conditions and asymptotic
properties of $\widehat{\psi}$. {The proof is relegated
to $\mathsection$\ref{sec:proof-thm2}. }

\begin{theorem}[Rate-double robustness]\label{thm:dr2}

Suppose the assumptions in Theorem \ref{Thm: semipar} hold. Assume
further the following regularity conditions hold:

\begin{condition}\label{cond:globalidentification} $\psi_{0}$ is
the unique solution to $\pr S_{\psi}(V;\vartheta_{0})=0$, and for
any sequence $\psi_{n},$ $||\pr S_{\psi_{n}}(V;\allowbreak\vartheta_{0})||_{2}\rightarrow0$
implies $||\psi_{n}-\psi_{0}||_{2}\rightarrow0$. \end{condition}

\begin{condition}\label{cond:partialDeri} (i) $\bbP_{N}\partial S_{\psi}(V;\vartheta)/\partial\psi$
exists and converges uniformly for $\psi$ and $\vartheta$ in the
neighborhoods of their true values, and (ii) $\Psi=\E[\partial S_{\psi_{0}}(V;\vartheta_{0})/\partial\psi]$
is non-singular. \end{condition}

\begin{condition}\label{cond:donsker} $S_{\psi_{0}}(V;\widehat{\vartheta})$
and $S_{\psi_{0}}(V;\vartheta_{0})$ belong to a Donsker class of
functions \citep{van1996weak}.\end{condition}

\begin{condition}\label{cond:cons4nuis} $|\partial\tau_{\varphi_{0}}(X)/\partial\varphi^{\T}|$,
$|\partial\lambda_{\phi_{0}}(X)/\partial\phi^{\T}|$, and $\{\widehat{\sigma}_{a}^{2}(X,S)\}^{-1}$
are uniformly bounded. \end{condition}

\begin{condition}\label{cond:rate4nuis} The convergence rates of
$\widehat{e}(X,S)$ and $\widehat{\mu}(X,S)$ satisfy $\alpha_{e}\geq1/4$
and $\alpha_{e}+\alpha_{\mu}\geq1/2.$\end{condition}

Then, we have $||\widehat{\psi}-\psi_{0}||=o_{\bbP}(1)$, and 
\begin{equation}
N^{1/2}(\widehat{\psi}-\psi_{0})\rightarrow\mathcal{N}\left\{ 0,\Sigma_{\Psi_{0}}=(\Psi^{-1})^{\T}\E[S_{\psi_{0}}(V;\vartheta_{0})^{\otimes2}]\Psi^{-1}\right\} ,\label{eq:asypV}
\end{equation}
in distribution, as $N\rightarrow\infty$. Moreover, if $\sum_{a=0}^{1}||\widehat{\sigma}_{a}^{2}(X,S)-\sigma_{a}^{2}(X,S)||=o_{\bbP}(1)$,
the asymptotic variance of $\widehat{\psi}$ achieves the semiparametric
efficiency bound $V_{\eff}$ in Theorem \ref{Thm: semipar}.

\end{theorem}

We discuss the implications of the regularity conditions. Condition
\ref{cond:globalidentification} is the identifiability condition.
Condition \ref{cond:partialDeri} is standard in the Z-estimation
literature \citep{van1998asymptotic}. {The Donsker
class condition in Condition \ref{cond:donsker} requires that the
nuisance functions should not be too complex without imposing independence
between the estimated nuisance functions and the data.} We refer the
interested readers to $\mathsection$4.2 of \citet{kennedy2016semiparametric}
for a thorough discussion of Donsker classes of functions. Relaxing
this condition is possible by using the sample splitting and cross
fitting technique for estimation \citep{chernozhukov2018double}.
See $\mathsection$S6.2 in the supplementary material for technical
details and empirical evidence. Condition \ref{cond:rate4nuis} requires
$\widehat{e}(X,S)$ and $\widehat{\mu}(X,S)$ to converge to $e(X,S)$
and $\mu(X,S)$ at the rates that make the remaining term in the empirical
process negligible; namely, 
\begin{align*}
||\bbP S_{\psi_{0}}(V;\widehat{\vartheta})||^{2} & \preceq||\widehat{e}(X,S)-e(X,S)||\times\{||\widehat{\mu}(X,S)-\mu(X,S)||+||\widehat{e}(X,S)-e(X,S)||\}\\
 & =o_{\bbP}(N^{-1/2}),
\end{align*}
where ``$A\preceq B$'' denote that $A$ is bounded by a constant
times $B$. Thus, the convergence rate of $\widehat{e}(X,S)$ should
be $o_{\bbP}(N^{-1/4})$, and the convergence rate of $\widehat{\mu}(X,S)$
combined with that of $\widehat{e}(X,S)$ should be $o_{\bbP}(N^{-1/2})$.
In general, there exist different combinations of convergence rates
of $\widehat{e}(X,S)$ and $\widehat{\mu}(X,S)$ that result in a
negligible error bound accommodating different smoothness conditions
of the underlying true nuisance functions, leading to the ``rate-double
robustness'' of $\widehat{\psi}$.{{} This result differs
from the mixed bias property of influence functions in \citet{rotnitzky2021characterization}.
However,} Condition \ref{cond:rate4nuis} appears similarly in the
R-learner of \citet{nie2021quasi} due to the similar residual formulation
as mentioned in %the end of 
$\mathsection$\ref{sec:Main-Theory}. 

Importantly, the consistency and asymptotic normality of $\widehat{\psi}$
do not require $\widehat{\sigma}_{a}^{2}(X,S)$ to be consistent for
$\sigma_{a}^{2}(X,S)$ but the efficiency of $\widehat{\psi}$ does.
For variance estimation of $\widehat{\psi}$, we approximate the variance
formula in (\ref{eq:asypV}) by replacing the analytical components
with their estimated counterparts, and the expectations with the empirical
averages.

\begin{remark}\label{rmk:mHuge}

{The asymptotic covariance matrix of the HTE estimator $\widehat{\varphi},$
denoted by $\Sigma_{\varphi_{0}}$, can be obtained from the upper
$p_{1}\times p_{1}$ block of $\Sigma_{\psi_{0}}$. This matrix allows
us to investigate the impact of the distribution of $S$ on $\widehat{\psi}$
and the potential efficiency gain due to data integration. Toward
this end, define $r_{A}^{2}=\left(A-\E\left[AW\mid X,S\right]\E[W\mid X,S]^{-1}\right)^{2}W$,
\begin{align*}
 & \Gamma_{1,\rct}=\E\left[S\left\{ \frac{\partial\tau_{\varphi_{0}}(X)}{\partial\varphi}\right\} ^{\otimes2}r_{A}^{2}\right], &  & \Gamma_{12}=\E\left[(1-S)\frac{\partial\tau_{\varphi_{0}}(X)}{\partial\varphi}\frac{\partial\lambda_{\phi_{0}}(X)}{\partial\phi^{\T}}r_{A}^{2}\right],\\
 & \Gamma_{1,\rwe}=\E\left[(1-S)\left\{ \frac{\partial\tau_{\varphi_{0}}(X)}{\partial\varphi}\right\} ^{\otimes2}r_{A}^{2}\right], &  & \Gamma_{2}=\E\left[(1-S)\left\{ \frac{\partial\lambda_{\phi_{0}}(X)}{\partial\phi}\right\} ^{\otimes2}r_{A}^{2}\right].
\end{align*}
Using these notations and some algebra ($\mathsection$S3 of the supplementary
material), we have $\Sigma_{\varphi_{0}}=\left(\Gamma_{1,\rct}+\Gamma_{1,\rwe}-\Gamma_{12}\Gamma_{2}^{-1}\Gamma_{12}^{\T}\right)^{-1}$.
We then obtain a general result that holds even when $n/m\rightarrow0$:
\[
N^{1/2}(\Gamma_{1,\rct}+\Gamma_{1,\rwe}-\Gamma_{12}\Gamma_{2}^{-1}\Gamma_{12}^{\T})^{1/2}(\widehat{\varphi}-\varphi_{0})\rightarrow\mathcal{N}(0,I_{p_{1}\times p_{1}}),
\]
where $I_{p_{1}\times p_{1}}$ is a $p_{1}\times p_{1}$ identity
matrix. }{

}

{The result sheds light on the advantages of using observational
studies for possible efficiency gains in treatment effect estimation.
The components $\Gamma_{1,\rct}$ and $\Gamma_{1,\rwe}-\Gamma_{12}\Gamma_{2}^{-1}\Gamma_{12}^{\T}$
in the precision matrix of $\widehat{\varphi}$ depend on the trial
sample ($S=1$) and the observational sample $(S=0)$, respectively.
If $\Gamma_{1,\rwe}-\Gamma_{12}\Gamma_{2}^{-1}\Gamma_{12}^{\T}$ is
nonzero, the precision of $\widehat{\varphi}$ is improved by using
the observational sample. The next subsection establishes the conditions
for achieving efficiency gains through the use of observational studies.
}

\end{remark}

\subsection{Efficiency gain of the treatment effect estimation by using the observational
studies}

{{}We now discuss the advantages of data integration. The trial
data grant a consistent estimator of $\varphi_{0}$. Under Assumptions
\ref{Asump:transp} and \ref{asmp:HTE}, following the same strategy
in $\mathsection$\ref{sec:Main-Theory} for the trial sample, the
efficient score of $\varphi_{0}$ is $S_{\rct,\varphi_{0}}(V;\vartheta)=S\{\partial\tau_{\varphi_{0}}(X)/\partial\varphi\}(A-\E\left[AW\mid X,S\right]\E[W\mid X,S]^{-1})W\epsilon_{H,\psi_{0}}$.
Then, the trial estimator $\widehat{\varphi}_{\rct}$ can be obtained
by solving $\mathbb{P}_{N}S_{\rct,\varphi}(V;\widehat{\vartheta})=0.$}

Theorem \ref{thm:efficiency gain} shows that combining trial and
observational studies has the advantage of gaining efficiency in the
estimation of $\varphi_{0}${.}

\begin{theorem}[Efficiency gain by combining trial and observational studies]\label{thm:efficiency gain}Suppose
the assumptions in Theorem \ref{thm:dr2} hold. The asymptotic variance
of $\widehat{\varphi}$ is equal to or less than the asymptotic variance
of $\widehat{\varphi}_{\rct}$, where the equality holds if 
\begin{eqnarray}
\frac{\partial\tau_{\varphi_{0}}(X_{i})}{\partial\varphi} & = & M\frac{\partial\lambda_{\phi_{0}}(X_{i})}{\partial\phi}\label{eq:condeff}
\end{eqnarray}
for some constant matrix $M$. Moreover, the gain in the asymptotic
precision, i.e., the inverse of the asymptotic variance, is 
\begin{equation}
\{\V_{a}(\widehat{\varphi})\}^{-1}-\{\V_{a}(\widehat{\varphi}_{\rct})\}^{-1}=m\times\left(\Omega_{\varphi\varphi}-\Omega_{\varphi\phi}\Omega_{\phi\phi}^{-1}\Omega_{\varphi\phi}^{\T}\right)\geq0,\label{eq:precision gain}
\end{equation}
where $\V_{a}$ denotes the asymptotic variance, $\Omega_{ab}$ is
a covariance matrix for $a,b\in\{\varphi,\phi\}$,{{}
and recalling $m$ is the sample size of the observational study.}

\end{theorem}Exact expressions of $\Omega_{ab}$ for $a,b\in\{\varphi,\phi\}$
are provided in the supplementary material. To gain intuition about
Theorem \ref{thm:efficiency gain}, it is helpful to discuss two scenarios.
If $\lambda_{\phi_{0}}(X)$ is known, $S_{\psi}(V;\vartheta)$ uses
the additional observational data$(1-S)\{\partial\tau_{\varphi}(X)/\partial\varphi\}(A-\E\left[AW\mid X,S\right]\E[W\mid X,S]^{-1})\{\sigma_{A}^{2}(X,S)\}^{-1}\epsilon_{H,\psi}$
for estimating $\varphi_{0}$, comparing with $S_{\rct,\varphi}(V;\vartheta)$;
therefore, $\widehat{\varphi}$ gains precision over $\widehat{\varphi}_{\rct}$.
Next, because $\lambda_{\phi_{0}}(X)$ is unknown, the estimation
of $\phi_{0}$ and $\varphi_{0}$ competes for the information in
the observational study. When (\ref{eq:condeff}) holds, the terms
in $\lambda_{\phi_{0}}(X)$ and that in $\tau_{\varphi_{0}}(X)$ are
collinear, and all observational data are used to estimate $\phi_{0}$.
In this case, $\widehat{\psi}$ and $\widehat{\psi}_{\rct}$ have
the same asymptotic precision. When (\ref{eq:condeff}) does not hold,
the terms in $\lambda_{\phi_{0}}(X)$ and that in $\tau_{\varphi_{0}}(X)$
are not entirely linearly dependent, and the observational data are
used to estimate both $\phi_{0}$ and $\varphi_{0}$. In this case,
$\widehat{\psi}$ gains precision over $\widehat{\psi}_{\rct}$, the
magnitude of the gain increases with the observational sample size.

\begin{remark}\label{rmk:proj} The integrative framework shows the
efficiency benefits of combining the trial and observational samples
over using only the trial sample under Assumptions \ref{Asump:transp}
and \ref{asmp:HTE}. {Drawn on the semiparametric
theory, the projection parameter $(\varphi_{0},\phi_{0})$ results
from a projection of the structural models on a constrained model
space, enjoys good theoretical properties in terms of consistency
and asymptotic efficiency, and is identical to the true model parameters
when the posited parametric models are correct.} If the underpinning
assumptions for the observational sample are violated, potential biases
may offset the efficiency benefits. Thus, it is important to scrutinize
the required assumptions in practice. 

When the putative models for $\tau(X)$ and $\lambda(X)$ are misspecified,
$\tau_{\varphi_{0}}(X)$ and $\lambda_{\phi_{0}}(X)$ are different
from the true estimands, and thus the estimators are biased for $\tau(X)$
and $\lambda(X)$. However, $\tau_{\varphi_{0}}(X)$ and $\lambda_{\phi_{0}}(X)$
can be interpreted as the best approximations of $\tau(X)$ and $\lambda(X)$
in the sense of 
\begin{eqnarray*}
(\varphi_{0},\phi_{0}) & = & \arg\min_{\varphi,\phi}\E_{W}[\omega(X,S)[\tau(X)-\tau_{\varphi}(X)+(1-S)\{\lambda(X)-\lambda_{\phi}(X)]^{2}],
\end{eqnarray*}
where $\E_{W}[g(V)\mid X,S]=\E[g(V)W\mid X,S]/\E[W\mid X,S]$ for
any $g(V)$ and $\omega(X,S)=\E_{W}[A\mid X,S](1-\E_{W}[A\mid X,S])$
is the overlap weight \citep{li2016balancing}; see a proof in $\mathsection$S2.
Additional simulations under model misspecification of $\tau(x)$
and $\lambda(x)$ are provided in $\mathsection$S6.3 of the supplementary
material, which confirms the above statement. 

In practice, a goodness-of-fit test can also be developed to assess
the adequacy of the structural models using over-identification restrictions;
see \citet{yang2015gof} and also $\mathsection$S4. 

\end{remark}

\section{Improve average treatment effect estimation\label{sec:Improve-ATE-Estimation}}

The HTE characterizes individual variations of the treatment effect,
while the average treatment effect $\tau_{0}$ summarizes the treatment
effect for the target patient population at large. Because the trial
assigns treatments randomly to the participants, $\tau_{\varphi_{0}}(X)$
is identifiable and can be estimated. However, due to the inclusion
and exclusion criteria for recruiting patients, the patient composition
in the trial may be different from the target population; i.e., $f(X\mid S=1)$
is different from $f(X)$ in general. Consequently, $\E[\tau_{\varphi_{0}}(X)\mid S=1]$
is different from $\tau_{0}$, and the estimator using the trial data
only is biased of $\tau_{0}$ generally. On the other hand, the observational
sample is conceivably more representative of the real patient population
because of the real-world data collection mechanisms. %Formally, we
%formulate the following assumption.

\begin{Assumption}\label{Asump:srs}$f(X\mid S=0)=f(X)$.

\end{Assumption}

We allow the support of $f(X\mid S=1)$ and $f(X)$ to be different,
and hence we allow the trial sample and the observational sample to
have non-overlapping covariate distributions. 

A byproduct of the proposed framework is the identification of $\tau_{0}$. 

\begin{Proposition}[Identification of $\tau_{0}$]\label{prop:2-1}

Under Assumptions \ref{Asump:transp} and \ref{Asump:srs}, $\tau_{0}$
is identified by $\tau_{0}=\E[\tau(X)]=\E[\tau(X)\mid S=0]$, where
$\tau(X)$ is identified by (\ref{eq:id1}). 

\end{Proposition}

The semiparametric efficient score of $\tau_{0}$ is presented in
the following theorem.

\begin{theorem}\label{Thm:ses of att}Suppose Assumptions \ref{Asump:transp}--\ref{Asump:srs}
hold. The semiparametric efficient score of $\tau_{0}$ is

\begin{equation}
S_{\tau_{0}}(V)=\frac{1-S}{\pi_{0}}\{\tau_{\varphi_{0}}(X)-\tau_{0}\}+\E\left[\left.\frac{\partial\tau_{\varphi_{0}}(X)}{\partial\varphi^{\T}}\right\vert S=0\right]S_{\varphi_{0}}(V),\label{eq:EIFatt}
\end{equation}
where $\pi_{0}=\bbP(S=0)$ and $S_{\varphi_{0}}(V)$ is the efficient
score of $\varphi_{0}$, i.e, the first $p_{1}$ components of $S_{\psi_{0}}(V)$
in (\ref{eq:effEE}).

\end{theorem}

Recall that $\varphi_{0}$ is the parameter in the HTE, $\phi_{0}$
is the parameter in the confounding function, and $\psi_{0}=(\varphi_{0}^{\T},\phi_{0}^{\T})^{\T}$
is the combined vector of parameters. From (\ref{eq:effEE}), $S_{\varphi_{0}}(V)$
depends on $\phi_{0}$ in general. Although $\tau_0$ depends only on
$\varphi_{0}$ in $\tau_{\varphi_{0}}(X)$, $S_{\tau_{0}}(V)$ can
depend on $\phi_{0}$ through $S_{\varphi_{0}}(V)$. From Theorem
\ref{Thm:ses of att}, the observational sample not only provides
a representative covariate distribution of the target population but
also can contribute to the estimation efficiency of $\tau_{0}$. {Under
Assumptions \ref{Asump:transp} and \ref{Asump:srs}, one can derive
the nonparametric efficiency score of $\tau_{0}$. However, this approach
solely utilizes the covariate distribution from the observational
sample to adjust for the selection bias present in the trial sample,
without incorporating outcome data from the observational sample.}

Once we obtain $\widehat{\varphi}$, a simple plug-in estimator of
$\tau_{0}$ is $\widehat{\tau}=m^{-1}\sum_{i=1}^{N}(1-S_{i})\tau_{\widehat{\varphi}}(X_{i})$.
The following theorem shows the rate-double robustness and local efficiency
of $\widehat{\tau}$.

\begin{theorem}\label{Thm:att-est}

Suppose the assumptions in Theorem \ref{thm:dr2} and Assumption \ref{Asump:srs}
hold. Then, 
\begin{equation}
N^{1/2}(\widehat{\tau}-\tau_{0})\rightarrow\mathcal{N}\left(0,V_{\tau_{0}}\right),\label{eq:asypV-1}
\end{equation}
in distribution, as $N\rightarrow\infty$, where 
\[
V_{\tau_{0}}=\frac{1}{\pi_{0}}\V\left[\tau_{\varphi_{0}}(X)\mid S=0\right]+\Psi_{0}^{\T}\Psi^{-1,\T}\E[S_{\psi_{0}}(V;\vartheta_{0})^{\otimes2}]\Psi^{-1}\Psi_{0},
\]
and $\Psi_{0}=\E[\partial\tau_{\varphi_{0}}(X)/\partial\varphi\mid S=0]$.
Moreover, the asymptotic variance of $\widehat{\tau}$ achieves the
semiparametric efficiency bound.

\end{theorem}

\section{Simulation study\label{sec:Simulation}}

{{}We conduct a simulation study to evaluate the finite sample performance
of the proposed estimators of the HTE and $\tau_{0}$. We follow a
similar strategy in \citet{kallus2018removing} to generate the data.
We first generate the trial data with sample size $n=1000$. For the
trial sample, we sample the covariates $X$ from the superpopulation,
where $X_{j}\sim$ ${\rm N}(0,1)$ and $j=1,\dots,5$, with the sampling
probability $p(X)$, where $\text{logit}\{p(X)\}=X_{1}/2-X_{2}$,
to characterize a difference between the patient composition in the
trial and the target population, and we generate $A\mid(X,S=1)\sim$Ber$({0.5})$
and $Y(a)\mid(X,S=1)=a\tau(X)+(X_{1}+X_{2}X_{3}/4-X_{4}X_{5}/4)+\exp(X_{1}/4-X_{2}/4)\epsilon(a)$,
where $\tau(X)=1+X_{1}+X_{1}^{2}+X_{2}+X_{2}^{2}$ and $\epsilon(a)\sim$
${\rm N}(0,1)$, for $a=0,1$. We then generate the observational
data with sample size $m=5000$. We sample the covariates $X$ directly
from the superpopulation, where $X_{j}\sim$ ${\rm N}(0,1)$ and $j=1,\dots,5$,
and we generate $A\mid(X,S=0)\sim$ Ber$\{e(X,0)\}$, where logit$\{e(X,0)\}=-(X_{1}X_{2}/3+X_{3}X_{4}/3-X_{5})$,
and $Y(a)\mid(X,S=0)=a\tau(X)+(X_{1}+X_{2}X_{3}/4-X_{4}X_{5}/4)+(X^{\T}\beta)U+\exp(X_{1}/4-\allowbreak X_{2}/4)\epsilon(a)$,
where $U$ is a latent variable and $\epsilon(a)\sim$ ${\rm N}(0,1)$,
for $a=0,1$. We generate $U$ according to a pattern mixture model
$U\mid(X,A,S=0)\sim{\rm N}(A-1/2,1)$, and thus the confounding function
is $\lambda(X)=\mu_{1}(X,S=0)-\mu_{0}(X,S=0)-\tau(X)=X^{\T}\beta$.
We consider two settings: Setting 1 with $\beta=0\times(1,\ldots,1)^{\T},$
in which $U$ does not confound the relationship between $A$ and
$Y$, and Setting 2 with $\beta=(1,\ldots,1)^{\T},$ such that $U$
is an unmeasured confounder of $A$ and $Y$. We consider four estimators:
i) $\widehat{\varphi}_{\rct}$ using only the randomized controlled
trial data, ii) }{}$\widehat{\varphi}_{\meta},$ meta-analysis of
the combined trial and observational studies by regressing the inverse
probability of treatment weighting adjusted outcome $Y_{i}^{\adj}=\{\widehat{e}(X_{i},S_{i})\}^{-1}A_{i}Y_{i}-\{1-\widehat{e}(X_{i},S_{i})\}^{-1}(1-A_{i})Y_{i}$
on $(1,X_{1},X_{1}^{2},X_{2},X_{2}^{2})$,{{} iii) $\widehat{\varphi}$,
the proposed integrative estimator{} with the correctly specified
variance model estimated by the linear regression, and iv) $\widehat{\varphi}_{\text{v1}}$,
the integrative estimator with the misspecified variance model estimated
by $1$. The rationale for $\widehat{\varphi}_{\meta}$ is that{}
it uses weighting adjusting for measured confounders and in the absence
of unmeasured confounders, $\E[Y_{i}^{\adj}\mid X_{i},S_{i}]=\tau_{\varphi_{0}}(X_{i})$.
For all estimators, we estimate the nuisance functions via the super
learner, including candidate learners as generalized linear models,
generalized additive models, and multivariate adaptive regression
splines. The $95\%$ confidence interval is calculated via parametric-t
wild bootstrap as $\Big(\tau_{\widehat{\varphi}}(x)-c^{*}\mathbb{\hat{\mathbb{V}}}^{1/2}\big\{\tau_{\widehat{\varphi}}(x)\big\},\tau_{\widehat{\varphi}}(x)+c^{*}\mathbb{\mathbb{\hat{\mathbb{V}}}}^{1/2}\big\{\tau_{\widehat{\varphi}}(x)\big\}\Big)$
at some specific values of $x$, where $c^{*}$ is the $95\%$ quantile
of the bootstrap t-values $\big\{|T^{*(b)}|:b=1,\cdots,B\big\}$ and
$T^{*(b)}=\big\{\tau_{\widehat{\varphi}^{(b)}}(x)-\tau_{\widehat{\varphi}}(x)\big\}\big/\mathbb{\mathbb{\hat{\mathbb{V}}}}^{1/2}\big\{\tau_{\widehat{\varphi}^{(b)}}(x)\big\}$
in each bootstrap iteration. We select the bootstrap size $B=500$.}{

}

{{}Table \ref{tab:sim1point} reports results for point estimation
for $\tau(x)$ at various values of $x$ and $\tau_{0}$, where Integrative
denotes the integrative estimator $\tau_{\widehat{\varphi}}(x)$ and
Integrative0 denotes $\tau_{\widehat{\varphi}_{\text{v1}}}(x)$. In
Setting 1 without unmeasured confounding in the observational study,
$\tau_{\widehat{\varphi}_{\meta}}(x)$ shows bias due to the use of
the flexible modeling strategy to approximate the propensity score.
Among all the estimators, $\tau_{\widehat{\varphi}}(x)$ has a smaller
variance than $\tau_{\widehat{\varphi}_{\meta}}(x)$ by capitalizing
on semiparametric efficiency theory; $\tau_{\widehat{\varphi}}(x)$
has a smaller variance than $\tau_{\widehat{\varphi}_{\rct}}(x)$
by leveraging the confounding function in the observational study.
Although $\tau_{\widehat{\varphi}_{\text{v1}}}(x)$ preserves consistency,
its variation is larger than $\tau_{\widehat{\varphi}}(x)$, indicating
a loss of efficiency due to the variance model misspecification. In
Setting 2 with unmeasured confounding in the observational study,
$\tau_{\widehat{\varphi}_{\meta}}(x)$ assuming no unmeasured confounding
is biased for $\tau(x)$, due to the unmeasured confounding biases
in the observational studies, $\tau_{\widehat{\varphi}_{\rct}}(x)$,
$\tau_{\widehat{\varphi}}(x)$, and $\tau_{\widehat{\varphi}_{\text{v1}}}(x)$
remain unbiased for $\tau(x)$, and $\tau_{\widehat{\varphi}}(x)$
has improved efficiency over $\tau_{\widehat{\varphi}_{\rct}}(x)$
and $\tau_{\widehat{\varphi}_{\text{v1}}}(x)$. From Table \ref{tab:sim1cvg},
the empirical coverage rates for $\tau_{\widehat{\varphi}}(x)$ and
$\tau_{\widehat{\varphi}_{\text{v1}}}(x)$ in both settings with and
without unmeasured confounding in the observational study are close
to the nominal level. For ${\tau_{0}}$, $\hat{\tau}_{\rct}$ is biased
when using only the trial data, as the covariate distribution in the
trial is different from that in the target population. In contrast,
the integrative estimators are consistent by leveraging the representativeness
of the covariate distribution in the observational sample. }{

}

{{}} 
\begin{table}[!ht]
{{}\centering\caption{\label{tab:sim1point}Simulation results for point estimation under
two settings with and without unmeasured confounding in the observational
study, where the biases are scaled by $10^{-2}$ and the variances
are scaled by $10^{-3}$. }
\resizebox{\textwidth}{!}{%%}%
\begin{tabular}{ccccccccccccccccccc}
\hline 
 & \multicolumn{2}{c}{{{}Meta}} & \multicolumn{2}{c}{{{}RCT}} & \multicolumn{2}{c}{{{}Integrative}} & \multicolumn{2}{c}{{{}Integrative0}} &  &  & \multicolumn{2}{c}{{{}Meta}} & \multicolumn{2}{c}{{{}RCT}} & \multicolumn{2}{c}{{{}Integrative}} & \multicolumn{2}{c}{{{}Integrative0}}\tabularnewline
 & {{}Bias }  & {{}Var }  & {{}Bias }  & {{}Var }  & {{}Bias }  & {{}Var }  & {{}Bias }  & {{}Var }  &  &  & {{}Bias }  & {{}Var }  & {{}Bias }  & {{}Var }  & {{}Bias }  & {{}Var }  & {{}Bias }  & {{}Var }\tabularnewline
\hline 
\multicolumn{19}{c}{{{}Setting 1 (without unmeasured confounding in the observational
study)}}\tabularnewline
{\scriptsize{}{}{}{}{}$\tau(-3,0)$}{{} }  & {{}-49 }  & {{}1024 }  & {{}-4 }  & {{}638 }  & {{}4 }  & {{}203 }  & {{}6 }  & {{}262 }  &  & {\scriptsize{}{}{}{}{}$\tau(0,-3)$}{{} }  & {{}110 }  & {{}703 }  & {{}4 }  & {{}828 }  & {{}5 }  & {{}227 }  & {{}3 }  & {{}279 }\tabularnewline
{\scriptsize{}{}{}{}{}$\tau(-1.5,0)$}{{} }  & {{}-26 }  & {{}100 }  & {{}-1 }  & {{}79 }  & {{}-0 }  & {{}45 }  & {{}1 }  & {{}52 }  &  & {\scriptsize{}{}{}{}{}$\tau(0,-1.5)$}{{} }  & {{}54 }  & {{}111 }  & {{}1 }  & {{}98 }  & {{}0 }  & {{}50 }  & {{}-0 }  & {{}58 }\tabularnewline
{\scriptsize{}{}{}{}{}$\tau(1.5,0)$}{{} }  & {{}22 }  & {{}100 }  & {{}1 }  & {{}79 }  & {{}2 }  & {{}45 }  & {{}1 }  & {{}52 }  &  & {\scriptsize{}{}{}{}{}$\tau(0,1.5)$}{{} }  & {{}-59 }  & {{}111 }  & {{}2 }  & {{}98 }  & {{}2 }  & {{}50 }  & {{}2 }  & {{}58 }\tabularnewline
{\scriptsize{}{}{}{}{}$\tau(3,0)$}{{} }  & {{}48 }  & {{}1024 }  & {{}0 }  & {{}638 }  & {{}8 }  & {{}203 }  & {{}7 }  & {{}262 }  &  & {\scriptsize{}{}{}{}{}$\tau(0,3)$}{{} }  & {{}-116 }  & {{}703 }  & {{}6 }  & {{}828 }  & {{}8 }  & {{}227 }  & {{}8 }  & {{}279 }\tabularnewline
{\scriptsize{}{}{}{}{}$\tau(0,0)$}{{} }  & {{}-2 }  & {{}14 }  & {{}0 }  & {{}22 }  & {{}-1 }  & {{}13 }  & {{}-1 }  & {{}14 }  &  & {\scriptsize{}{}{}{}{}$\tau_0$}{{} }  & {{}-2 }  & {{}5 }  & {{}60 }  & {{}17 }  & {{}1 }  & {{}13 }  & {{}1 }  & {{}13 }\tabularnewline
\multicolumn{19}{c}{{{}Setting 2 (with unmeasured confounding in the observational study)}}\tabularnewline
{\scriptsize{}{}{}{}{}$\tau(-3,0)$}{{} }  & {{}-324 }  & {{}1060 }  & {{}-4 }  & {{}638 }  & {{}4 }  & {{}203 }  & {{}7 }  & {{}266 }  &  & {\scriptsize{}{}{}{}{}$\tau(0,-3)$}{{} }  & {{}-95 }  & {{}691 }  & {{}4 }  & {{}828 }  & {{}5 }  & {{}227 }  & {{}3 }  & {{}282 }\tabularnewline
{\scriptsize{}{}{}{}{}$\tau(-1.5,0)$}{{} }  & {{}-157 }  & {{}107 }  & {{}-1 }  & {{}79 }  & {{}-0 }  & {{}44 }  & {{}1 }  & {{}52 }  &  & {\scriptsize{}{}{}{}{}$\tau(0,-1.5)$}{{} }  & {{}-59 }  & {{}108 }  & {{}1 }  & {{}98 }  & {{}0 }  & {{}50 }  & {{}-0 }  & {{}58 }\tabularnewline
{\scriptsize{}{}{}{}{}$\tau(1.5,0)$}{{} }  & {{}141 }  & {{}107 }  & {{}1 }  & {{}79 }  & {{}2 }  & {{}44 }  & {{}1 }  & {{}52 }  &  & {\scriptsize{}{}{}{}{}$\tau(0,1.5)$}{{} }  & {{}79 }  & {{}108 }  & {{}2 }  & {{}98 }  & {{}2 }  & {{}50 }  & {{}2 }  & {{}58 }\tabularnewline
{\scriptsize{}{}{}{}{}$\tau(3,0)$}{{} }  & {{}272 }  & {{}1060 }  & {{}0 }  & {{}638 }  & {{}8 }  & {{}203 }  & {{}7 }  & {{}266 }  &  & {\scriptsize{}{}{}{}{}$\tau(0,3)$}{{} }  & {{}182 }  & {{}691 }  & {{}6 }  & {{}828 }  & {{}8 }  & {{}227 }  & {{}8 }  & {{}282 }\tabularnewline
{\scriptsize{}{}{}{}{}$\tau(0,0)$}{{} }  & {{}-1 }  & {{}15 }  & {{}0 }  & {{}22 }  & {{}-1 }  & {{}13 }  & {{}-1 }  & {{}14 }  &  & {\scriptsize{}{}{}{}{}$\tau_0$}{{} }  & {{}1 }  & {{}6 }  & {{}60 }  & {{}17 }  & {{}1 }  & {{}13 }  & {{}1 }  & {{}13 }\tabularnewline
\hline 
\end{tabular}{{}}} } 
\end{table}

{

}

{{}} 
\begin{table}[!ht]
{{}\centering \caption{\label{tab:sim1cvg}Simulation results for variance estimation and
coverage rate for the integrative estimator under two settings with
and without unmeasured confounding in the observational study, where
the variances are scaled by $10^{-3}$ and the coverage rates are
scaled by $10^{-2}$.}
\resizebox{\textwidth}{!}{%%}%
\begin{tabular}{ccccccccccc}
\hline 
 & \multicolumn{2}{c}{{{}Integrative}} & \multicolumn{2}{c}{{{}Integrative0}} &  &  & \multicolumn{2}{c}{{{}Integrative}} & \multicolumn{2}{c}{{{}Integrative0}}\tabularnewline
 & {{}Var }  & {{}CVG }  & {{}Var }  & {{}CVG }  &  &  & {{}Var }  & {{}CVG }  & {{}Var }  & {{}CVG }\tabularnewline
\hline 
\multicolumn{11}{c}{{{}Setting 1 (without unmeasured confounding in the observational
study)}}\tabularnewline
{{}$\tau(-3,0)$ }  & {{}203 }  & {{}93.3 }  & {{}262 }  & {{}93.1 }  &  & {{}$\tau(0,-3)$ }  & {{}227 }  & {{}94.4 }  & {{}279 }  & {{}93.9 }\tabularnewline
{{}$\tau(-1.5,0)$ }  & {{}45 }  & {{}94.8 }  & {{}52 }  & {{}94.3 }  &  & {{}$\tau(0,-1.5)$ }  & {{}50 }  & {{}95.1 }  & {{}58 }  & {{}94.7 }\tabularnewline
{{}$\tau(1.5,0)$ }  & {{}45 }  & {{}93.2 }  & {{}52 }  & {{}94.4 }  &  & {{}$\tau(0,1.5)$ }  & {{}50 }  & {{}93.5 }  & {{}58 }  & {{}93.4 }\tabularnewline
{{}$\tau(3,0)$ }  & {{}203 }  & {{}94.1 }  & {{}262 }  & {{}93.1 }  &  & {{}$\tau(0,3)$ }  & {{}227 }  & {{}93.5 }  & {{}279 }  & {{}93.4 }\tabularnewline
{{}$\tau(0,0)$ }  & {{}13 }  & {{}92.6 }  & {{}14 }  & {{}94.2 }  &  & {{}$\tau_0$ }  & {{}13 }  & {{}93.9 }  & {{}13 }  & {{}93.6 }\tabularnewline
\multicolumn{11}{c}{{{}Setting 2 (with unmeasured confounding in the observational study)}}\tabularnewline
{{}$\tau(-3,0)$ }  & {{}203 }  & {{}93.0 }  & {{}266 }  & {{}92.8 }  &  & {{}$\tau(0,-3)$ }  & {{}227 }  & {{}94.7 }  & {{}282 }  & {{}94.3 }\tabularnewline
{{}$\tau(-1.5,0)$ }  & {{}44 }  & {{}95.0 }  & {{}52 }  & {{}94.1 }  &  & {{}$\tau(0,-1.5)$ }  & {{}50 }  & {{}95.2 }  & {{}58 }  & {{}94.6 }\tabularnewline
{{}$\tau(1.5,0)$ }  & {{}44 }  & {{}93.6 }  & {{}52 }  & {{}94.6 }  &  & {{}$\tau(0,1.5)$ }  & {{}50 }  & {{}93.1 }  & {{}58 }  & {{}93.3 }\tabularnewline
{{}$\tau(3,0)$ }  & {{}203 }  & {{}94.0 }  & {{}266 }  & {{}93.0 }  &  & {{}$\tau(0,3)$ }  & {{}227 }  & {{}92.9 }  & {{}282 }  & {{}93.0 }\tabularnewline
{{}$\tau(0,0)$ }  & {{}13 }  & {{}93.8 }  & {{}14 }  & {{}93.8 }  &  & {{}$\tau_0$ }  & {{}13 }  & {{}93.6 }  & {{}13 }  & {{}93.6 }\tabularnewline
\hline 
\end{tabular}{{}}} } 
\end{table}

{

}

\section{Real data application\label{sec:Real-data-application}}

We apply the proposed estimators to evaluate the effect of adjuvant
chemotherapy for early-stage resected non-small-cell lung cancer using
the CALGB 9633 trial data and a large clinical oncology observational
database -- the national cancer database. The CALGB 9633 trial used
a set of patient eligibility criteria, including disease stage, age,
performance status, resection methods, to enroll patients. In the
trial sample, $319$ patients were randomly assigned to observation
versus chemotherapy, resulting $163$ on observation, $A=0$, and
$156$ on chemotherapy, $A=1$. The same set of patient eligibility
criteria defines the target patient population and is used to select
the comparable patients in the national cancer database. The comparable
observational sample consists of $15166$ patients diagnosed with
the same disease between years 2004 -- 2016 in stage IB disease with
$10903$ on observation and $4263$ received chemotherapy after surgery.
As the treatments for the trial patients were randomly assigned and
the treatments for the observational patients were chosen by physicians
and patients, the numbers of treated and controls are relatively balanced
in the trial sample while they are unbalanced in the observational
sample. The outcome $Y$ is the indicator of cancer recurrence within
three years after the surgery.

We are interested in estimating the heterogeneous treatment effects
of chemotherapy varying by tumor size. The original trial analysis
did not show any clinical improvement for chemotherapy, possibly because
of its small sample size \citep{strauss2008adjuvant}. Some exploratory
analysis, however, showed that tumor size might modify the treatment
effect and that patients with larger tumor sizes may benefit more
from the chemotherapy (\citealp{strauss2008adjuvant}, \citealp{speicher2015adjuvant},
\citealp{morgensztern2016adjuvant}). Thus, we formulate the HTE of
interest to be $\tau_{\varphi}(X)=\varphi_{1}+\varphi_{2}\mbox{tumor size}^{*}+\varphi_{3}(\mbox{tumor size}^{*})^{2}$,
where $\text{tumor size}^{*}$ standardizes $\text{tumor size}$ by
subtracting the mean $4.8$ and dividing the standard error $1.7$,
and $\varphi=(\varphi_{1},\varphi_{2},\varphi_{3})^{\T}$. In the
analysis, we include five covariates to adjust for in both samples:
age, tumor size, sex, histology, and race, and we use generalized
additive models for approximating the nuisance functions. Table \ref{tab:covbal-2}
reports the covariate means by treatment group in the two samples.
Due to treatment randomization, all covariates are balanced between
the treated and the control in the trial sample. While due to a lack
of treatment randomization, some covariates are highly unbalanced
in the observational sample. It can be seen that older patients with
smaller tumor sizes and histology are likely to choose a conservative
treatment, on observation. Moreover, we cannot rule out the possibility
of unmeasured confounders in the observational sample. To formulate
the confounding function, possible unmeasured confounders include
disease status at diagnosis, financial status, and accessibility to
health care facilities that affect the decision of receiving adjuvant
chemotherapy after surgery and clinical outcomes (\citealp{speicher2015adjuvant,yang2016role,morgensztern2016adjuvant,speicher2017traveling}).
The \textcolor{black}{linear} confounding function $\lambda_{\phi}(X)$
includes age, tumor size, gender, race, histology, Charlson co-morbidity
score, income level, insurance coverage, and travel range to large
health care facilities \textcolor{black}{as predictors}.

\begin{figure}
\centering

\includegraphics[width=0.6\textwidth]{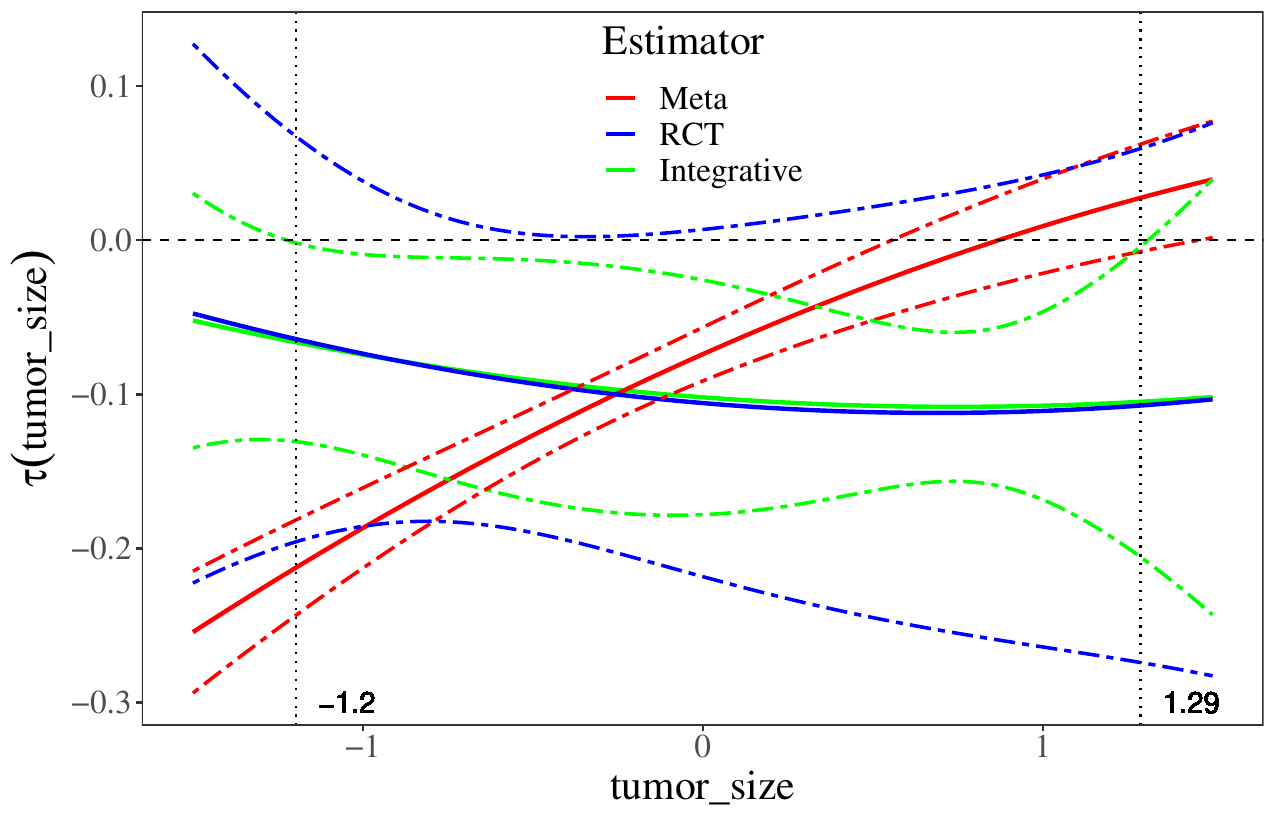}

\vspace{0.5cm}

\caption{\label{fig:app}Estimated treatment effect as a function of tumor
size$^{*}$. The solid lines represent the estimators, and the bands
represent the solid lines $\pm1.96$ standard errors of the estimators.}
\end{figure}

We compare the trial, Meta, and integrative estimators. Figure \ref{fig:app}
displays the estimated treatment effect as a function of tumor size$^{*}$.
Table \ref{tab:covbal-2} reports the results for \citet{lu2014asimplemethod}
the estimated parameters. Due to the small sample size, the trial
estimator is not statistically significant. By pooling all information
from the trial and observational sample, the Meta and integrative
estimators gain efficiency and both show that the tumor size is a
significant treatment effect modifier. Interestingly, the two combining
approaches produce different conclusions. The difference between the
Meta and integrative estimators may be attributable to the no unmeasured
confounding. The Meta estimators assume that there are no unmeasured
confounders, while the integrative estimators take into account the
possible unmeasured confounders in the observational sample. The results
in the supplementary material show that age, gender and histology
are significant in the confounding function, suggesting that the no
unmeasured confounding assumption is not plausible in the observational
sample. For the integrative estimator, we carry out the over-identification
restrictions test to assess the goodness-of-fit of $\tau_{\psi}(X)$
and $\lambda_{\phi}(X)$. The test is directed at the alternative
model specifications $\tau_{\psi}^{{\rm alt}}(X)=$ a quadratic function
of age$^{*}$ and tumor size$^{*}$, and $\lambda_{\phi}^{{\rm alt}}(X)=\text{\ensuremath{\lambda_{\phi}(X)}}$
augmented with $($tumor size$^{*})^{2}$. The test statistics is
$5.5$ with p-value $0.14$ based on a $\chi_{3}^{2}$ null reference
distribution. Therefore, there is no strong evidence to reject the
model specifications of $\tau_{\psi}(X)$ and $\lambda_{\phi}(X)$
in this application. From the integrative approach, chemotherapy has
significant benefits for patients with tumor size in $[-0.71,1.2]\times1.7+4.8=[3.6,6.8]$cm.

\begin{table}[!ht]
\centering\caption{\label{tab:covbal-2}Sample sizes and covariate means by $A$ in the
trial and observational samples.}

\resizebox{\textwidth}{!}{%
\begin{tabular}{lccccccc}
\hline 
 &  &  & Age  & Tumor Size  & Sex  & Histology  & Race \tabularnewline
 &  &  &  &  & (1=Male)  & (1=Squamous)  & (1=White)\tabularnewline
 & $N$  & $A$  & (years)  & (cm)  & (1/0)  & (1/0)  & (1/0)\tabularnewline
\hline 
Trial  & 156  & $A$=1  & 60.6  & 4.62  & 64.1\%  & 40.4\%  & 90.4\%\tabularnewline
 & 163  & $A$=0  & 61.1  & 4.57  & 63.8\%  & 39.3\%  & 88.3\%\tabularnewline
Observational study  & 4263  & $A$=1  & 63.9  & 5.19  & 54.3\%  & 35.6\%  & 88.6\%\tabularnewline
 & 10903  & $A$=0  & 69.4  & 4.67  & 54.8\%  & 40.5\%  & 90.0\%\tabularnewline
\hline 
\end{tabular}} 
\end{table}

\section{Proofs of the main results\label{sec:proofs}}

{As Assumptions \ref{Asump:transp} and \ref{asmp:HTE}
induce the conditional moment restriction, it becomes crucial to take
into account such constraint when obtaining the semiparametric efficient
score and assessing the asymptotic properties of the integrative estimator.
In this section, we provide a detailed derivation of $S_{\psi_{0}}(V)$
in Theorem \ref{Thm: semipar} and the rate-double robustness in Theorem
\ref{thm:dr2}. Those explorations establish the main theoretical
contribution of the paper.}

\subsection{Proof of Theorem \ref{Thm: semipar} \label{subsec:proof-thm1}}

{We present a roadmap and Propositions \ref{Thm: The-nuisance-tangent}--\ref{thm: tangent bot},
whose proofs have been included in the supplementary material, to
facilitate the construction of the semiparametric efficient score
$S_{\psi_{0}}(V)$ in Theorem \ref{Thm: semipar}.}

\subsubsection{A roadmap}

We consider a regular asymptotically linear (RAL) estimator $\widehat{\psi}$
of $\psi_{0}$: 
\begin{equation}
N^{1/2}(\widehat{\psi}-\psi_{0})=N^{-1/2}\bbP_{N}\iif(V)+o_{\bbP}(1),\label{eq:RAL-1}
\end{equation}
where $\iif(V)$ is the influence function of $\widehat{\psi}$, which
has zero mean and finite and nonsingular variance. By (\ref{eq:RAL-1}),
the asymptotic variance of $N^{1/2}(\widehat{\psi}-\psi_{0})$ is
equal to the variance of $\iif(V)$. Consider the Hilbert space $\mathcal{H}$
of all $p$-dimensional, mean-zero finite variance squared integrable
functions of $V$, $h(V)$, equipped with the covariance inner product
$<h_{1},h_{2}>=\E[h_{1}(V)^{\T}h_{2}(V)]$ and the $\mathcal{L}_{2}$-norm
$||h||^{2}=\E[h(V)^{\T}h(V)]<\infty$. To construct the efficient
estimator for $\psi_{0}$, we follow the geometric approach of \citet{bickel1993efficient}
to derive the semiparametric efficient score for $\psi_{0}$ following
the road map below.

The density function of a single variable $V=(A,X,Y,S)$ is $f(V)=f(Y\mid A,X,S)f(A\mid X,S)\allowbreak f(X,S)$.
The parameter of interest $\psi_{0}$ satisfies restriction (10) with
$H=H_{\psi_{0}}$, and the nuisance parameter is the nonparametric
density functions $f(Y\mid A,X,S)$, $f(A\mid X,S),$ and $f(X,S)$.
In order to incorporate restriction \eqref{eq:model-part} into the
likelihood function directly, we consider an equivalent re-parameterization,
and re-express the semiparametric likelihood function; see (\ref{eq:full-data lik}).
Based on the likelihood function, we characterize the nuisance tangent
space $\Lambda$ in the Hilbert space $\mathcal{H}$; see Proposition
 \ref{Thm: The-nuisance-tangent}. We then express $\Lambda$ as a
direct sum of orthogonal subspaces; see Proposition  \ref{Thm: direct sum of nuis tang space}.
This effort will be valuable in characterizing the orthogonal complement
space of the nuisance tangent space $\Lambda^{\bot}$, which consists
of all influence functions; see Proposition  \ref{thm: tangent bot}.
The semiparametric efficient score of $\psi_{0}$ is thus derived
as the projection of the score of $\psi_{0}$ onto $\Lambda^{\bot}$;
see Theorem \ref{Thm: semipar}.

\subsubsection{Re-parameterization of likelihood function}

We consider an equivalent re-parameterization, in order to incorporate
restriction \eqref{eq:model-part} into the likelihood function directly.
Toward that end, we decompose $H$ as follows: 
\begin{equation}
H=\underbrace{H-\E[H\mid A,X,S]}_{\epsilon_{H}=\epsilon_{H}(H,A,X,S)}+\underbrace{\E[H\mid A,X,S]-\E[H]}_{Q=Q(A,X,S)}+\E[H],\label{eq:reparamter}
\end{equation}
where $\E[\epsilon_{H}\mid A,X,S]=0$, $\E[Q]=0$, and $\epsilon_{H}$
and $Q$ are squared integrable. Note that ``squared integrable''
is a technical condition to ensure that the nuisance score vectors
lie in the Hilbert space $\mathcal{H}$. Then, the semiparametric
model defined by restriction \eqref{eq:model-part} is equivalent
to the following re-parameterization 
\begin{equation}
H=\epsilon_{H}+q(X,S)-\E[q(X,S)]+\E[H],\ \E\left[\epsilon_{H}\mid A,X,S\right]=0.\label{eq:reparamter1}
\end{equation}
On the one hand, if restriction \eqref{eq:model-part} holds, it implies
$Q$ depends only on $(X,S)$, but not on $A$. Because $\E[Q]=0$,
we can then express $Q=q(X,S)-\E[q(X,S)]$ with $q(X,S)$ a squared
integrable function of $(X,S)$, so the re-parameterization (\ref{eq:reparamter1})
exists. On the other hand, if $H$ can be expressed in (\ref{eq:reparamter1}),
$H$ satisfies the restriction \eqref{eq:model-part}.

We can write the likelihood function based on a single variable $V$
as 
\begin{eqnarray}
\mathcal{L}(\psi,\theta;V) & = & f(Y\mid A,X,S)f(A\mid X,S)f(X,S)\nonumber \\
 & = & f(\epsilon_{H}\mid A,X,S)f(A\mid X,S)f(X,S)\frac{\partial\epsilon_{H}}{\partial Y}\nonumber \\
 & = & f(\epsilon_{H}\mid A,X,S)f(A\mid X,S)f(X,S),\label{eq:full-data lik}
\end{eqnarray}
where the last equality follows from 
\begin{eqnarray*}
\epsilon_{H} & = & Y-\{\tau_{\varphi_{0}}(X)+(1-S)\lambda_{\phi_{0}}(X)\}A\\
 &  & -\E[H]-\left\{ q(X,S)-\int q(X,S)f(X,S)\de\nu(X,S)\right\} ,
\end{eqnarray*}
with $q(X,S)$ a nonparametric function of $(X,S)$. Because $\E[\epsilon_{H}\mid A,X,S]=0$,
we require $\int\epsilon_{H}\allowbreak f(\epsilon_{H}\mid A,X,S)\de\nu(\epsilon_{H})=0$,
where $\nu(\cdot)$ is a generic measure. After re-parameterization,
the nuisance parameter becomes the infinite dimensional set $\theta$
consisting of $f(\epsilon_{H}\mid A,X,S)$, $f(A\mid X,S),$ $f(X,S)$,
$\E[H]$, and $q(X,S)$.

We assume all the regularity conditions to ensure the existence of
the efficient score function of $\psi_{0}$ are satisfied, which are
mainly continuity conditions for the parameter and the semiparametric
model; e.g., we need $\psi=\psi(\theta)$ to be pathwise differentiable
with respect to $\theta$ \citep{bickel1993efficient,tsiatis2007semiparametric}.
These conditions are not restrictive for a typical application problem.

To distinguish nuisance parameters, we re-write the likelihood function
as 
\begin{equation}
\mathcal{L}(\psi_{0},\theta;V)=f_{1}(\epsilon_{H}\mid A,X,S)f_{5}(A\mid X,S)f_{3}(X,S),\label{eq:semiparlik}
\end{equation}
where 
\vspace{-1em}
\begin{eqnarray*}
\epsilon_{H} & = & Y-\{\tau_{\varphi_{0}}(X)+(1-S)\lambda_{\phi_{0}}(X)\}A -c_{4}-\left\{ q_{2}(X,S)-\int q_{2}(X,S)f_{3}(X,S)\de\nu(X,S)\right\} ,
\end{eqnarray*}
and $\theta=(\theta_{1},\ldots,\theta_{5})$ consists of the nuisance
parameters $\theta_{1}=f_{1}(\epsilon_{H}\mid A,X,S),$ $\theta_{2}=q_{2}(X,S)$,
$\theta_{3}=f_{3}(X,S)$, $\theta_{4}=c_{4}$, and $\theta_{5}=f_{5}(A\mid X,S).$
{Then, $\epsilon_{H}=\epsilon_{H}(\psi_{0},\theta_{2},\theta_{3},\theta_{4})$
depends on the parameter of interest $\psi_{0}$ and the nuisance
parameters $(\theta_{2},\theta_{3},\theta_{4})$.} This order for
indexing the nuisance parameters makes the characterization of the
nuisance tangent space easier.

Propositions \ref{Thm: The-nuisance-tangent} and \ref{Thm: direct sum of nuis tang space}
present the characterizations of the nuisance tangent space and its
orthogonal complement, respectively. The proofs are presented in $\mathsection$S1
of the supplementary material.

\begin{Proposition}\label{Thm: The-nuisance-tangent}The nuisance
tangent space corresponding to $\theta=(\theta_{1},\ldots,\theta_{5})$
is 
\[
\Lambda=\Lambda^{(1)}+\Lambda^{(2)}+\Lambda^{(3)}+\Lambda^{(4)}+\Lambda^{(5)},
\]
where $\Lambda^{(j)}$ is the nuisance tangent space with respect
to $\theta_{j}$, for $j=1,\ldots,5$. Define $\Lambda^{*}=\{\Gamma^{*}=\Gamma^{*}(X,S)\in\R^{p}:\E[\Gamma^{*}]=0\}$
and $S_{\epsilon}=S_{\epsilon}(\epsilon_{H},A,X,S)=\partial\log f_{1}(\epsilon_{H}\mid A,X,S)/\partial\epsilon_{H}\in\R^{1}$
evaluated at the truth. Then,

\begin{eqnarray*}
\Lambda^{(1)} & = & \{\Gamma^{(1)}=\Gamma^{(1)}(\epsilon_{H},A,X,S)\in\R^{p}:\E[\Gamma^{(1)}\mid A,X,S]=0,\ \text{and}\ \E[\Gamma^{(1)}\epsilon_{H}\mid A,X,S]=0\},\\
\Lambda^{(2)} & = & \{\Gamma^{(2)}=\Gamma^{(2)}(\epsilon_{H},A,X,S)=\Gamma^{(2)}(\Gamma^{*})=\Gamma^{*}S_{\epsilon}\in\R^{p}:\Gamma^{*}\in\Lambda^{*}\},\\
\Lambda^{(3)} & = & \{\Gamma^{(3)}=\Gamma^{(3)}(\epsilon_{H},A,X,S)=\Gamma^{(3)}(\Gamma^{*})=\Gamma^{*}+\E[Q\Gamma^{*}]S_{\epsilon}\in\R^{p}:\Gamma^{*}\in\Lambda^{*}\},\\
\Lambda^{(4)} & = & \{cS_{\epsilon}:S_{\epsilon}=S_{\epsilon}(\epsilon_{H},A,X,S),c\in\R^{p}\},\\
\Lambda^{(5)} & = & \left\{ \Gamma^{(5)}=\Gamma^{(5)}(A,X,S)\in\R^{p}:\E[\Gamma^{(5)}(A,X,S)\mid X,S]=0\right\} .
\end{eqnarray*}
Here and throughout in a slight abuse of notation, we use $\Gamma^{(2)}(\cdot)$
and $\Gamma^{(3)}(\cdot)$ as functions of $(\epsilon_{H},A,X,S)$
and also as operators on $\Gamma^{*}$, but their meaning should be
clear in the context.

\end{Proposition}

\begin{remark}\label{rmk: 1oplus3}It is important to note that $\Gamma^{(5)}(A,X,S)$
with $\E[\Gamma^{(5)}(A,X,S)\mid X,S]=0$ is orthogonal to all other
subspaces in $\Lambda$. \end{remark}

For simplicity, we define the following notation.

\begin{definition} \label{Def1}Let 
\begin{eqnarray}
W & = & W(A,X,S)=(\V[\epsilon_{H}\mid A,X,S])^{-1},\label{eq:Wmk}\\
T & = & T(X,S)=\E[W\mid X,S],\label{eq:Tmk}\\
\epsilon_{0} & = & \epsilon_{0}(\epsilon_{H},A,X,S)=\E[W\mid X,S]^{-1}W\epsilon_{H}+Q,\label{eq:epsilon}\\
T^{*} & = & \E[T^{-1}]=\E[\E[W\mid X,S]^{-1}].\label{eq:T*mk}
\end{eqnarray}
\end{definition}

We now express $\Lambda$ as a direct sum of orthogonal subspaces.
This effort will be valuable in characterizing the orthogonal complement
space of the nuisance tangent space $\Lambda^{\bot}$.

\begin{Proposition}\label{Thm: direct sum of nuis tang space} 
The
space $\Lambda$ can be written as a direct sum of orthogonal subspaces:
\begin{equation}
\Lambda=\widetilde{\Lambda}^{(1)}\oplus\widetilde{\Lambda}^{(2)}\oplus\widetilde{\Lambda}^{(3)}\oplus\widetilde{\Lambda}^{(4)}\oplus\widetilde{\Lambda}^{(5)},\label{eq:orthog}
\end{equation}
where $\oplus$ denotes a direct sum, and using the notation in Proposition
\ref{Thm: The-nuisance-tangent} and Definition \ref{Def1}, $\widetilde{\Lambda}^{(1)}=\Lambda^{(1)},$

\begin{eqnarray}
\widetilde{\Lambda}^{(2)} & = & \left\{ \widetilde{\Gamma}^{(2)}=\widetilde{\Gamma}^{(2)}(\Gamma^{*})=\Gamma^{*}W\epsilon_{H}:\Gamma^{*}\in\Lambda^{*}\right\} ,\label{eq:tilde2}\\
\widetilde{\Lambda}^{(3)} & = & \left\{ \widetilde{\Gamma}^{(3)}=\widetilde{\Gamma}^{(3)}(\Gamma^{*})=\Gamma^{*}-\E[Q\Gamma^{*}](T^{*}T)^{-1}W\epsilon_{H}:\Gamma^{*}\in\Lambda^{*}\right\} ,\label{eq:tilde3}\\
\widetilde{\Lambda}^{(4)} & = & \left\{ \widetilde{\Gamma}^{(4)}=c\epsilon_{0}:c\in\R^{p}\right\} ,\label{eq:tilde4}\\
\widetilde{\Lambda}^{(5)} & = & \left\{ \widetilde{\Gamma}^{(5)}=\Gamma^{(5)}(A,X,S):\E[\Gamma^{(5)}(A,X,S)\mid X,S]=0\right\} .\label{eq:tilde5}
\end{eqnarray}

\end{Proposition}

\begin{Proposition}\label{thm: tangent bot}Suppose Assumptions \ref{Asump:transp}
and \ref{asmp:HTE} hold. The space of the influence function space
of $\psi_{0}$ is 
\begin{equation}
\Lambda^{\bot}=\left\{ G(A,X,S;\psi_{0},c)=c(A,X,S)\epsilon_{H,\psi_{0}}:\E[c(A,X,S)\mid X,S]=0\right\} .\label{eq:lambda-bot}
\end{equation}

\end{Proposition}

\subsubsection{Proof of Theorem \ref{Thm: semipar}}

Based on Proposition \ref{thm: tangent bot}, we show that the projection
of any $B\in\mathcal{H}$, $\prod\left[B\mid\Lambda^{\bot}\right]$,
is of the form $c(A,X,S)\epsilon_{H,\psi_{0}}$, where $\E[c(A,X,S)\mid X,S]=0$.
Let the score vector of $\psi_{0}$ be $s_{\psi_{0}}(V)$. Then, the
semiparametric efficient score is the projection of $s_{\psi_{0}}(V)$
onto $\Lambda^{\bot}$, given by 
\begin{multline*}
S_{\psi_{0}}(V)=\prod\left[s_{\psi_{0}}(V)\mid\Lambda^{\bot}\right]=\left(\E[s_{\psi_{0}}(V)\epsilon_{H,\psi_{0}}\mid A,X,S]\right.\\
\left.-\E[\E[s_{\psi_{0}}(V)\epsilon_{H,\psi_{0}}\mid A,X,S]W\mid X,S]\E[W\mid X,S]^{-1}\right)W\epsilon_{H,\psi_{0}}\coloneqq c^{*}(A,X,S)\epsilon_{H,\psi_{0}}.
\end{multline*}
To evaluate $c^{*}(A,X,S)$ further, we note that $\E[\epsilon_{H,\psi_{0}}\mid A,X,S]=0$.
We differentiate this equality with respect to $\psi_{0}$. By the
generalized information equality \citep{newey1990semiparametric},
we have $\E[-\partial\epsilon_{H,\psi_{0}}/\partial\psi\mid A,X,S]+\E[s_{\psi_{0}}(V)\epsilon_{H,\psi_{0}}\mid A,X,S]=0$.
Therefore, ignoring the negative sign, we have $c^{*}(A,X,S)$ as
given by 
\begin{eqnarray*}
c^{*}(A,X,S) & = & \left(\E\left[\frac{\partial\epsilon_{H,\psi_{0}}}{\partial\psi^{\T}}\mid A,X,S\right]-\E\left[\frac{\partial\epsilon_{H,\psi_{0}}}{\partial\psi^{\T}}W\mid X,S\right]\E[W\mid X,S]^{-1}\right)W\\
 & = & \left(\begin{array}{c}
\frac{\partial\tau_{\varphi_{0}}(X)}{\partial\varphi}\\
(1-S)\frac{\partial\lambda_{\phi_{0}}(X)}{\partial\phi}
\end{array}\right)\left(A-\E\left[AW\mid X,S\right]\E[W\mid X,S]^{-1}\right)W.
\end{eqnarray*}

\subsection{Proof of Theorem \ref{thm:dr2} \label{sec:proof-thm2} }

\subsubsection{Preliminaries }

\textcolor{black}{We introduce more notations and useful results to
prepare for the proof of Theorem \ref{thm:dr2}. Let ``$\rightsquigarrow$''
denote weak convergence, and let ``$A\preceq B$'' denote that $A$
is bounded by a constant times $B$. Denote $\dot{S}_{\psi}(V;\vartheta)=\partial S_{\psi}(V;\vartheta)/\partial\psi.$
Denote a set of nuisance functions as $\mathcal{G}_{\vartheta_{0}}=\{\vartheta:||\vartheta-\vartheta_{0}||<\delta\}$
for some $\delta>0$ and denote $l^{\infty}(\mathcal{G}_{\vartheta_{0}})$
as the collection of all bounded functions $f:\mathcal{G}_{\vartheta_{0}}\rightarrow\R^{p}.$}

\textcolor{black}{The following lemmas show the asymptotic properties
of functions belong to Donsker classes. }

\begin{lemma}\textcolor{black}{\label{lemma:A1} Suppose Conditions
\ref{cond:partialDeri} and \ref{cond:donsker} hold. Then, we have
$
\sup_{\psi\in\Theta,\vartheta\in\mathcal{G}_{\vartheta_{0}}}||\bbP_{N}S_{\psi}(V;$ $\vartheta)-\bbP S_{\psi}(V;\vartheta)||_{2}\rightarrow0
$
in probability as $N\rightarrow\infty$, and 
$
\sup_{\psi\in\Theta,\vartheta\in\mathcal{G}_{\vartheta_{0}}}||\bbP_{N}\dot{S}_{\psi}(V;\vartheta)-\bbP\dot{S}_{\psi}(V;\vartheta)||_{2}\rightarrow0
$
in probability as $N\rightarrow\infty$.}

\end{lemma}

\begin{lemma}\textcolor{black}{\label{lemma:A2} Suppose Conditions
\ref{cond:partialDeri} and \ref{cond:donsker} hold. Then, we have
\[
N^{1/2}(\bbP_{N}-\bbP)S_{\psi_{0}}(V;\vartheta)\rightsquigarrow Z\in l^{\infty}(\mathcal{G}_{\vartheta_{0}}),
\]
where the limiting process $Z=\{Z(\vartheta):\vartheta\in\mathcal{G}_{\vartheta_{0}}\}$
is a mean-zero multivariate Gaussian process, and the sample paths
of $Z$ belong to $\{z\in l^{\infty}(\mathcal{G}_{\vartheta_{0}}):z$
is uniformly continuous with respect to $||\cdot||\}$. }

\end{lemma}

\subsubsection{Proof of Theorem \ref{thm:efficiency gain}}

\textit{\textcolor{black}{First, we show the consistency of $\widehat{\psi}$
.}}\textcolor{black}{{} Toward this end, we show $||\bbP S_{\widehat{\psi}}(V;\vartheta_{0})||_{2}\rightarrow0$.
We bound $||\bbP S_{\widehat{\psi}}(V;\vartheta_{0})||_{2}$ by
\begin{eqnarray}
||\bbP S_{\widehat{\psi}}(V;\vartheta_{0})||_{2} & \leq & ||\bbP S_{\widehat{\psi}}(V;\vartheta_{0})-\bbP S_{\widehat{\psi}}(V;\widehat{\vartheta})||_{2}+||\bbP S_{\widehat{\psi}}(V;\widehat{\vartheta})||_{2}\nonumber \\
 & = & ||\bbP S_{\widehat{\psi}}(V;\vartheta_{0})-\bbP S_{\widehat{\psi}}(V;\widehat{\vartheta})||_{2}+||\bbP S_{\widehat{\psi}}(V;\widehat{\vartheta})-\bbP_{N}S_{\widehat{\psi}}(V;\widehat{\vartheta})||_{2}\nonumber \\
 & \leq & ||\bbP S_{\widehat{\psi}}(V;\vartheta_{0})-\bbP S_{\widehat{\psi}}(V;\widehat{\vartheta})||_{2}+\sup_{\psi\in\Theta,\vartheta\in\mathcal{G}_{\vartheta_{0}}}||\bbP_{N}S_{\psi}(V;\vartheta)-\bbP S_{\psi}(V;\vartheta)||_{2}.\label{eq:A1}
\end{eqnarray}
Both terms in (\ref{eq:A1}) are $o_{\bbP}(1)$ as shown below. By
the Taylor expansion, we have 
\begin{eqnarray*}
||S_{\widehat{\psi}}(V;\vartheta_{0})-S_{\widehat{\psi}}(V;\widehat{\vartheta})||_{2} & = & \left\vert \left\vert \left.\frac{\partial S_{\psi}(V;\vartheta)}{\partial\psi^{\T}}\right\vert _{\psi=\widehat{\psi},\vartheta=\widetilde{\vartheta}}(\widehat{\vartheta}-\vartheta_{0})\right\vert \right\vert _{2}\\
 & \leq & \left\vert \left\vert \left.\frac{\partial S_{\psi}(V;\vartheta)}{\partial\psi^{\T}}\right\vert _{\psi=\widehat{\psi},\vartheta=\widetilde{\vartheta}}\right\vert \right\vert _{2}\times||\widehat{\vartheta}-\vartheta_{0}||_{2},
\end{eqnarray*}
where $\widetilde{\vartheta}$ lies in the segment between $\widehat{\vartheta}$
and $\vartheta_{0}$. By the Cauchy--Schwartz inequality, we have
\begin{eqnarray}
||\bbP S_{\widehat{\psi}}(V;\vartheta_{0})-\bbP S_{\widehat{\psi}}(V;\widehat{\vartheta})||_{2} & \leq & \bbP||S_{\widehat{\psi}}(V;\vartheta_{0})-S_{\widehat{\psi}}(V;\widehat{\vartheta})||_{2}\nonumber \\
 & \leq & \bbP\left\{ \left\vert \left\vert \left.\frac{\partial S_{\widehat{\psi}}(V;\vartheta)}{\partial\vartheta^{\T}}\right\vert _{\vartheta=\widetilde{\vartheta}}\right\vert \right\vert _{2}\times||\widehat{\vartheta}-\vartheta_{0}||_{2}\right\} \nonumber \\
 & \leq & \left\{ \E\left\vert \left\vert \left.\frac{\partial S_{\widehat{\psi}}(V;\vartheta)}{\partial\vartheta^{\T}}\right\vert _{\vartheta=\widetilde{\vartheta}}\right\vert \right\vert _{2}^{2}\right\} ^{1/2}\times\left\{ \E||\widehat{\vartheta}-\vartheta_{0}||_{2}^{2}\right\} ^{1/2}\nonumber \\
 & \preceq & ||\widehat{\vartheta}-\vartheta_{0}||_{2}\nonumber = o_{\bbP}(1).\label{eq:A2}
\end{eqnarray}
By Lemma \ref{lemma:A1}, we have 
\begin{equation}
\sup_{\psi\in\Theta,\vartheta\in\mathcal{G}_{\vartheta_{0}}}||\bbP_{N}S_{\psi}(V;\vartheta)-\bbP S_{\psi}(V;\vartheta)||_{2}\rightarrow0\label{eq:A2-1}
\end{equation}
in probability as $N\rightarrow\infty.$ Plugging (\ref{eq:A2}) and
(\ref{eq:A2-1}) into (\ref{eq:A1}) leads to $||\bbP S_{\widehat{\psi}}(V;\vartheta_{0})||_{2}=o_{\bbP}(1)$.
Now, by Condition 1, $||\widehat{\psi}-\psi_{0}||_{2}=o_{\bbP}(1)$. }

\noindent \textit{\textcolor{black}{Second, we show the asymptotic
distribution of $\widehat{\psi}$. }}\textcolor{black}{By the Taylor
expansion of $N^{1/2}\bbP_{N}S_{\widehat{\psi}}(V;\widehat{\vartheta})=0$,
we have 
\[
0=N^{1/2}\bbP_{N}S_{\psi_{0}}(V;\widehat{\vartheta})+\{\bbP_{N}\dot{S}_{\widetilde{\psi}}(V;\widehat{\vartheta})\}N^{1/2}(\widehat{\psi}-\psi_{0}),
\]
where $\widetilde{\psi}$ lies in the segment between $\widehat{\psi}$
and $\psi_{0}$. By Lemma \ref{lemma:A1}, we have 
\[
\sup_{\psi\in\Theta,\vartheta\in\mathcal{G}_{\vartheta_{0}}}||\bbP_{N}\dot{S}_{\psi}(V;\vartheta)-\bbP\dot{S}_{\psi}(V;\vartheta)||_{2}\rightarrow0
\]
in probability as $N\rightarrow\infty$. Because $\widetilde{\psi}\rightarrow\psi_{0}$
and $\widehat{\vartheta}\rightarrow\vartheta_{0}$, we have 
\[
\bbP_{N}\dot{S}_{\widetilde{\psi}}(V;\widehat{\vartheta})\rightarrow\Psi=\bbP\dot{S}_{\psi_{0}}(V;\vartheta_{0})
\]
in probability as $N\rightarrow\infty$. Thus, we have 
\begin{equation}
N^{1/2}(\widehat{\psi}-\psi_{0})=-\Psi^{-1}N^{1/2}\bbP_{N}S_{\psi_{0}}(V;\widehat{\vartheta})+o_{\bbP}(1).\label{eq:A3-0}
\end{equation}
We express 
\begin{equation}
\bbP_{N}S_{\psi_{0}}(V;\widehat{\vartheta})=(\bbP_{N}-\bbP)S_{\psi_{0}}(V;\widehat{\vartheta})+\bbP S_{\psi_{0}}(V;\widehat{\vartheta}),\label{eq:A3}
\end{equation}
and show that 
\begin{eqnarray}
\bbP S_{\psi_{0}}(V;\widehat{\vartheta}) & = & o_{\bbP}(N^{-1/2}),\label{eq:A3-1}\\
(\bbP_{N}-\bbP)S_{\psi_{0}}(V;\widehat{\vartheta}) & = & (\bbP_{N}-\bbP)S_{\psi_{0}}(V;\vartheta_{0})+o_{\bbP}(N^{-1/2}).\label{eq:A3-2}
\end{eqnarray}
To show (\ref{eq:A3-1}), we denote $c(X,S)=(\partial\tau_{\varphi_{0}}(X)/\partial\varphi^{\T},(1-S)\partial\lambda_{\phi_{0}}(X)/\partial\phi^{\T})^{\T}$
for simplicity and evaluate $\bbP S_{\psi_{0}}(V;\widehat{\vartheta})$
explicitly as 
\begin{eqnarray*}
\bbP S_{\psi_{0}}(V;\widehat{\vartheta}) & = & \E\left[c(X,S)\left(A\{\widehat{\sigma}_{A}^{2}(X,S)\}^{-1}-\{\widehat{\sigma}_{1}^{2}(X,S)\}^{-1}\widehat{e}(X,S)\widehat{\E}[\widehat{W}\mid X,S]^{-1}\widehat{W}\right)\widehat{\epsilon}_{H,\psi_{0}}\right]\\
 & = & \E\left[c(X,S)\left(\{\widehat{\sigma}_{1}^{2}(X,S)\}^{-1}e(X,S)-\{\widehat{\sigma}_{1}^{2}(X,S)\}^{-1}\widehat{e}(X,S)\widehat{\E}[\widehat{W}\mid X,S]^{-1}\widehat{W}\right)\right.\\
 &  & \times\left.[\mu_{0}(X,S)-\widehat{\mu}_{0}(X,S)-(1-S)\lambda_{\phi_{0}}(X)\{e(X,S)-\widehat{e}(X,S)\}]\right]\\
 & = & \E\left[c(X,S)\left(\{\widehat{\sigma}_{1}^{2}(X,S)\}^{-1}\{e(X,S)-\widehat{e}(X,S)\}\right.\right.\\
 &  & \left.-\{\widehat{\sigma}_{1}^{2}(X,S)\}^{-1}\widehat{e}(X,S)\widehat{\E}[\widehat{W}\mid X,S]^{-1}(\widehat{W}-\widehat{\E}[\widehat{W}\mid X,S])\right)\\
 &  & \times\left.[\mu_{0}(X,S)-\widehat{\mu}_{0}(X,S)-(1-S)\lambda_{\phi_{0}}(X)\{e(X,S)-\widehat{e}(X,S)\}]\right]\\
 & = & \E\left[c(X,S)\left(\{\widehat{\sigma}_{1}^{2}(X,S)\}^{-1}\{e(X,S)-\widehat{e}(X,S)\}\right.\right.\\
 &  & \left.-\{\widehat{\sigma}_{1}^{2}(X,S)\}^{-1}\widehat{e}(X,S)\widehat{\E}[\widehat{W}\mid X,S]^{-1}\{\widehat{\sigma}_{1}^{2}(X,S)\}^{-1}\{e(X,S)-\widehat{e}(X,S)\}\right.\\
 &  & \left.+\{\widehat{\sigma}_{1}^{2}(X,S)\}^{-1}\widehat{e}(X,S)\widehat{\E}[\widehat{W}\mid X,S]^{-1}\{\widehat{\sigma}_{0}^{2}(X,S)\}^{-1}\{e(X,S)-\widehat{e}(X,S)\}\right)\\
 &  & \left.\times[\mu_{0}(X,S)-\widehat{\mu}_{0}(X,S)-(1-S)\lambda_{\phi_{0}}(X)\{e(X,S)-\widehat{e}(X,S)\}]\right].
\end{eqnarray*}
Applying the Cauchy--Schwartz inequality and Condition \ref{cond:rate4nuis},
we have 
\begin{eqnarray*}
 &  & ||\bbP S_{\psi_{0}}(V;\widehat{\vartheta})||_{2}\\
 & \leq & \E\left[\left\vert \left\vert c(X,S)\left(\{\widehat{\sigma}_{1}^{2}(X,S)\}^{-1}\{e(X,S)-\widehat{e}(X,S)\}\right.\right.\right.\right.\\
 &  & \left.-\{\widehat{\sigma}_{1}^{2}(X,S)\}^{-1}\widehat{e}(X,S)\widehat{\E}[\widehat{W}\mid X,S]^{-1}\{\widehat{\sigma}_{1}^{2}(X,S)\}^{-1}\{e(X,S)-\widehat{e}(X,S)\}\right.\\
 &  & \left.+\{\widehat{\sigma}_{1}^{2}(X,S)\}^{-1}\widehat{e}(X,S)\widehat{\E}[\widehat{W}\mid X,S]^{-1}\{\widehat{\sigma}_{0}^{2}(X,S)\}^{-1}\{e(X,S)-\widehat{e}(X,S)\}\right)\\
 &  & \left.\left.\left.\times[\mu_{0}(X,S)-\widehat{\mu}_{0}(X,S)-(1-S)\lambda_{\phi_{0}}(X)\{e(X,S)-\widehat{e}(X,S)\}]\right\vert \right\vert _{2}\right].\\
 & \preceq & (\E[\{e(X,S)-\widehat{e}(X,S)\}^{2}]\times\E\left[\{\mu_{0}(X,S)-\widehat{\mu}_{0}(X,S)\}^{2}+\{e(X,S)-\widehat{e}(X,S)\}^{2}\right])^{1/2}\\
 & = & \{||\widehat{\mu}_{0}(X,S)-\mu_{0}(X,S)||\times||\widehat{e}(X,S)-e(X,S)||+||\widehat{e}(X,S)-e(X,S)||^{2}\} =  o_{\bbP}(N^{-1/2}).
\end{eqnarray*}
}
\vspace{-1em}
\noindent \textcolor{black}{To show (\ref{eq:A3-2}), Lemma \ref{lemma:A2}
leads to 
\[
N^{1/2}(\bbP_{N}-\bbP)S_{\psi_{0}}(V;\vartheta)\rightsquigarrow Z\in l^{\infty}(\mathcal{G}_{\vartheta_{0}}),
\]
as $N\rightarrow\infty.$ Combining with the fact that $||\widehat{\vartheta}-\vartheta_{0}||=o_{\bbP}(1)$,
we have 
\[
\left(\begin{array}{c}
N^{1/2}(\bbP_{N}-\bbP)S_{\psi_{0}}(V;\vartheta)\\
\widehat{\vartheta}
\end{array}\right)\rightsquigarrow\left(\begin{array}{c}
Z\\
\vartheta_{0}
\end{array}\right)
\]
in $l^{\infty}(\mathcal{G}_{\vartheta_{0}})\times\mathcal{G}_{\vartheta_{0}}$
as $N\rightarrow\infty$. Define a function $s:l^{\infty}(\mathcal{G}_{\vartheta_{0}})\times\mathcal{G}_{\vartheta_{0}}\mapsto\R^{p}$
by $s(z,\vartheta)=z(\vartheta)-z(\vartheta_{0})$, which is continuous
for all $(z,\vartheta)$ where $\vartheta\mapsto z(\vartheta)$ is
continuous. By Lemma \ref{lemma:A2}, all sample paths of $Z$ are
continuous on $\mathcal{G}_{\vartheta_{0}}$, and thus, $s(z,\vartheta)$
is continuous for $(Z,\vartheta).$ By the Continuous-Mapping Theorem,
\[
s(Z,\widehat{\vartheta})=(\bbP_{N}-\bbP)S_{\psi_{0}}(V;\widehat{\vartheta})-(\bbP_{N}-\bbP)S_{\psi_{0}}(V;\vartheta_{0})\rightsquigarrow s(Z,\vartheta_{0})=0.
\]
Thus, (\ref{eq:A3-2}) holds. Plugging (\ref{eq:A3})--(\ref{eq:A3-2})
into (\ref{eq:A3-0}), we have 
\begin{eqnarray}
N^{1/2}(\widehat{\psi}-\psi_{0}) & = & -\Psi^{-1}N^{1/2}\{(\bbP_{N}-\bbP)S_{\psi_{0}}(V;\vartheta_{0})\}+o_{\bbP}(1).\nonumber \\
 & \rightarrow & \text{\ensuremath{\mathcal{N}}\{0,\ensuremath{(\Psi^{-1})^{\T}\E[S_{\psi_{0}}(V;\vartheta_{0})^{\otimes2}]\Psi^{-1}}\},}\label{eq:asympN}
\end{eqnarray}
in distribution as $N\rightarrow\infty.$ If $\sum_{a=0}^{1}||\widehat{\sigma}_{a}(X,S)-\sigma_{a}(X,S)||=o_{\bbP}(1)$,
$S_{\psi_{0}}(V;\vartheta_{0})$ becomes the efficient score $S_{\psi_{0}}(V)$.
Thus, the asymptotic variance in (\ref{eq:asympN}) achieves the efficiency
bound. This completes the proof of Theorem }\ref{thm:dr2}\textcolor{black}{. }

\section*{Acknowledgment}
This project is supported by the Food and Drug Administration (FDA) of the U.S. Department of Health and Human Services (HHS) as part of a financial assistance award U01FD007934 totaling $\$1,674,013$ over two years funded by FDA/HHS. It is also supported by the National Institute On Aging of the National Institutes of Health under Award Number R01AG06688, totaling $\$1,565,763$ over four years and the National Science Foundation under Award Number SES 2242776, totaling $\$225,000$ over three years. The contents are those of the authors and do not necessarily represent the official views of, nor an endorsement by, FDA/HHS, the National Institutes of Health, or the U.S. Government.

\section*{Supplementary material}

Supplementary material includes proofs
and simulation and application results.
\bibliographystyle{Chicago}
\bibliography{short_ci}       % Bibliography file (usually '*.bib')

%% or include bibliography directly:
% \begin{thebibliography}{}
% \bibitem[\protect\citeauthoryear{???}{???}]{b1}
% \end{thebibliography}

% \include{supp_comb} 
\newpage

\begin{center}
\title{{\Large{}\textbf{Supplementary material for "Data fusion methods for the heterogeneity of treatment effect and confounding function"}}}

\vspace{2em}
%\iffalse 
\author{Shu Yang$^1$, Siyi Liu$^1$, Donglin Zeng$^2$, Xiaofei Wang$^{3}$} 

\vspace{1em}

\maketitle
% \begin{center}
$^{1}$Department of Statistics, North Carolina State University, Raleigh, NC, USA \\
$^{2}$Department of Biostatistics, University of North Carolina, Chapel Hill, NC, USA \\
$^{3}$Department of Biostatistics and Bioinformatics, Duke University, Durham, NC, USA
\end{center}
\spacingset{1.5} 

\pagenumbering{arabic} %reset page counter to 1
\global\long\def\thelemma{\textup{S}\arabic{lemma}}%
\setcounter{equation}{0} 
\global\long\def\theequation{S\arabic{equation}}%
\setcounter{section}{0} 
\global\long\def\thesection{S\arabic{section}}%
\global\long\def\thesubsection{S\arabic{section}.\arabic{subsection}}%
\setcounter{table}{0} 
\global\long\def\thetable{\textup{S}\arabic{table}}%
\setcounter{figure}{0} 
\global\long\def\thefigure{\textup{S}\arabic{figure}}%
\setcounter{theorem}{0} 
\global\long\def\thetheorem{\textup{S}\arabic{theorem}}%

Section \ref{sec:The-semiparametric-efficiency-HTE} develops the
semiparametric efficiency theory of the treatment effect. Sections \ref{sec:Proof-of-Remark1},
\ref{sec:Proof-of-Theorem3}, and \ref{sec:S3} provide the proofs
of Remark in Section 4.2, Theorems 2 and 3, respectively.
Section \ref{sec:S2Goodness-of-fit-test} develops a goodness-of-fit
test of the structural model assumptions based on the over-identification
restrictions.
Section \ref{sec:The-semiparametric-efficiency-att} illustrates the semiparametric theory of the average treatment effect estimation. 
Section \ref{sec:AdditionalSim} provides figures and complementary results from
the simulation study. Section \ref{sec:Application} presents additional
results from the application.

\section{The semiparametric efficiency theory \label{sec:The-semiparametric-efficiency-HTE}}

\subsection{A roadmap}

{We consider a regular asymptotically linear (RAL) estimator
$\widehat{\psi}$ of $\psi_{0}$: 
\begin{equation}
N^{1/2}(\widehat{\psi}-\psi_{0})=N^{-1/2}\bbP_{N}\iif(V)+o_{\bbP}(1),\label{eq:RAL-1}
\end{equation}
where $\iif(V)$ is the influence function of $\widehat{\psi}$, which
has zero mean and finite and nonsingular variance. By (\ref{eq:RAL-1}),
the asymptotic variance of $N^{1/2}(\widehat{\psi}-\psi_{0})$ is
equal to the variance of $\iif(V)$. Consider the Hilbert space $\mathcal{H}$
of all $p$-dimensional, mean-zero finite variance }{squared
integrable functions}{ of $V$, $h(V)$, equipped with the
covariance inner product $<h_{1},h_{2}>=\E[h_{1}(V)^{\T}h_{2}(V)]$
and the $\mathcal{L}_{2}$-norm $||h||^{2}=\E[h(V)^{\T}h(V)]<\infty$.
To construct the efficient estimator for $\psi_{0}$, we follow the
geometric approach of \citet{bickel1993efficient} to derive the semiparametric
efficient score for $\psi_{0}$ following the road map below.}{\par}

The density function of a single variable $V=(A,X,Y,S)$ is $f(V)=f(Y\mid A,X,S)f(A\mid X,S)\allowbreak f(X,S)$.
The parameter of interest $\psi_{0}$ satisfies restriction (10)
with $H=H_{\psi_{0}}$, and the nuisance parameter is the nonparametric
density functions $f(Y\mid A,X,S)$, $f(A\mid X,S),$ and $f(X,S)$.
In order to incorporate restriction (10) into the
likelihood function directly, we consider an equivalent re-parameterization,
and re-express the semiparametric likelihood function; see (\ref{eq:full-data lik}).
Based on the likelihood function, we characterize the nuisance tangent
space $\Lambda$ in the Hilbert space $\mathcal{H}$; see Theorem
\ref{Thm: The-nuisance-tangent}. We then express $\Lambda$ as a
direct sum of orthogonal subspaces; see Theorem \ref{Thm: direct sum of nuis tang space}.
This effort will be valuable in characterizing the orthogonal complement
space of the nuisance tangent space $\Lambda^{\bot}$, which consists
of all influence functions; see Theorem \ref{thm: tangent bot}. The
semiparametric efficient score of $\psi_{0}$ is thus derived as the
projection of the score of $\psi_{0}$ onto $\Lambda^{\bot}$; see
Theorem 1.

\subsection{Re-parameterization of likelihood function}

We consider an equivalent re-parameterization, in order to incorporate
restriction (10) into the likelihood function directly.
Toward that end, we decompose $H$ as follows: 
\begin{equation}
H=\underbrace{H-\E[H\mid A,X,S]}_{\epsilon_{H}=\epsilon_{H}(H,A,X,S)}+\underbrace{\E[H\mid A,X,S]-\E[H]}_{Q=Q(A,X,S)}+\E[H],\label{eq:reparamter}
\end{equation}
where $\E[\epsilon_{H}\mid A,X,S]=0$, $\E[Q]=0$, and $\epsilon_{H}$
and $Q$ are squared integrable. Note that ``squared integrable''
is a technical condition to ensure that the nuisance score vectors
lie in the Hilbert space $\mathcal{H}$. Then, the semiparametric
model defined by restriction (10) is equivalent
to the following re-parameterization 
\begin{equation}
H=\epsilon_{H}+q(X,S)-\E[q(X,S)]+\E[H],\ \E\left[\epsilon_{H}\mid A,X,S\right]=0.\label{eq:reparamter1}
\end{equation}
On the one hand, if restriction (10) holds, it implies
$Q$ depends only on $(X,S)$, but not on $A$. Because $\E[Q]=0$,
we can then express $Q=q(X,S)-\E[q(X,S)]$ with $q(X,S)$ a squared
integrable function of $(X,S)$, so the re-parameterization (\ref{eq:reparamter1})
exists. On the other hand, if $H$ can be expressed in (\ref{eq:reparamter1}),
$H$ satisfies the restriction (10).

We can write the likelihood function based on a single variable $V$
as 
\begin{eqnarray}
\mathcal{L}(\psi,\theta;V) & = & f(Y\mid A,X,S)f(A\mid X,S)f(X,S)\nonumber \\
 & = & f(\epsilon_{H}\mid A,X,S)f(A\mid X,S)f(X,S)\frac{\partial\epsilon_{H}}{\partial Y}\nonumber \\
 & = & f(\epsilon_{H}\mid A,X,S)f(A\mid X,S)f(X,S),\label{eq:full-data lik}
\end{eqnarray}
where the last equality follows from 
\begin{eqnarray*}
\epsilon_{H} & = & Y-\{\tau_{\varphi_{0}}(X)+(1-S)\lambda_{\phi_{0}}(X)\}A\\
 &  & -\E[H]-\left\{ q(X,S)-\int q(X,S)f(X,S)\de\nu(X,S)\right\} ,
\end{eqnarray*}
with $q(X,S)$ a nonparametric function of $(X,S)$. Because $\E[\epsilon_{H}\mid A,X,S]=0$,
we require $\int\epsilon_{H}\allowbreak f(\epsilon_{H}\mid A,X,S) \de\nu(\epsilon_{H})=0$,
where $\nu(\cdot)$ is a generic measure. After re-parameterization,
the nuisance parameter becomes the infinite dimensional set $\theta$
consisting of $f(\epsilon_{H}\mid A,X,S)$, $f(A\mid X,S),$ $f(X,S)$,
$\E[H]$, and $q(X,S)$.

We assume all the regularity conditions to ensure the existence of
the efficient score function of $\psi_{0}$ are satisfied, which are
mainly continuity conditions for the parameter and the semiparametric
model; e.g., we need $\psi=\psi(\theta)$ to be pathwise differentiable
with respect to $\theta$ \citep{bickel1993efficient,tsiatis2007semiparametric}.
These conditions are not restrictive for a typical application problem.

To distinguish nuisance parameters, we re-write the likelihood function
as 
\begin{equation}
\mathcal{L}(\psi_{0},\theta;V)=f_{1}(\epsilon_{H}\mid A,X,S)f_{5}(A\mid X,S)f_{3}(X,S),\label{eq:semiparlik}
\end{equation}
where 
\begin{eqnarray*}
\epsilon_{H} & = & Y-\{\tau_{\varphi_{0}}(X)+(1-S)\lambda_{\phi_{0}}(X)\}A\\
 &  & -c_{4}-\left\{ q_{2}(X,S)-\int q_{2}(X,S)f_{3}(X,S)\de\nu(X,S)\right\} ,
\end{eqnarray*}
and $\theta=(\theta_{1},\ldots,\theta_{5})$ consists of the nuisance
parameters $\theta_{1}=f_{1}(\epsilon_{H}\mid A,X,S),$ $\theta_{2}=q_{2}(X,S)$,
$\theta_{3}=f_{3}(X,S)$, $\theta_{4}=c_{4}$, and $\theta_{5}=f_{5}(A\mid X,S).$
Then, $\epsilon_{H}=\epsilon_{H}(\psi_{0},\theta_{2},\theta_{3},\theta_{4},\theta_{5})$
depends on the parameter of interest $\psi_{0}$ and the nuisance
parameters $(\theta_{2},\theta_{3},\theta_{4},\theta_{5})$. This
order for indexing the nuisance parameters makes the characterization
of the nuisance tangent space easier.

Theorems \ref{Thm: The-nuisance-tangent} and \ref{Thm: direct sum of nuis tang space}
present the characterizations of the nuisance tangent space and its
orthogonal complement, respectively. The proofs are followed after
stating these results.

\begin{theorem}\label{Thm: The-nuisance-tangent}The nuisance tangent
space corresponding to $\theta=(\theta_{1},\ldots,\theta_{5})$ is
\[
\Lambda=\Lambda^{(1)}+\Lambda^{(2)}+\Lambda^{(3)}+\Lambda^{(4)}+\Lambda^{(5)},
\]
where $\Lambda^{(j)}$ is the nuisance tangent space with respect
to $\theta_{j}$, for $j=1,\ldots,5$. Define $\Lambda^{*}=\{\Gamma^{*}=\Gamma^{*}(X,S)\in\R^{p}:\E[\Gamma^{*}]=0\}$
and $S_{\epsilon}=S_{\epsilon}(\epsilon_{H},A,X,S)=\partial\log f_{1}(\epsilon_{H}\mid A,X,S)/\partial\epsilon_{H}\in\R^{1}$
evaluated at the truth. Then,

\begin{eqnarray*}
\Lambda^{(1)} & = & \{\Gamma^{(1)}=\Gamma^{(1)}(\epsilon_{H},A,X,S)\in\R^{p}:\E[\Gamma^{(1)}\mid A,X,S]=0,\ \text{and}\ \E[\Gamma^{(1)}\epsilon_{H}\mid A,X,S]=0\},\\
\Lambda^{(2)} & = & \{\Gamma^{(2)}=\Gamma^{(2)}(\epsilon_{H},A,X,S)=\Gamma^{(2)}(\Gamma^{*})=\Gamma^{*}S_{\epsilon}\in\R^{p}:\Gamma^{*}\in\Lambda^{*}\},\\
\Lambda^{(3)} & = & \{\Gamma^{(3)}=\Gamma^{(3)}(\epsilon_{H},A,X,S)=\Gamma^{(3)}(\Gamma^{*})=\Gamma^{*}+\E[Q\Gamma^{*}]S_{\epsilon}\in\R^{p}:\Gamma^{*}\in\Lambda^{*}\},\\
\Lambda^{(4)} & = & \{cS_{\epsilon}:S_{\epsilon}=S_{\epsilon}(\epsilon_{H},A,X,S),c\in\R^{p}\},\\
\Lambda^{(5)} & = & \left\{ \Gamma^{(5)}=\Gamma^{(5)}(A,X,S) \in\R^{p}:\E[\Gamma^{(5)}(A,X,S)\mid X,S]=0\right\} .
\end{eqnarray*}
Here and throughout in a slight abuse of notation, we use $\Gamma^{(2)}(\cdot)$
and $\Gamma^{(3)}(\cdot)$ as functions of $(\epsilon_{H},A,X,S)$
and also as operators on $\Gamma^{*}$, but their meaning should be
clear in the context.

\end{theorem}

\begin{remark}\label{rmk: 1oplus3}It is important to note that $\Gamma^{(5)}(A,X,S)$
with $\E[\Gamma^{(5)}(A,X,S)\mid X,S]=0$ is orthogonal to all other
subspaces in $\Lambda$. \end{remark}

For simplicity, we define the following notation.

\begin{definition} \label{Def1}Let 
\begin{eqnarray}
W & = & W(A,X,S)=(\V[\epsilon_{H}\mid A,X,S])^{-1},\label{eq:Wmk}\\
T & = & T(X,S)=\E[W\mid X,S],\label{eq:Tmk}\\
\epsilon_{0} & = & \epsilon_{0}(\epsilon_{H},A,X,S)=\E[W\mid X,S]^{-1}W\epsilon_{H}+Q,\label{eq:epsilon}\\
T^{*} & = & \E[T^{-1}]=\E[\E[W\mid X,S]^{-1}].\label{eq:T*mk}
\end{eqnarray}
\end{definition}

We now express $\Lambda$ as a direct sum of orthogonal subspaces.
This effort will be valuable in characterizing the orthogonal complement
space of the nuisance tangent space $\Lambda^{\bot}$.

\begin{theorem}\label{Thm: direct sum of nuis tang space} The space
$\Lambda$ can be written as a direct sum of orthogonal subspaces:
\begin{equation}
\Lambda=\widetilde{\Lambda}^{(1)}\oplus\widetilde{\Lambda}^{(2)}\oplus\widetilde{\Lambda}^{(3)}\oplus\widetilde{\Lambda}^{(4)}\oplus\widetilde{\Lambda}^{(5)},\label{eq:orthog}
\end{equation}
where $\oplus$ denotes a direct sum, and using the notation in Theorem
\ref{Thm: The-nuisance-tangent} and Definition \ref{Def1}, $\widetilde{\Lambda}^{(1)}=\Lambda^{(1)},$

\begin{eqnarray}
\widetilde{\Lambda}^{(2)} & = & \left\{ \widetilde{\Gamma}^{(2)}=\widetilde{\Gamma}^{(2)}(\Gamma^{*})=\Gamma^{*}W\epsilon_{H}:\Gamma^{*}\in\Lambda^{*}\right\} ,\label{eq:tilde2}\\
\widetilde{\Lambda}^{(3)} & = & \left\{ \widetilde{\Gamma}^{(3)}=\widetilde{\Gamma}^{(3)}(\Gamma^{*})=\Gamma^{*}-\E[Q\Gamma^{*}](T^{*}T)^{-1}W\epsilon_{H}:\Gamma^{*}\in\Lambda^{*}\right\} ,\label{eq:tilde3}\\
\widetilde{\Lambda}^{(4)} & = & \left\{ \widetilde{\Gamma}^{(4)}=c\epsilon_{0}:c\in\R^{p}\right\} ,\label{eq:tilde4}\\
\widetilde{\Lambda}^{(5)} & = & \left\{ \widetilde{\Gamma}^{(5)}=\Gamma^{(5)}(A,X,S):\E[\Gamma^{(5)}(A,X,S)\mid X,S]=0\right\} .\label{eq:tilde5}
\end{eqnarray}

\end{theorem}

\begin{theorem}[Influence function space]\label{thm: tangent bot}Suppose
Assumptions 1 and 2 hold. The space
of the influence function space of $\psi_{0}$ is 
\begin{equation}
\Lambda^{\bot}=\left\{ G(A,X,S;\psi_{0},c)=c(A,X,S)\epsilon_{H,\psi_{0}}:\E[c(A,X,S)\mid X,S]=0\right\} .\label{eq:lambda-bot}
\end{equation}

\end{theorem}

\subsection{Proof of Theorem \ref{Thm: The-nuisance-tangent}}

For the semiparametric model, we specify the likelihood function as
given by (\ref{eq:semiparlik}). In the following, we characterize
the tangent space of $\theta_{k}$ for $k=1,\ldots,5$.

For the nuisance parameter $\theta_{1}$, we require that (i) $f_{1}(\epsilon_{H}\mid A,X,S)$
is a nonparametric distribution, and (ii) $\int\epsilon_{H}f_{1}(\epsilon_{H}\mid A,X,S)\de\nu(\epsilon_{H})=0$
because $\E[\epsilon_{H}\mid A,X,S]=0$. This leads to a semiparametric
restricted moment model for $\theta_{1}$. Following \citet{brown1998efficient},
the tangent space of $\theta_{1}$ is $\Lambda^{(1)}$.

For the nuisance parameter $\theta_{2}$ in the likelihood through
$\epsilon_{H}=\epsilon_{H}(\psi_{0},\theta_{2},\theta_{3},\theta_{4},\theta_{5})$
only, the score function is 
\begingroup\makeatletter\def\f@size{8}\check@mathfonts
\begin{align*}
\Gamma^{(2)}=\Gamma^{(2)}(\epsilon_{H},A,X,S)=\{\partial\log f(\epsilon_{H}\mid A,X,S;\theta_{1})/\partial\epsilon_{H}\}
\times\{-\dot{q}_{2}(X,S)+\int\dot{q}_{2}(X,S)f_{3}(X,S)\de\nu(X,S)\},
\end{align*}
where $\dot{q}_{2}(X,S)$ is an arbitrary function of $(X,S)$. In
a different notation, the score function $\Gamma^{(2)}$ can be written
as $\Gamma^{*}S_{\epsilon},$ where $S_{\epsilon}=\partial\log f(\epsilon_{H}\mid A,X,S;\theta_{1})/\partial\epsilon_{H}$,
$\Gamma^{*}=\Gamma^{*}(X,S)=-\dot{q}_{2}(X,S)+\int\dot{q}_{2}(X,S)f_{3}(X,S)\de\nu(X,\allowbreak S)$,
and $\E[\Gamma^{*}]=0$. Therefore, the tangent space of $\theta_{2}$
is $\Lambda^{(2)}$.

For the nuisance parameter $\theta_{3}$, the score function is 
\begin{eqnarray*}
\Gamma^{(3)} & = & \Gamma^{(3)}(\epsilon_{H},A,X,S)=\Gamma^{*}+\{\partial\log f_{1}(\epsilon_{H}\mid A,X,S)/\partial\epsilon_{H}\}\times\int q_{2}(X,S)\Gamma^{*}f_3(X,S)\de\nu(X,S)\\
 & = & \Gamma^{*}+\{\partial\log f_{1}(\epsilon_{H}\mid A,X,S)/\partial\epsilon_{H}\}\times\int\left\{ q_{2}(X,S)-\E[q_{2}(X,S)]\right\} \Gamma^{*}f_3(X,S)\de\nu(X,S)\\
 & = & \Gamma^{*}+\{\partial\log f_{1}(\epsilon_{H}\mid A,X,S)/\partial\epsilon_{H}\}\times\int Q\Gamma^{*}f_3(X,S)\de\nu(X,S),
\end{eqnarray*}
where $\Gamma^{*}=\Gamma^{*}(X,S)$ satisfies $\E[\Gamma^{*}]=0$,
the second equality follows because $\E[\Gamma^{*}]=0$. In a different
notation, the score function $\Gamma^{(3)}$ is of the form $\Gamma^{*}+S_{\epsilon}\E[Q\Gamma^{*}]$.
Therefore, the tangent space of $\theta_{3}$ is $\Lambda^{(3)}$.

For the nuisance parameter $\theta_{4}$ in the likelihood through
$\epsilon_{H}=\epsilon_{H}(\psi_{0},\theta_{2},\theta_{3},\theta_{4},\theta_{5})$
only, the tangent space of $\theta_{4}$ is $\Lambda^{(4)}$.

For the nuisance parameter $\theta_{5}$, 
%we require that (i) $f_{5}(A\mid X,S)=e(X,S)^{A}\{1-e(X,S)\}^{1-A}$ is a nonparametric distribution, and (ii) $\epsilon_{H}=\epsilon_{H}(\psi,\theta_{2},\theta_{3},\theta_{4},\theta_{5})$ depends on $\theta_{5}$ through $e(X,S)$. Then, 
the score function is $\Gamma^{(5)}(A,X,S)$,
where $\Gamma^{(5)}(A,X,S)\in\R^{p}$ satisfies $\E[\Gamma^{(5)}(A,X,S)\mid X,S]=0$.
%The relationship between $\Gamma^{(5)}(X,S)$ and $\Gamma^{(5)}(A,X,S)$ is derived based on $f_{5}(A\mid X,S)=e(X,S)^{A}\{1-e(X,S)\}^{1-A}$.
Therefore, the tangent space of $\theta_{5}$ is $\Lambda^{(5)}$.

\subsection{Remarks}

\begin{remark}\label{rmk:Sk} Because $S_{\epsilon}$ is the score
vector with respect to $f_{1}(\epsilon_{H}\mid A,X,S),$ $\E[S_{\epsilon}\mid A,X,S]=0$.
In addition, because $\E[\epsilon_{H}\mid A,X,S]=0$, $\E[S_{\epsilon}\epsilon_{H}\mid A,X,S]=-1$.

\end{remark}

\begin{remark}\label{rmk:Wmk0}Because $\E[\epsilon_{H}\mid A,X,S]=0$,
\begin{equation}
W=(\V[\epsilon_{H}\mid A,X,S])^{-1}=(\E[\epsilon_{H}^{2}\mid A,X,S])^{-1}.\label{eq:Wmk-1}
\end{equation}
\end{remark}

\subsection{\label{subsec:S8.4}Overview and Lemmas for the proof of Theorem
\ref{Thm: direct sum of nuis tang space}}

For $l\geq1$, let $\oplus_{k=1}^{l}\widetilde{\Lambda}^{(k)}=\widetilde{\Lambda}^{(1)}\oplus\cdots\oplus\widetilde{\Lambda}^{(l)}$
for orthogonal spaces $\widetilde{\Lambda}^{(1)},\cdots,\widetilde{\Lambda}^{(l)}$.
To express $\Lambda=\sum_{k=1}^{5}\Lambda^{(k)}$ as a direct sum
of orthogonal subspaces, we provide a road map: Define $\widetilde{\Lambda}_{1}=\Lambda_{1}$. 
\begin{description}
\item [{(i)}] Define $\widetilde{\Lambda}^{(2)}\equiv\prod\left[\Lambda^{(2)}\mid\widetilde{\Lambda}^{(1),\bot}\right]$.
We show that $\widetilde{\Lambda}^{(2)}$ is the same as in (\ref{eq:tilde2}). 
\item [{(ii)}] Define $\widetilde{\Lambda}^{(3)}\equiv\prod\left[\Lambda^{(3)}\mid\left(\oplus_{k=1}^{2}\widetilde{\Lambda}^{(k)}\right)^{\bot}\right]$.
We show that $\widetilde{\Lambda}^{(3)}$ is the same as in (\ref{eq:tilde3}). 
\item [{(iii)}] Define $\widetilde{\Lambda}^{(4)}\equiv\prod\left[\Lambda^{(3)}\mid\left(\oplus_{k=1}^{3}\widetilde{\Lambda}^{(k)}\right)^{\bot}\right]$.
We show that $\widetilde{\Lambda}^{(4)}$ is the same as in (\ref{eq:tilde4}). 
\item [{(iv)}] Define $\widetilde{\Lambda}^{(5)}\equiv\prod\left[\Lambda^{(5)}\mid\left(\oplus_{k=1}^{4}\widetilde{\Lambda}^{(k)}\right)^{\bot}\right]$.
We show that $\widetilde{\Lambda}^{(5)}$ is the same as in (\ref{eq:tilde5}). 
\end{description}
To calculate these projections in the above process, we need the following
Lemmas.

\begin{lemma}[Normal equations]\label{(Normal-equations)}Let $\Lambda$
be a subspace of a Hilbert space $\mathcal{H}$ equipped with an inner
product $<\cdot,\cdot>$. Suppose that $\Lambda$ is defined as $\Lambda=\{\op(\Gamma),\Gamma\in\mathcal{H}\}$,
where $\op:\mathcal{H}\rightarrow\mathcal{H}$ is a bounded linear
operator. Let $\op^{*}$ be the adjoint of $O$, i.e., for any vectors
$h_{1},h_{2}\in\mathcal{H}$, $<\op(h_{1}),h_{2}>=<h_{1},\op^{*}(h_{2})>$.
For a given vector $B\in\mathcal{H}$, because $\prod[B\mid\Lambda]\in\Lambda$,
by the definition of $\Lambda$, there exists $\Gamma\in\mathcal{H}$
such that $\prod[B\mid\Lambda]=\op(\Gamma)$. Then, $\Gamma$ satisfies
the normal equation $\op^{*}(B)=\op^{*}\op(\Gamma)$.

\end{lemma}

See, e.g., \citet{weidmann2012linear}. Lemma \ref{(Normal-equations)}
is useful to express a projection $\prod[B\mid\Lambda]$. By Lemma
\ref{(Normal-equations)}, we only need to find the unique $\Gamma\in\mathcal{H}$
that satisfies the normal equation. Then, $\prod[B\mid\Lambda]=\op(\Gamma)$.

\begin{lemma}\label{Lambda2-bot-4} (a) For $\Gamma^{*}S_{\epsilon}\in\Lambda^{(2)}$,
where $S_{\epsilon}=\partial\log f_{1}(\epsilon_{H}\mid A,X,S$$)/\partial\epsilon_{H}$,
$\Gamma^{*}=\Gamma^{*}(X,S)$ and $\E[\Gamma^{*}]=0$, we have $\prod\left[\Gamma^{*}S_{\epsilon}\mid\Lambda^{(1),\bot}\right]=-\Gamma^{*}W\epsilon_{H}$.
(b) For $cS,$ where $c$ is a constant, we have $\prod\left[cS_{\epsilon}\mid\Lambda^{(1),\bot}\right]=-cW\epsilon_{H}$.

\end{lemma}

To show (a), we can verify (i) $\Gamma^{*}W\epsilon_{H}\in\Lambda^{(1),\bot}$;
and (ii) $\Gamma^{*}S_{\epsilon}+\Gamma^{*}W\epsilon_{H}\in\Lambda^{(1)}$.
Similarly, we can show (b).

\begin{lemma}\label{Lambda2-bot-5} For $\Gamma^{(3)}(\Gamma^{*})\in\Lambda^{(3)}$,
$\prod\left[\Gamma^{(3)}(\Gamma^{*})\mid\Lambda^{(1),\bot}\right]=\Gamma^{*}-\E[Q\Gamma^{*}]W\epsilon_{H}$.

\end{lemma}

To show $\prod\left[\Gamma^{(3)}(\Gamma^{*})\mid\Lambda^{(1),\bot}\right]=\prod\left[\Gamma^{*}+S\E[Q\Gamma^{*}]\mid\Lambda^{(1),\bot}\right]=\Gamma^{*}-\E[Q\Gamma^{*}]$$W\epsilon_{H}$,
we can verify (i) $\Gamma^{*}-\E[Q\Gamma^{*}]W\epsilon_{H}\in\Lambda^{(1),\bot},$
and (ii) $S_{\epsilon}\E[Q\Gamma^{*}]+\E[Q\Gamma^{*}]W\epsilon_{H}\in\Lambda^{(1)}.$

\begin{lemma}\label{Lemma: Theorem(A4.2c)}Let $D^{*}=d^{*}(X,S)$,
then 
\[
\prod\left[D^{*}T^{-1}W\epsilon_{H}\mid\widetilde{\Lambda}^{(2),\bot}\right]=\E[D^{*}T^{-1}](T^{*}T)^{-1}W\epsilon_{H},
\]
where all the terms are defined in Theorem \ref{Thm: direct sum of nuis tang space}.
In particular, by choosing $D^{*}=cT$, where $c$ is a constant,
we have 
\[
\prod\left[cW\epsilon_{H}\mid\widetilde{\Lambda}^{(2),\bot}\right]=c(T^{*}T)^{-1}W\epsilon_{H}.
\]
\end{lemma}

For any $\widetilde{\Gamma}^{(2)}\in\widetilde{\Lambda}^{(2)}$, we
have $\widetilde{\Gamma}^{(2)}=\widetilde{\Gamma}^{(2)}(\Gamma^{*})=\Gamma^{*}W\epsilon_{H}$.
By definition of $\Gamma^{*}$, $\Gamma^{*}=\Gamma^{*}(X,S)$ and
$\E[\Gamma^{*}]=0$. Thus, $\Gamma^{*}$ can be written as ${\op}_{1}(B)\coloneqq\E[B\mid X,S]-\E[B]$,
for some $B=b(V)\in\mathcal{H}$. For convenience of notation, let
${\op}_{2}({\cdot})$ denote $\widetilde{\Gamma}^{(2)}(\cdot)$. We
can show that the adjoint operators are ${\op}_{1}^{*}={{\op}_{1}}$
and ${\op}_{2}^{*}={{\op}_{2}}$ Then, we can rewrite $\widetilde{\Gamma}^{(2)}={\op}_{2}{{\op}_{1}}(B)$,
for some $B=b(V)\in\mathcal{H}$. The adjoint operator of $\op\coloneqq{\op}_{2}{{\op}_{1}}$
is $\op^{*}={\op}_{1}^{*}{{\op}_{2}^{*}}={\op}_{1}{\op}_{2}$.

By Lemma \ref{(Normal-equations)} with $\op={\op}_{2}{{\op}_{1}}$,
we have 
\begin{equation}
\prod\left[D^{*}T^{-1}W\epsilon_{H}\mid\widetilde{\Lambda}^{(2)}\right]=\Gamma^{*}W\epsilon_{H},\label{eq:proj1}
\end{equation}
where $\Gamma^{*}$ satisfies the normal equation: 
\begin{equation}
{\op}_{1}^{*}{{\op}_{2}^{*}}{\op}_{2}{{\op}_{1}}(\Gamma^{*})={\op}_{1}^{*}{{\op}_{2}^{*}}(D^{*}T^{-1}W\epsilon_{H}).\label{eq:norm equ1}
\end{equation}
The solution $\Gamma^{*}$ to the normal equation (\ref{eq:norm equ1})
is $\Gamma^{*}=T^{-1}\{D^{*}-(T^{*})^{-1}\E[T^{-1}D^{*}]\}.$

Finally, plugging in the expression of $\Gamma^{*}$ into (\ref{eq:proj1})
proves Lemma \ref{Lemma: Theorem(A4.2c)}.

\begin{lemma} \label{Lemma: Theorem (A4.2a)}Let $D=d(X,S)$ with
$\E[D]=0$. Then, $\prod\left[D\mid\widetilde{\Lambda}^{(3),\bot}\right]=\E[QD](\V[\epsilon_{0}])^{-1}\epsilon_{0}$.\end{lemma}

For any $\widetilde{\Gamma}^{(3)}\in\widetilde{\Lambda}^{(3)}$, we
have $\widetilde{\Gamma}^{(3)}=\widetilde{\Gamma}^{(3)}(\Gamma^{*})=\Gamma^{*}-\E[Q\Gamma^{*}](T^{*}T)^{-1}W{\epsilon_{H}}$,
where $\Gamma^{*}=\Gamma^{*}(X,S)$ and $\E[\Gamma^{*}]=0$. Thus,
we can rewrite $\widetilde{\Gamma}^{(3)}=\op(B)$ for some $B=b(V)\in\mathcal{H}$,
where $\op(B)={\op}_{3}{{\op}_{1}}(B)$ with ${\op}_{1}(B)\coloneqq\E[B\mid X,S]-\E[B]$
and ${\op}_{3}(B)\coloneqq\widetilde{\Gamma}^{(3)}(B)$. Similar in
the proof of Lemma \ref{Lemma: Theorem(A4.2c)} , we can show that
${\op}_{1}^{*}={{\op}_{1}}$. Moreover, for any $B_{2}\in\mathcal{H},$
\begin{equation}
{\op}_{3}^{*}(B_{2})=B_{2}-\E[B_{2}(T^{*}T)^{-1}W\epsilon_{H}]Q.\label{eq:O*5}
\end{equation}

Now by Lemma \ref{(Normal-equations)}, $\prod\left[D\mid\widetilde{\Lambda}^{(3)}\right]=\widetilde{\Gamma}^{(3)}(\Gamma^{*})$,
where $\Gamma^{*}$ satisfies the normal equation: 
\begin{equation}
\op^{*}\op(\Gamma^{*})=\op^{*}(D).\label{eq:normal equ2}
\end{equation}
By the normal equation (\ref{eq:normal equ2}), we then have $D=\Gamma^{*}+\E[Q\Gamma^{*}](T^{*})^{-1}Q,$
which leads to 
\begin{eqnarray}
\widetilde{\Gamma}^{(3)}(\Gamma^{*}) & = & D-\E[Q\Gamma^{*}](T^{*})^{-1}\epsilon_{0}.\label{eq:Dm}
\end{eqnarray}
Therefore, from (\ref{eq:normal equ2}), 
\[
\prod\left[D\mid\widetilde{\Lambda}^{(3),\bot}\right]=D-\widetilde{\Gamma}^{(3)}(\Gamma^{*})=\E[QD](\V[\epsilon_{0}])^{-1}\epsilon_{0},
\]
where the second equality follows from (\ref{eq:Dm}).

\begin{lemma}\label{lemma: Theorem(A4.2b)} Let $G=g(A,X,S)$, then
\[
\prod\left[GW\epsilon_{H}\mid(\widetilde{\Lambda}^{(2)}+\widetilde{\Lambda}^{(3)})^{\bot}\right]=\mathcal{L}_{1}+\mathcal{L}_{2},
\]
where $\mathcal{L}_{1}=\left(G-T^{-1}\E[GW\mid X,S]\right)W\epsilon_{H}$
and $\mathcal{L}_{2}=G_{0}(\V[\epsilon_{0}])^{-1}\epsilon_{0}$ with
\begin{equation}
G_{0}=\E[GWT^{-1}].\label{eq:G_m-1k}
\end{equation}

Moreover, if $G$ is not a function of $A$, then $\mathcal{L}_{1}=0$.
In particular with $G=\Gamma^{*}$, which is not a function of $A$,
then 
\[
\prod\left[\Gamma^{*}W\epsilon_{H}\mid(\widetilde{\Lambda}^{(2)}+\widetilde{\Lambda}^{(3)})^{\bot}\right]=0.
\]
\end{lemma}

Let $\widehat{\Lambda}^{(2)}\equiv\{\widehat{\Gamma}^{(2)}(B)=BW\epsilon_{H}:B=b(X,S)\}$.
Because $\widetilde{\Lambda}^{(2)}\subset\widehat{\Lambda}^{(2)}$,
we have $\widehat{\Lambda}^{(2),\bot}\subset\widetilde{\Lambda}^{(2),\bot}$.
We will show in the proof of Theorem \ref{Thm: direct sum of nuis tang space}
that any vector in $\widetilde{\Lambda}^{(2)}$ is constructed as
a projection onto $\widetilde{\Lambda}^{(3),\bot}$, and therefore
by construction, $\widetilde{\Lambda}^{(2)}\bot\widetilde{\Lambda}^{(3)}$.
Now, we can express 
\begin{eqnarray}
\prod\left[GW\epsilon_{H}\mid(\widetilde{\Lambda}^{(2)}+\widetilde{\Lambda}^{(3)})^{\bot}\right] & = & \underbrace{\prod\left[\prod\left[GW\epsilon_{H}\mid\widehat{\Lambda}^{(2),\bot}\right]\mid\widetilde{\Lambda}^{(3),\bot}\right]}_{\mathcal{L}_{1}}\nonumber \\
 &  & +\underbrace{\prod\left[\prod\left[\prod\left[GW\epsilon_{H}\mid\widehat{\Lambda}^{(2)}\right]\mid\widetilde{\Lambda}^{(2),\bot}\right]\mid\widetilde{\Lambda}^{(3),\bot}\right]}_{\mathcal{L}_{2}}.\label{eq:eq10}
\end{eqnarray}

To evaluate $\mathcal{L}_{1}$, note that 
\begin{equation}
\prod\left[GW\epsilon_{H}\mid\widehat{\Lambda}^{(2)}\right]=\E[GW\mid X,S]T^{-1}W\epsilon_{H},\label{eq:A4.4}
\end{equation}
because (i) $\E[GW\mid X,S]T^{-1}$ is a function of $(X,S)$, so
$\E[GW\mid X,S]T^{-1}W\epsilon_{H}\in\widehat{\Lambda}^{(2)}$, and
(ii) for any $\widehat{\Gamma}^{(2)}(B)=BW\epsilon_{H}\in\widehat{\Lambda}^{(2)}$,
where $B=b(X,S)$, $\E[\{G-\E[GW\mid X,S]T^{-1}\}W\epsilon_{H}\times\widehat{\Gamma}^{(2)}(B)]=0.$
Consequently, 
\begin{eqnarray}
\mathcal{L}_{1} & = & \prod\left[GW\epsilon_{H}-\prod\left[GW\epsilon_{H}\mid\widehat{\Lambda}^{(2)}\right]\mid\widetilde{\Lambda}^{(3),\bot}\right]\nonumber \\
 & = & \prod\left[GW\epsilon_{H}-\E[GW\mid X,S]T^{-1}W\epsilon_{H}\mid\widetilde{\Lambda}^{(3),\bot}\right]\label{eq:L1-1}\\
 & = & GW\epsilon_{H}-\E[GW\mid X,S]T^{-1}W\epsilon_{H},\label{eq:L1}
\end{eqnarray}
where the last equality follows because $GW\epsilon_{H}-\E[GW\mid X,S]T^{-1}W\epsilon_{H}\in\widetilde{\Lambda}^{(3),\bot}.$

To evaluate $\mathcal{L}_{2}$, we have 
\begin{eqnarray}
\mathcal{L}_{2} & = & \prod\left[\prod\left[\prod\left[GW\epsilon_{H}\mid\widehat{\Lambda}^{(2)}\right]\mid\widetilde{\Lambda}^{(2),\bot}\right]\mid\widetilde{\Lambda}^{(3),\bot}\right]\nonumber \\
 & = & \prod\left[\E[\E[GW\mid X,S]T^{-1}](T^{*}T)^{-1}W\epsilon_{H}\mid\widetilde{\Lambda}^{(3),\bot}\right]\nonumber \\
 & = & \prod\left[\E[GWT^{-1}](T^{*}T)^{-1}W\epsilon_{H}\mid\widetilde{\Lambda}^{(3),\bot}\right]\nonumber \\
 & \coloneqq & \E[GWT^{-1}](\V[\epsilon_{0}])^{-1}\epsilon_{0}\label{eq:L2}
\end{eqnarray}
where the first equality follows from (\ref{eq:A4.4}), the second
equality follows by applying Lemma \ref{Lemma: Theorem(A4.2c)} with
$D^{*}=\E[GW\mid X,S]$, and the last equality follows by applying
the normal equation technique.

Together, (\ref{eq:eq10}), (\ref{eq:L1}) and (\ref{eq:L2}) prove
Lemma \ref{lemma: Theorem(A4.2b)}.

\subsection{\label{subsec:S8.4-1}Proof of Theorem \ref{Thm: direct sum of nuis tang space}}

We show (\ref{eq:orthog}) step by step following the road map in
Section \ref{subsec:S8.4}.

First, we show that $\prod\left[\Lambda^{(2)}\mid\widetilde{\Lambda}^{(1),\bot}\right]$
is the same as $\widetilde{\Lambda}^{(2)}$ in (\ref{eq:tilde2}).
By definition of $\Lambda^{(2)}$, for any $\Gamma^{(2)}\in\Lambda^{(2)}$,
by Lemma \ref{Lambda2-bot-4}, $\prod\left[\Gamma^{(2)}\mid\widetilde{\Lambda}^{(1),\bot}\right]=\prod\left[\Gamma^{*}S\mid\widetilde{\Lambda}^{(1),\bot}\right]=-\Gamma^{*}W\epsilon_{H}$.
Therefore, $\widetilde{\Lambda}^{(2)}$ is given by (\ref{eq:tilde2}).

Second, we show that $\prod\left[\Lambda^{(3)}\mid\left(\oplus_{k=1}^{2}\widetilde{\Lambda}^{(k)}\right)^{\bot}\right]$
is the same as $\widetilde{\Lambda}^{(3)}$ in (\ref{eq:tilde3}).
For any $\Gamma^{(3)}=\Gamma^{(3)}(\Gamma^{*})\in\Lambda^{(3)}$,
(a) by Lemma \ref{Lambda2-bot-5}, we have $\prod\left[\Gamma^{(3)}(\Gamma^{*})\mid\widetilde{\Lambda}^{(1),\bot}\right]=\Gamma^{*}-\E[Q\Gamma^{*}]W\epsilon_{H}$;
then (b) by the fact that $\Gamma^{*}\bot\widetilde{\Lambda}^{(3),\bot}$
and $\prod\Big[\Gamma^{*}-\E[Q\Gamma^{*}]W\epsilon_{H}\mid\widetilde{\Lambda}^{(2),\bot}\Big]=\Gamma^{*}-\E[Q\Gamma^{*}](T^{*}T)^{-1}W\epsilon_{H}$
by Lemma \ref{Lemma: Theorem(A4.2c)}. Therefore, $\widetilde{\Lambda}^{(3)}$
is given by (\ref{eq:tilde3}).

Third, we show that $\left\{ \prod\left[\Gamma^{(4)}\mid\left(\oplus_{k=1}^{3}\widetilde{\Lambda}^{(k)}\right)^{\bot}\right]:\Gamma^{(4)}\in\Lambda^{(4)}\right\} $
is the same as $\widetilde{\Lambda}^{(4)}$ in (\ref{eq:tilde4}).
For any $cS\in\Lambda^{(4)}$, (a) by Lemma \ref{Lambda2-bot-4},
$\prod\left[cS\mid\widetilde{\Lambda}^{(1),\bot}\right]=-cW\epsilon_{H}$;
then (b) by Lemma \ref{lemma: Theorem(A4.2b)}, $\prod\left[cW\epsilon_{H}\mid\left(\widetilde{\Lambda}^{(2)}\oplus\widetilde{\Lambda}^{(3)}\right)^{\bot}\right]=cT^{*}(\V[\epsilon_{0}])^{-1}\epsilon_{0}$.
Therefore, $\widetilde{\Lambda}^{(4)}$ is given by (\ref{eq:tilde4}).

Finally, we show that $\left\{ \prod\left[\Gamma^{(5)}\mid\left(\oplus_{k=1}^{4}\widetilde{\Lambda}^{(k)}\right)^{\bot}\right]:\Gamma^{(5)}\in\Lambda^{(5)}\right\} $
is the same as $\widetilde{\Lambda}^{(5)}$ in (\ref{eq:tilde5}).
For any $\Gamma^{(5)}\in\Lambda^{(5)},$ by Lemma \ref{lemma: A4.1}
in Section \ref{subsec:Lemma A4.1} below, we can show that $(1-S)\lambda(X)\Gamma^{(5)}(X,S)S_{\epsilon}\in\left(\oplus_{k=1}^{4}\widetilde{\Lambda}^{(k)}\right)^{\bot}$,
so $\prod\left[\Gamma^{(5)}\mid\left(\oplus_{k=1}^{4}\widetilde{\Lambda}^{(k)}\right)^{\bot}\right]=\Gamma^{(5)}(A,X,S)$.
Therefore, $\widetilde{\Lambda}^{(5)}$ is given by (\ref{eq:tilde5}).

\subsection{Lemma \ref{lemma: A4.1} and its proof\label{subsec:Lemma A4.1}}

For simplicity, we define the following notation.

\begin{definition}\label{Def2}Let 
\begin{eqnarray}
\dot{B} & = & \dot{B}({\epsilon_{H}},A,X,S)=B-\E[B\mid A,X,S],\label{eq:Bk}\\
B^{*} & = & B^{*}(X,S)=\E[B\mid X,S]-\E[B],\label{eq:Bm}\\
R & = & R(A,X,S)=\E[B{\epsilon_{H}}\mid A,X,S].\label{eq:R}
\end{eqnarray}

\end{definition}

\begin{lemma}\label{lemma: A4.1}Following the notation in Definitions
\ref{Def1} and \ref{Def2}, for any $B=b(V)$, 
\begin{equation}
\prod\left[B\mid\left(\oplus_{k=1}^{4}\widetilde{\Lambda}^{(k)}\right)^{\bot}\right]=\left(R-T^{-1}\E[RW\mid X,S]\right)W{\epsilon_{H}}.\label{eq:A4.1-2}
\end{equation}
\end{lemma}

To compute $\prod\left[B\mid\left(\oplus_{k=1}^{4}\widetilde{\Lambda}^{(k)}\right)^{\bot}\right]$,
decompose $B$ into 
\begin{eqnarray*}
B & = & \underbrace{B-\E[B\mid A,X,S]}_{\dot{B}=\dot{B}({\epsilon_{H}},A,X,S)}+\underbrace{\E[B\mid A,X,S]-\E[B\mid X,S]}_{B^{(2)}=B^{(2)}(A,X,S)}\\
 &  & +\underbrace{\E[B\mid X,S]-\E[B]}_{B^{*}=B^{*}(X,S)}.
\end{eqnarray*}
By Remark \ref{rmk: 1oplus3}, we have $\prod\left[B^{(2)}\mid\left(\oplus_{k=1}^{4}\widetilde{\Lambda}^{(k)}\right)^{\bot}\right]=0$.
By Lemma \ref{Lemma: Theorem (A4.2a)}, $\prod\left[B^{*}\mid\left(\oplus_{k=1}^{4}\widetilde{\Lambda}^{(k)}\right)^{\bot}\right]=0$.
It then suffices to obtain $\prod\left[\dot{B}\mid\left(\oplus_{k=1}^{4}\widetilde{\Lambda}^{(k)}\right)^{\bot}\right]$
in (\ref{eq:A4.1-2}) in the following steps: 
\begin{description}
\item [{Step$\ $1.}] We show that $\prod\left[\dot{B}\mid\Lambda^{(1),\bot}\right]=\E[B\epsilon_{H}\mid A,X,S]W\epsilon_{H}$
using the following two facts: (i) $\E[B\epsilon_{H}\mid A,X,S]W\epsilon_{H}\in\Lambda^{(1),\bot}$,
and (ii) $\dot{B}-\E[B\epsilon_{H}\mid A,X,S]W\epsilon_{H}\in\Lambda^{(1)}$. 
\item [{Step$\ $2.}] We show that 
\begin{multline*}
\prod\left[\E[B{\epsilon_{H}}\mid A,X,S]W{\epsilon_{H}}\mid(\widetilde{\Lambda}^{(2)}+\widetilde{\Lambda}^{(3)})^{\bot}\right]=\prod\left[RW{\epsilon_{H}}\mid(\widetilde{\Lambda}^{(2)}+\widetilde{\Lambda}^{(3)})^{\bot}\right]\\
=(R-T^{-1}\E[RW\mid X,S])W{\epsilon_{H}}+\E[RWT^{-1}](\V[\epsilon_{0}])^{-1}\epsilon_{0},
\end{multline*}
where first equality follows by the definition of $R$ in (\ref{eq:R}),
and the second equality follows from Lemma \ref{lemma: Theorem(A4.2b)}. 
\item [{Step$\ $3.}] Because $(R-T^{-1}\E[RW\mid X,S])W{\epsilon_{H}}\indep\widetilde{\Lambda}^{(4)}$
and $\E[RWT^{-1}](\V[\epsilon_{0}])^{-1}\epsilon_{0}\in\widetilde{\Lambda}^{(4)}$,
we have $\prod\Big[\dot{B}\mid\big(\oplus_{k=1}^{4}\widetilde{\Lambda}^{(k)}\big)^{\bot}\Big]=(R-T^{-1}\E[RW\mid X,S])W{\epsilon_{H}}$. 
\end{description}

\subsection{Proof of Theorem \ref{thm: tangent bot}}

By Lemma \ref{lemma: A4.1}, for any $B=b(V)$, 
\begin{eqnarray*}
\prod\left[B\mid\Lambda^{\bot}\right] & = & \prod\left[\prod\left[B\mid\left(\oplus_{k=1}^{4}\widetilde{\Lambda}^{(k)}\right)^{\bot}\right]\mid\widetilde{\Lambda}^{(5),\bot}\right]\\
 & = & \prod\left[R-T^{-1}\E[RW\mid X,S])W\epsilon_{H}\mid\widetilde{\Lambda}^{(5),\bot}\right]\\
 & = & R-T^{-1}\E[RW\mid X,S])W\epsilon_{H},
\end{eqnarray*}
where the last equality follows by Remark \ref{rmk: 1oplus3}. Therefore,
$\prod\left[B\mid\Lambda^{\bot}\right]$ is of the form $c(A,X,S)\epsilon_{H}$,
with $c(A,X,S)=\left(R-T^{-1}\E[RW\mid X,S]\right)W$ and $\E[c(A,X,S)\mid X,S]=0$.

\subsection{Proof of Theorem 1}

Based on Theorem \ref{thm: tangent bot}, we show that the projection
of any $B\in\mathcal{H}$, $\prod\left[B\mid\Lambda^{\bot}\right]$,
is of the form $c(A,X,S)\epsilon_{H,\psi_{0}}$, where $\E[c(A,X,S)\mid X,S]=0$.
Let the score vector of $\psi_{0}$ be $s_{\psi_{0}}(V)$. Then, the
semiparametric efficient score is the projection of $s_{\psi_{0}}(V)$
onto $\Lambda^{\bot}$, given by 
\begin{multline*}
S_{\psi_{0}}(V)=\prod\left[s_{\psi_{0}}(V)\mid\Lambda^{\bot}\right]=\left(\E[s_{\psi_{0}}(V)\epsilon_{H,\psi_{0}}\mid A,X,S]\right.\\
\left.-\E[\E[s_{\psi_{0}}(V)\epsilon_{H,\psi_{0}}\mid A,X,S]W\mid X,S]\E[W\mid X,S]^{-1}\right)W\epsilon_{H,\psi_{0}}\coloneqq c^{*}(A,X,S)\epsilon_{H,\psi_{0}}.
\end{multline*}
To evaluate $c^{*}(A,X,S)$ further, we note that $\E[\epsilon_{H,\psi_{0}}\mid A,X,S]=0$.
We differentiate this equality with respect to $\psi_{0}$. By the
generalized information equality \citep{newey1990semiparametric},
we have $\E[-\partial\epsilon_{H,\psi_{0}}/\partial\psi\mid A,X,S]+\E[s_{\psi_{0}}(V)\epsilon_{H,\psi_{0}}\mid A,X,S]=0$.
Therefore, ignoring the negative sign, we have $c^{*}(A,X,S)$ as
given by 
\begin{eqnarray*}
c^{*}(A,X,S) & = & \left(\E\left[\frac{\partial\epsilon_{H,\psi_{0}}}{\partial\psi^{\T}}\mid A,X,S\right]-\E\left[\frac{\partial\epsilon_{H,\psi_{0}}}{\partial\psi^{\T}}W\mid X,S\right]\E[W\mid X,S]^{-1}\right)W\\
 & = & \left(\begin{array}{c}
\frac{\partial\tau_{\varphi_{0}}(X)}{\partial\varphi}\\
(1-S)\frac{\partial\lambda_{\phi_{0}}(X)}{\partial\phi}
\end{array}\right)\left(A-\E\left[AW\mid X,S\right]\E[W\mid X,S]^{-1}\right)W.
\end{eqnarray*}

\section{Proof of Remark in Section 4.2\label{sec:Proof-of-Remark1}}

\textcolor{black}{With the definition of $\tau(X)$ and $\lambda(X)$,
we write 
\begin{equation}
Y=\mu_{0}(X,S)+\{\tau(X)+(1-S)\lambda(X)\}A+\epsilon,\label{eq:Y}
\end{equation}
where $\E[\epsilon\mid A,X,S]=\E[\epsilon\mid X,S]=0$. Define the
loss function of $(\varphi,\phi)$ as 
\begin{equation}
L(\varphi,\phi)=\E_{W}\{Y-\tau_{\varphi}(X)A-(1-S)\lambda_{\phi}(X)A-\E_{W}[H_{\psi}\mid X,S]\}^{2}.\label{eq:Loss1}
\end{equation}
By the definition of }$H_{\psi}=Y-\{\tau_{\varphi}(X)-(1-S)\lambda_{\phi}(X)\}A$
\textcolor{black}{and (\ref{eq:Y}), we have 
\begin{equation}
L(\varphi,\phi)=\E_{W}[Y-\E_{W}[Y\mid X,S]-\tau_{\varphi}(X)\{A-\E_{W}[A\mid X,S]\}-(1-S)\lambda_{\phi}(X)\{A-\E_{W}[A\mid X,S]\}]^{2}.\label{eq:Loss2}
\end{equation}
Based on (\ref{eq:Loss1}) and (\ref{eq:Loss2}), we can verify that
$\partial L(\varphi,\phi)/\partial(\varphi^{\T},\phi^{\T})=\E[S_{\psi}(V)]$.
Because $\psi_{0}=(\varphi_{0}^{\T},\phi_{0}^{\T})^{\T}$ satisfies
$\E[S_{\psi_{0}}(V)]=0$, we have $(\varphi_{0},\phi_{0})=\arg\min_{\varphi,\phi}L(\varphi,\phi)$.
Lastly, continuing with (\ref{eq:Loss2}), we evaluate $L(\varphi,\phi)$
as 
\[
L(\varphi,\phi)=\E_{W}[(A-\E_{W}[A\mid X,S])^{2}[\tau(X)-\tau_{\varphi}(X)+(1-S)\{\lambda(X)-\lambda_{\phi}(X)]^{2}].
\]
This completes the proof of Remark }in Section 4.2.

\section{Proof of Theorem 2 \label{sec:Proof-of-Theorem3} }

\subsection{Preliminaries }

\textcolor{black}{We introduce more notations and useful results to
prepare for the proof of Theorem 2. Let ``$\rightsquigarrow$''
denote weak convergence, and let ``$A\preceq B$'' denote that $A$
is bounded by a constant times $B$. Denote $\dot{S}_{\psi}(V;\vartheta)=\partial S_{\psi}(V;\vartheta)/\partial\psi.$
Denote a set of nuisance functions as $\mathcal{G}_{\vartheta_{0}}=\{\vartheta:||\vartheta-\vartheta_{0}||<\delta\}$
for some $\delta>0$ and denote $l^{\infty}(\mathcal{G}_{\vartheta_{0}})$
as the collection of all bounded functions $f:\mathcal{G}_{\vartheta_{0}}\rightarrow\R^{p}.$}

\textcolor{black}{The following lemmas show the asymptotic properties
of functions belong to Donsker classes. }

\begin{lemma}\textcolor{black}{\label{lemma:A1} Suppose Condition
3 and 2 hold. Then, we have
\[
\sup_{\psi\in\Theta,\vartheta\in\mathcal{G}_{\vartheta_{0}}}||\bbP_{N}S_{\psi}(V;\vartheta)-\bbP S_{\psi}(V;\vartheta)||_{2}\rightarrow0
\]
in probability as $N\rightarrow\infty$, and 
\[
\sup_{\psi\in\Theta,\vartheta\in\mathcal{G}_{\vartheta_{0}}}||\bbP_{N}\dot{S}_{\psi}(V;\vartheta)-\bbP\dot{S}_{\psi}(V;\vartheta)||_{2}\rightarrow0
\]
in probability as $N\rightarrow\infty$.}

\end{lemma}

\begin{lemma}\textcolor{black}{\label{lemma:A2} Suppose Condition
3 and 2 hold. Then, we have
\[
N^{1/2}(\bbP_{N}-\bbP)S_{\psi_{0}}(V;\vartheta)\rightsquigarrow Z\in l^{\infty}(\mathcal{G}_{\vartheta_{0}}),
\]
where the limiting process $Z=\{Z(\vartheta):\vartheta\in\mathcal{G}_{\vartheta_{0}}\}$
is a mean-zero multivariate Gaussian process, and the sample paths
of $Z$ belong to $\{z\in l^{\infty}(\mathcal{G}_{\vartheta_{0}}):z$
is uniformly continuous with respect to $||\cdot||\}$. }

\end{lemma}

\subsection{Proof of Theorem 2}

\textit{\textcolor{black}{First, we show the consistency of $\widehat{\psi}$
.}}\textcolor{black}{{} Toward this end, we show $||\bbP S_{\widehat{\psi}}(V;\vartheta_{0})||_{2}\rightarrow0$.
We bound $||\bbP S_{\widehat{\psi}}(V;\vartheta_{0})||_{2}$ by
\begingroup\makeatletter\def\f@size{7.5}\check@mathfonts
\begin{eqnarray}
||\bbP S_{\widehat{\psi}}(V;\vartheta_{0})||_{2} & \leq & ||\bbP S_{\widehat{\psi}}(V;\vartheta_{0})-\bbP S_{\widehat{\psi}}(V;\widehat{\vartheta})||_{2}+||\bbP S_{\widehat{\psi}}(V;\widehat{\vartheta})||_{2}\nonumber \\
 & = & ||\bbP S_{\widehat{\psi}}(V;\vartheta_{0})-\bbP S_{\widehat{\psi}}(V;\widehat{\vartheta})||_{2}+||\bbP S_{\widehat{\psi}}(V;\widehat{\vartheta})-\bbP_{N}S_{\widehat{\psi}}(V;\widehat{\vartheta})||_{2}\nonumber \\
 & \leq & ||\bbP S_{\widehat{\psi}}(V;\vartheta_{0})-\bbP S_{\widehat{\psi}}(V;\widehat{\vartheta})||_{2}+\sup_{\psi\in\Theta,\vartheta\in\mathcal{G}_{\vartheta_{0}}}||\bbP_{N}S_{\psi}(V;\vartheta)-\bbP S_{\psi}(V;\vartheta)||_{2}.\label{eq:A1}
\end{eqnarray}
\endgroup
Both terms in (\ref{eq:A1}) are $o_{\bbP}(1)$ as shown below. By
the Taylor expansion, we have 
\begin{eqnarray*}
||S_{\widehat{\psi}}(V;\vartheta_{0})-S_{\widehat{\psi}}(V;\widehat{\vartheta})||_{2} & = & \left\vert \left\vert \left.\frac{\partial S_{\psi}(V;\vartheta)}{\partial\psi^{\T}}\right\vert _{\psi=\widehat{\psi},\vartheta=\widetilde{\vartheta}}(\widehat{\vartheta}-\vartheta_{0})\right\vert \right\vert _{2}\\
 & \leq & \left\vert \left\vert \left.\frac{\partial S_{\psi}(V;\vartheta)}{\partial\psi^{\T}}\right\vert _{\psi=\widehat{\psi},\vartheta=\widetilde{\vartheta}}\right\vert \right\vert _{2}\times||\widehat{\vartheta}-\vartheta_{0}||_{2},
\end{eqnarray*}
where $\widetilde{\vartheta}$ lies in the segment between $\widehat{\vartheta}$
and $\vartheta_{0}$. By the Cauchy--Schwartz inequality, we have
\begin{eqnarray}
||\bbP S_{\widehat{\psi}}(V;\vartheta_{0})-\bbP S_{\widehat{\psi}}(V;\widehat{\vartheta})||_{2} & \leq & \bbP||S_{\widehat{\psi}}(V;\vartheta_{0})-S_{\widehat{\psi}}(V;\widehat{\vartheta})||_{2}\nonumber \\
 & \leq & \bbP\left\{ \left\vert \left\vert \left.\frac{\partial S_{\widehat{\psi}}(V;\vartheta)}{\partial\vartheta^{\T}}\right\vert _{\vartheta=\widetilde{\vartheta}}\right\vert \right\vert _{2}\times||\widehat{\vartheta}-\vartheta_{0}||_{2}\right\} \nonumber \\
 & \leq & \left\{ \E\left\vert \left\vert \left.\frac{\partial S_{\widehat{\psi}}(V;\vartheta)}{\partial\vartheta^{\T}}\right\vert _{\vartheta=\widetilde{\vartheta}}\right\vert \right\vert _{2}^{2}\right\} ^{1/2}\times\left\{ \E||\widehat{\vartheta}-\vartheta_{0}||_{2}^{2}\right\} ^{1/2}\nonumber \\
 & \preceq & ||\widehat{\vartheta}-\vartheta_{0}||_{2}\nonumber \\
 & = & o_{\bbP}(1).\label{eq:A2}
\end{eqnarray}
By Lemma \ref{lemma:A1}, we have 
\begin{equation}
\sup_{\psi\in\Theta,\vartheta\in\mathcal{G}_{\vartheta_{0}}}||\bbP_{N}S_{\psi}(V;\vartheta)-\bbP S_{\psi}(V;\vartheta)||_{2}\rightarrow0\label{eq:A2-1}
\end{equation}
in probability as $N\rightarrow\infty.$ Plugging (\ref{eq:A2}) and
(\ref{eq:A2-1}) into (\ref{eq:A1}) leads to $||\bbP S_{\widehat{\psi}}(V;\vartheta_{0})||_{2}=o_{\bbP}(1)$.
Now, by Condition 1, $||\widehat{\psi}-\psi_{0}||_{2}=o_{\bbP}(1)$. }

\noindent \textit{\textcolor{black}{Second, we show the asymptotic
distribution of $\widehat{\psi}$. }}\textcolor{black}{By the Taylor
expansion of $N^{1/2}\bbP_{N}S_{\widehat{\psi}}(V;\widehat{\vartheta})=0$,
we have 
\[
0=N^{1/2}\bbP_{N}S_{\psi_{0}}(V;\widehat{\vartheta})+\{\bbP_{N}\dot{S}_{\widetilde{\psi}}(V;\widehat{\vartheta})\}N^{1/2}(\widehat{\psi}-\psi_{0}),
\]
where $\widetilde{\psi}$ lies in the segment between $\widehat{\psi}$
and $\psi_{0}$. By Lemma \ref{lemma:A1}, we have 
\[
\sup_{\psi\in\Theta,\vartheta\in\mathcal{G}_{\vartheta_{0}}}||\bbP_{N}\dot{S}_{\psi}(V;\vartheta)-\bbP\dot{S}_{\psi}(V;\vartheta)||_{2}\rightarrow0
\]
in probability as $N\rightarrow\infty$. Because $\widetilde{\psi}\rightarrow\psi_{0}$
and $\widehat{\vartheta}\rightarrow\vartheta_{0}$, we have 
\[
\bbP_{N}\dot{S}_{\widetilde{\psi}}(V;\widehat{\vartheta})\rightarrow\Psi=\bbP\dot{S}_{\psi_{0}}(V;\vartheta_{0})
\]
in probability as $N\rightarrow\infty$. Thus, we have 
\begin{equation}
N^{1/2}(\widehat{\psi}-\psi_{0})=-\Psi^{-1}N^{1/2}\bbP_{N}S_{\psi_{0}}(V;\widehat{\vartheta})+o_{\bbP}(1).\label{eq:A3-0}
\end{equation}
We express 
\begin{equation}
\bbP_{N}S_{\psi_{0}}(V;\widehat{\vartheta})=(\bbP_{N}-\bbP)S_{\psi_{0}}(V;\widehat{\vartheta})+\bbP S_{\psi_{0}}(V;\widehat{\vartheta}),\label{eq:A3}
\end{equation}
and show that 
\begin{eqnarray}
\bbP S_{\psi_{0}}(V;\widehat{\vartheta}) & = & o_{\bbP}(N^{-1/2}),\label{eq:A3-1}\\
(\bbP_{N}-\bbP)S_{\psi_{0}}(V;\widehat{\vartheta}) & = & (\bbP_{N}-\bbP)S_{\psi_{0}}(V;\vartheta_{0})+o_{\bbP}(N^{-1/2}).\label{eq:A3-2}
\end{eqnarray}
To show (\ref{eq:A3-1}), we denote $c(X,S)=(\partial\tau_{\varphi_{0}}(X)/\partial\varphi^{\T},(1-S)\partial\lambda_{\phi_{0}}(X)/\partial\phi^{\T})^{\T}$
for simplicity and evaluate $\bbP S_{\psi_{0}}(V;\widehat{\vartheta})$
explicitly as 
\begin{eqnarray*}
\bbP S_{\psi_{0}}(V;\widehat{\vartheta}) & = & \E\left[c(X,S)\left(A\{\widehat{\sigma}_{A}^{2}(X,S)\}^{-1}-\{\widehat{\sigma}_{1}^{2}(X,S)\}^{-1}\widehat{e}(X,S)\widehat{\E}[\widehat{W}\mid X,S]^{-1}\widehat{W}\right)\widehat{\epsilon}_{H,\psi_{0}}\right]\\
 & = & \E\left[c(X,S)\left(\{\widehat{\sigma}_{1}^{2}(X,S)\}^{-1}e(X,S)-\{\widehat{\sigma}_{1}^{2}(X,S)\}^{-1}\widehat{e}(X,S)\widehat{\E}[\widehat{W}\mid X,S]^{-1}\widehat{W}\right)\right.\\
 &  & \times\left.[\mu_{0}(X,S)-\widehat{\mu}_{0}(X,S)-(1-S)\lambda_{\phi_{0}}(X)\{e(X,S)-\widehat{e}(X,S)\}]\right]\\
 & = & \E\left[c(X,S)\left(\{\widehat{\sigma}_{1}^{2}(X,S)\}^{-1}\{e(X,S)-\widehat{e}(X,S)\}\right.\right.\\
 &  & \left.-\{\widehat{\sigma}_{1}^{2}(X,S)\}^{-1}\widehat{e}(X,S)\widehat{\E}[\widehat{W}\mid X,S]^{-1}(\widehat{W}-\widehat{\E}[\widehat{W}\mid X,S])\right)\\
 &  & \times\left.[\mu_{0}(X,S)-\widehat{\mu}_{0}(X,S)-(1-S)\lambda_{\phi_{0}}(X)\{e(X,S)-\widehat{e}(X,S)\}]\right]\\
 & = & \E\left[c(X,S)\left(\{\widehat{\sigma}_{1}^{2}(X,S)\}^{-1}\{e(X,S)-\widehat{e}(X,S)\}\right.\right.\\
 &  & \left.-\{\widehat{\sigma}_{1}^{2}(X,S)\}^{-1}\widehat{e}(X,S)\widehat{\E}[\widehat{W}\mid X,S]^{-1}\{\widehat{\sigma}_{1}^{2}(X,S)\}^{-1}\{e(X,S)-\widehat{e}(X,S)\}\right.\\
 &  & \left.+\{\widehat{\sigma}_{1}^{2}(X,S)\}^{-1}\widehat{e}(X,S)\widehat{\E}[\widehat{W}\mid X,S]^{-1}\{\widehat{\sigma}_{0}^{2}(X,S)\}^{-1}\{e(X,S)-\widehat{e}(X,S)\}\right)\\
 &  & \left.\times[\mu_{0}(X,S)-\widehat{\mu}_{0}(X,S)-(1-S)\lambda_{\phi_{0}}(X)\{e(X,S)-\widehat{e}(X,S)\}]\right].
\end{eqnarray*}
Applying the Cauchy--Schwartz inequality and Condition 5,
we have 
\begin{eqnarray*}
 &  & ||\bbP S_{\psi_{0}}(V;\widehat{\vartheta})||_{2}\\
 & \leq & \E\left[\left\vert \left\vert c(X,S)\left(\{\widehat{\sigma}_{1}^{2}(X,S)\}^{-1}\{e(X,S)-\widehat{e}(X,S)\}\right.\right.\right.\right.\\
 &  & \left.-\{\widehat{\sigma}_{1}^{2}(X,S)\}^{-1}\widehat{e}(X,S)\widehat{\E}[\widehat{W}\mid X,S]^{-1}\{\widehat{\sigma}_{1}^{2}(X,S)\}^{-1}\{e(X,S)-\widehat{e}(X,S)\}\right.\\
 &  & \left.+\{\widehat{\sigma}_{1}^{2}(X,S)\}^{-1}\widehat{e}(X,S)\widehat{\E}[\widehat{W}\mid X,S]^{-1}\{\widehat{\sigma}_{0}^{2}(X,S)\}^{-1}\{e(X,S)-\widehat{e}(X,S)\}\right)\\
 &  & \left.\left.\left.\times[\mu_{0}(X,S)-\widehat{\mu}_{0}(X,S)-(1-S)\lambda_{\phi_{0}}(X)\{e(X,S)-\widehat{e}(X,S)\}]\right\vert \right\vert _{2}\right].\\
 & \preceq & (\E[\{e(X,S)-\widehat{e}(X,S)\}^{2}]\times\E\left[\{\mu_{0}(X,S)-\widehat{\mu}_{0}(X,S)\}^{2}+\{e(X,S)-\widehat{e}(X,S)\}^{2}\right])^{1/2}\\
 & = & \{||\widehat{\mu}_{0}(X,S)-\mu_{0}(X,S)||\times||\widehat{e}(X,S)-e(X,S)||+||\widehat{e}(X,S)-e(X,S)||^{2}\}\\
 & = & o_{\bbP}(N^{-1/2}).
\end{eqnarray*}
}

\noindent \textcolor{black}{To show (\ref{eq:A3-2}), Lemma \ref{lemma:A2}
leads to 
\[
N^{1/2}(\bbP_{N}-\bbP)S_{\psi_{0}}(V;\vartheta)\rightsquigarrow Z\in l^{\infty}(\mathcal{G}_{\vartheta_{0}}),
\]
as $N\rightarrow\infty.$ Combining with the fact that $||\widehat{\vartheta}-\vartheta_{0}||=o_{\bbP}(1)$,
we have 
\[
\left(\begin{array}{c}
N^{1/2}(\bbP_{N}-\bbP)S_{\psi_{0}}(V;\vartheta)\\
\widehat{\vartheta}
\end{array}\right)\rightsquigarrow\left(\begin{array}{c}
Z\\
\vartheta_{0}
\end{array}\right)
\]
in $l^{\infty}(\mathcal{G}_{\vartheta_{0}})\times\mathcal{G}_{\vartheta_{0}}$
as $N\rightarrow\infty$. Define a function $s:l^{\infty}(\mathcal{G}_{\vartheta_{0}})\times\mathcal{G}_{\vartheta_{0}}\mapsto\R^{p}$
by $s(z,\vartheta)=z(\vartheta)-z(\vartheta_{0})$, which is continuous
for all $(z,\vartheta)$ where $\vartheta\mapsto z(\vartheta)$ is
continuous. By Lemma \ref{lemma:A2}, all sample paths of $Z$ are
continuous on $\mathcal{G}_{\vartheta_{0}}$, and thus, $s(z,\vartheta)$
is continuous for $(Z,\vartheta).$ By the Continuous-Mapping Theorem,
\[
s(Z,\widehat{\vartheta})=(\bbP_{N}-\bbP)S_{\psi_{0}}(V;\widehat{\vartheta})-(\bbP_{N}-\bbP)S_{\psi_{0}}(V;\vartheta_{0})\rightsquigarrow s(Z,\vartheta_{0})=0.
\]
Thus, (\ref{eq:A3-2}) holds. Plugging (\ref{eq:A3})--(\ref{eq:A3-2})
into (\ref{eq:A3-0}), we have 
\begin{eqnarray}
N^{1/2}(\widehat{\psi}-\psi_{0}) & = & -\Psi^{-1}N^{1/2}\{(\bbP_{N}-\bbP)S_{\psi_{0}}(V;\vartheta_{0})\}+o_{\bbP}(1).\nonumber \\
 & \rightarrow & \text{\ensuremath{\mathcal{N}}\{0,\ensuremath{(\Psi^{-1})^{\T}\E[S_{\psi_{0}}(V;\vartheta_{0})^{\otimes2}]\Psi^{-1}}\},}\label{eq:asympN}
\end{eqnarray}
in distribution as $N\rightarrow\infty.$ If $\sum_{a=0}^{1}||\widehat{\sigma}_{a}(X,S)-\sigma_{a}(X,S)||=o_{\bbP}(1)$,
$S_{\psi_{0}}(V;\vartheta_{0})$ becomes the efficient score $S_{\psi_{0}}(V)$.
Thus, the asymptotic variance in (\ref{eq:asympN}) achieves the efficiency
bound. This completes the proof of Theorem 2. }

\section{Proof of Theorem 3\label{sec:S3}}

We express the semiparametric efficient score as $S_{\eff,\psi_{0}}=c^{*}(A,X,S)\left(H_{\psi_{0}}-\E[H_{\psi_{0}}\mid X,S]\right)$
with $H_{\psi_{0}}=Y-\{\tau_{\varphi_{0}}(X)+(1-S)\lambda_{\phi_{0}}(X)\}A.$
First, we have 
\[
H_{\psi_{0}}-\E[H_{\psi_{0}}\mid X,S]=Y-\E[Y\mid X,S]-\tau_{\varphi_{0}}(X)\{A-e(X,S)\}-(1-S)\lambda_{\phi_{0}}(X)\{A-e(X,S)\},
\]
\begin{eqnarray*}
c^{*}(A,X,S) & = & \left(\E\left[\frac{\partial\epsilon_{H,\psi_{0}}}{\partial\psi^{\T}}\mid A,X,S\right]-\E\left[\frac{\partial\epsilon_{H,\psi_{0}}}{\partial\psi^{\T}}W\mid X,S\right]\E[W\mid X,S]^{-1}\right)W\\
 & = & \left(\begin{array}{c}
\frac{\partial\tau_{\varphi_{0}}(X)}{\partial\varphi}\\
(1-S)\frac{\partial\lambda_{\phi_{0}}(X)}{\partial\phi}
\end{array}\right)\left(A-\E\left[AW\mid X,S\right]\E[W\mid X,S]^{-1}\right)W,
\end{eqnarray*}
and $W=(\V[Y\mid A,X,S])^{-1}.$

We compare the asymptotic variance of $\widehat{\varphi}_{\rct}$
and $\widehat{\varphi}$. To simplify the proof, define the following
expression: 
\begin{eqnarray*}
\Gamma_{1,\rct} & = & \E\left[S\left\{ \frac{\partial\tau_{\varphi_{0}}(X)}{\partial\varphi}\right\} ^{\otimes2}\left(A-\E\left[AW\mid X,S\right]\E[W\mid X,S]^{-1}\right)\{A-e(X,S)\}W\right],\\
 & = & \E\left[S\left\{ \frac{\partial\tau_{\varphi_{0}}(X)}{\partial\varphi}\right\} ^{\otimes2}\left(A-\E\left[AW\mid X,S\right]\E[W\mid X,S]^{-1}\right)AW\right]\\
 &  & -\E\left[S\left\{ \frac{\partial\tau_{\varphi_{0}}(X)}{\partial\varphi}\right\} ^{\otimes2}\left(AW-\E\left[AW\mid X,S\right]\E[W\mid X,S]^{-1}W\right)e(X,S)\right]\\
 & = & \E\left[S\left\{ \frac{\partial\tau_{\varphi_{0}}(X)}{\partial\varphi}\right\} ^{\otimes2}\left(A-\E\left[AW\mid X,S\right]\E[W\mid X,S]^{-1}\right)^{2}W\right],\\
\Gamma_{1,\rwe} & = & \E\left[(1-S)\left\{ \frac{\partial\tau_{\varphi_{0}}(X)}{\partial\varphi}\right\} ^{\otimes2}\left(A-\E\left[AW\mid X,S\right]\E[W\mid X,S]^{-1}\right)^{2}W\right],
\end{eqnarray*}
\begin{eqnarray*}
\Gamma_{1} & = & \Gamma_{1,\rct}+\Gamma_{1,\rwe},\\
\Gamma_{12} & = & \E\left[(1-S)\frac{\partial\tau_{\varphi_{0}}(X)}{\partial\varphi}\frac{\partial\lambda_{\phi_{0}}(X)}{\partial\phi^{\T}}\left(A-\E\left[AW\mid X,S\right]\E[W\mid X,S]^{-1}\right)^{2}W\right],\\
\Gamma_{2} & = & \E\left[(1-S)\left\{ \frac{\partial\lambda_{\phi_{0}}(X)}{\partial\phi}\right\} ^{\otimes2}\left(A-\E\left[AW\mid X,S\right]\E[W\mid X,S]^{-1}\right)^{2}W\right].
\end{eqnarray*}
Also, we denote 
\begin{multline*}
S_{N,1}(\psi)=N^{-1}\sum_{i\in\mathcal{A}\cup\mathcal{B}}\frac{\partial\tau_{\varphi}(X_{i})}{\partial\varphi}\left(A_{i}-\E\left[AW\mid X_{i},S_{i}\right]\E[W\mid X_{i},S_{i}]^{-1}\right)W_{i}\\
\times\left(Y_{i}-\E[Y_{i}\mid X_{i},S_{i}]-\tau_{\varphi}(X_{i})\left\{ A_{i}-e(X_{i},S_{i})\right\} -(1-S_{i})\lambda_{\phi}(X_{i})\left\{ A_{i}-e(X_{i},S_{i})\right\} \right),
\end{multline*}
and 
\begin{multline*}
S_{N,2}(\psi)=N^{-1}\sum_{i\in\mathcal{A}\cup\mathcal{B}}\frac{\partial\lambda_{\phi}(X_{i})}{\partial\phi}(1-S_{i})\left(A_{i}-\E\left[AW\mid X_{i},S_{i}\right]\E[W\mid X_{i},S_{i}]^{-1}\right)W_{i}\\
\times\left(Y_{i}-\E[Y_{i}\mid X_{i},S_{i}]-\tau_{\varphi}(X_{i})\left\{ A_{i}-e(X_{i},S_{i})\right\} -(1-S_{i})\lambda_{\phi}(X_{i})\left\{ A_{i}-e(X_{i},S_{i})\right\} \right).
\end{multline*}
Then, 
\begin{equation}
\V\left[S_{N,1}(\psi_{0})\right]=N^{-1}\times\Gamma_{1},\ \V\left[S_{N,2}(\psi_{0})\right]=N^{-1}\times\Gamma_{2},\ \cov\left[S_{N,1}(\psi_{0}),S_{N,2}(\psi_{0})^{\T}\right]=N^{-1}\times\Gamma_{12}.\label{eq:VS}
\end{equation}

By the Taylor expansion, we have 
\begin{eqnarray}
\widehat{\psi}-\psi_{0} & = & \left(\begin{array}{cc}
\Gamma_{1} & \Gamma_{12}\\
\Gamma_{12}^{\T} & \Gamma_{2}
\end{array}\right)^{-1}\left(\begin{array}{c}
S_{N,1}(\psi_{0})\\
S_{N,2}(\psi_{0})
\end{array}\right)+o_{\bbP}(n^{-1/2})\nonumber \\
 & = & \left(\begin{array}{cc}
\Sigma_{1} & \Sigma_{12}\\
\Sigma_{12}^{\T} & \Sigma_{2}
\end{array}\right)\left(\begin{array}{c}
S_{N,1}(\psi_{0})\\
S_{N,2}(\psi_{0})
\end{array}\right)+o_{\bbP}(n^{-1/2}),\label{eq:varphi-1}
\end{eqnarray}
where 
\[
\left(\begin{array}{cc}
\Sigma_{1} & \Sigma_{12}\\
\Sigma_{12}^{\T} & \Sigma_{2}
\end{array}\right)=\left(\begin{array}{cc}
\Gamma_{1} & \Gamma_{12}\\
\Gamma_{12}^{\T} & \Gamma_{2}
\end{array}\right)^{-1}.
\]
Equivalently, we have 
\begin{eqnarray}
\Sigma_{1}\Gamma_{1}+\Sigma_{12}\Gamma_{12}^{\T} & = & I_{p\times p},\label{eq:m2-1}\\
\Sigma_{1}\Gamma_{12}+\Sigma_{12}\Gamma_{2} & = & 0_{p\times p},\label{eq:m3-1}
\end{eqnarray}
and therefore, 
\begin{eqnarray}
\Sigma_{1}\Gamma_{1}+\Sigma_{12}\Gamma_{12}^{\T} & = & I_{p\times p},\nonumber \\
\Sigma_{1}\Gamma_{12}\Gamma_{2}^{-1}\Gamma_{12}^{\T}+\Sigma_{12}\Gamma_{12}^{\T} & = & 0_{p\times p}.\label{eq:m1-1}
\end{eqnarray}
By (\ref{eq:m1-1}), we can solve $\Sigma_{1}=\left(\Gamma_{1}-\Gamma_{12}\Gamma_{2}^{-1}\Gamma_{12}^{\T}\right)^{-1}$.

Now continuing from (\ref{eq:varphi-1}), we have 
\begin{eqnarray*}
\widehat{\varphi}-\varphi_{0} & = & \Sigma_{1}S_{N,1}(\psi_{0})+\Sigma_{12}S_{N,2}(\psi_{0})+o_{\bbP}(n^{-1/2}),
\end{eqnarray*}
and therefore, 
\begin{eqnarray*}
\V\left[\widehat{\varphi}-\varphi_{0}\right] & = & \Sigma_{1}\V\left[S_{N,1}(\psi_{0})\right]\Sigma_{1}+\Sigma_{12}\cov\left[S_{N,2}(\psi_{0}),S_{N,1}(\psi_{0})^{\T}\right]\Sigma_{1}\\
 &  & +\Sigma_{1}\cov\left[S_{N,1}(\psi_{0}),S_{N,2}(\psi_{0})^{\T}\right]\Sigma_{12}+\Sigma_{12}\V\left[S_{N,2}(\psi_{0})\right]\Sigma_{12}+o(n^{-1}).
\end{eqnarray*}
Then, by (\ref{eq:VS}), we have 
\begin{eqnarray}
\V\left[\widehat{\varphi}-\varphi_{0}\right] & = & N^{-1}\times\left(\Sigma_{1}\Gamma_{1}\Sigma_{1}+\Sigma_{12}\Gamma_{12}^{\T}\Sigma_{1}\right.\nonumber \\
 &  & +\left.\Sigma_{1}\Gamma_{12}\Sigma_{12}+\Sigma_{12}\Gamma_{2}\Sigma_{12}\right)+o(n^{-1})\nonumber \\
 & = & N^{-1}\times\Sigma_{1}+0+o(n^{-1})\label{eq:v1-1}\\
 & = & N^{-1}\times\left(\Gamma_{1}-\Gamma_{12}\Gamma_{2}^{-1}\Gamma_{12}^{\T}\right)^{-1}+o(n^{-1})\nonumber \\
 & = & N^{-1}\times\left(\Gamma_{1,\rct}+\Gamma_{1,\rwe}-\Gamma_{12}\Gamma_{2}^{-1}\Gamma_{12}^{\T}\right)^{-1}+o(n^{-1}),\label{eq:V1-forcomparison}
\end{eqnarray}
where (\ref{eq:v1-1}) follows by (\ref{eq:m2-1}) and (\ref{eq:m3-1}).

For the semiparametric efficient score based only on the trial data,
we denote

\begin{eqnarray*}
S_{N,1,\rct}(\psi) & = & N^{-1}\sum_{i\in\mathcal{A}\cup\mathcal{B}}S_{i}\frac{\partial\tau_{\varphi}(X_{i})}{\partial\varphi}\left(A_{i}-\E\left[AW\mid X_{i},S_{i}\right]\E[W\mid X_{i},S_{i}]^{-1}\right)W_{i}\\
 &  & \times\left(Y_{i}-\E[Y_{i}\mid X_{i},S_{i}]-\tau_{\varphi}(X_{i})\left\{ A_{i}-e(X_{i},S_{i})\right\} \right).
\end{eqnarray*}
Then, 
\[
\V\left[S_{N,1,\rct}(\psi_{0})\right]=N^{-1}\times\Gamma_{1,\rct}.
\]
By the Taylor expansion, we have 
\begin{equation}
\V\left[\widehat{\varphi}_{\rct}-\varphi_{0}\right]=N^{-1}\times\Gamma_{1,\rct}^{-1}\Gamma_{1,\rct}\Gamma_{1,\rct}^{-1}+o(n^{-1})=N^{-1}\times\Gamma_{1,\rct}^{-1}+o(n^{-1}).\label{eq:V2-forcomparison}
\end{equation}

By the matrix extension of Cauchy-Schwarz's inequality \citep{tripathi1999matrix},
$\Gamma_{1,\rwe}-\Gamma_{12}\Gamma_{2}^{-1}\Gamma_{12}^{\T}$ is non-negative
definitive; i.e., for any $v\in\R^{p},$ 
\begin{eqnarray*}
v^{\T}\left(\Gamma_{1,\rwe}-\Gamma_{12}\Gamma_{2}^{-1}\Gamma_{12}^{\T}\right)v & \geq & 0,
\end{eqnarray*}
with equality holds when $\partial\tau_{\varphi_{0}}(X_{i})/\partial\varphi=M\partial\lambda_{\phi_{0}}(X_{i})/\partial\phi$
holds for some constant matrix $M$.

Comparing (\ref{eq:V1-forcomparison}) and (\ref{eq:V2-forcomparison})
leads to our result in Theorem 3.

\section{Goodness-of-fit test using over-identification restrictions\label{sec:S2Goodness-of-fit-test}}

The model assumptions for the treatment effect and confounding function,
i.e., Assumption 2, are crucial in our proposed framework.
In this section, we extend the goodness-of-fit test of \citet{yang2015gof}
to evaluate the model specifications of the treatment effect and confounding
function based on over-identification restrictions tests.

The insight is that we can create a larger number of estimating functions
than the number of parameters. Let $c_{1}(X)\in\R^{q_{1}}$ and $c_{2}(X)\in\R^{q_{2}}$
be some functions of $X$ that are different from $c_{1}^{*}(X)$
and $c_{2}^{*}(X)$, where $q_{1},q_{2}>0$. As with (11),
it is easy to show that 
\begin{equation}
G_{\psi_{0}}(V)=\left(\begin{array}{c}
c_{1}(X)\\
c_{2}(X)(1-S)
\end{array}\right)\epsilon_{H,\psi_{0}}\epsilon_{A}\label{eq:simpleEE-1}
\end{equation}
is unbiased. As a result, the number of unbiased estimating functions
is larger than the dimension of $\psi_{0}$, resulting in over-identification
restrictions. Under Assumptions 1 and 2,
combining the unbiasedness of $G_{\psi_{0}}(V)$ and the consistency
of $\widehat{\psi}$, we expect that the sample quantity $\bbP_{N}G_{\widehat{\psi}}(V)$
to be close to zero. We show that 
\[
N^{-1/2}\bbP_{N}G_{\widehat{\psi}}(V)=N^{-1/2}\bbP_{N}\left\{ G_{\psi_{0}}(V)+\Gamma^{\T}S_{\psi_{0}}(V)\right\} +o_{\bbP}(1),
\]
where $\Gamma=-\left\{ \E\left[\partial S_{\psi_{0}}(V)/\partial\psi\right]\right\} ^{-1}\E\left[\partial G_{\psi_{0}}(V)/\partial\psi\right]$.
Therefore, $N^{-1/2}\bbP_{N}G_{\widehat{\psi}}(V)$ is asymptotically
normal with mean $0$ and variance $\Sigma_{GG}=\V[G_{\psi_{0}}(V)+\Gamma^{\T}S_{\psi_{0}}(V)]$.

Based on the above results, we construct a goodness-of-fit test statistic
\[
T=\left\{ \bbP_{N}G_{\widehat{\psi}}(V)\right\} ^{\T}\widehat{\Sigma}_{GG}^{-1}\left\{ \bbP_{N}G_{\widehat{\psi}}(V)\right\} ,
\]
where $\widehat{\Sigma}_{GG}$ is a consistent estimator of $\Sigma_{GG}$.
Theorem \ref{Thm:T} provides the reference distribution to gauge
the plausibility of the values $T$ takes.

\begin{theorem}\label{Thm:T}

Suppose Assumptions 1 and 2 hold.
We have $T\rightarrow\chi_{q}^{2},$ a Chi-square distribution with
degrees of freedom $q=q_{1}+q_{2}$, in distribution, as $n\rightarrow\infty.$

\end{theorem}

Let $\chi_{q,\alpha}^{2}$ be the $(1-\alpha)$ quantile of $\chi_{q}^{2}$.
From Theorem \ref{Thm:T}, if $T<\chi_{q,\alpha}^{2}$, there is no
significant evidence to reject the model specifications of the treatment
effect and confounding function, and if $T\geq\chi_{q,\alpha}^{2}$,
there is strong evidence to reject the model specifications.

The choice of $c_{1}(X)$ and $c_{2}(X)$ determines the power of
the goodness-of-fit test. For the power consideration, we can specify
alternative plausible models for the treatment effect and confounding
function. Theorem 1 provides the optimal $c_{1}(X)$
and $c_{2}(X)$ under the alternative models. We then adopt this choice
for the goodness-of-fit test.

\section{The semiparametric efficiency theory of the average treatment effect
$\tau_{0}$\label{sec:The-semiparametric-efficiency-att}}

\subsection{Preliminary}

\textcolor{black}{Recall that $\pi_{0}=\pr(S=0)$. We express $\tau_{0}$
as}

\textcolor{black}{
\begin{equation}
\tau_{0}=\E[\tau_{\varphi_{0}}(X)]=\E[\tau_{\varphi_{0}}(X)\mid S=0]=\frac{\E[(1-S)\tau_{\varphi_{0}}(X)]}{\pi_{0}}.\label{eq:att}
\end{equation}
To derive the efficient score of $\tau_{0}$, we can follow a similar
approach in Section \ref{sec:The-semiparametric-efficiency-HTE}.
Because $\tau_{0}$ depends on $\varphi_{0}$ and $S_{\varphi_{0}}(V)$
is available, we consider another approach. Consider a one-dimensional
parametric submodel, $f_{\theta}(V)$, which contains the true model
$f(V)$ at $\theta=0$, i.e., $f_{\theta}(V)\vert_{\theta=0}=f(V)$.
We use $\theta$ in the subscript to denote the quantity evaluated
with respect to the submodel, e.g., $\tau_{\theta}$ is the value
of $\tau_{0}$ with respect to the submodel. We use dot to denote
the partial derivative with respect to $\theta$, e.g., $\dot{\tau}_{\theta}=\tau_{\theta}/\partial\theta$,
and use $s(\cdot)$ to denote the score function induced from the
likelihood function.}

\textcolor{black}{Recall that, the tangent space $\Lambda$ is given
in Theorem \ref{Thm: direct sum of nuis tang space}. The efficient
score of $\tau_{0}$, denoted by $S_{\tau_{0}}(V)\in\Lambda$, must
satisfy 
\[
\left.\dot{\tau}_{\theta}\right\rvert _{\theta=0}=\E[S_{\tau_{0}}(V)s(V)].
\]
We will derive the efficient score by calculating $\left.\dot{\tau}_{\theta}\right\rvert _{\theta=0}$.
To simplify the proof, we introduce some lemmas.}

\textcolor{black}{\begin{lemma}\label{lem:ratio} Consider a ratio-type
parameter $R=N/D$. If $\dot{N}_{\theta}|_{\theta=0}=\E[S_{N}(V)s(V)]$
and $\dot{D}_{\theta}|_{\theta=0}=\E[S_{D}(V)s(V)]$, then $\dot{R}_{\theta}|_{\theta=0}=\E[S_{R}(V)s(V)]$
where 
\begin{equation}
S_{R}(V)=\frac{1}{D}S_{N}(V)-\frac{R}{D}S_{D}(V).\label{eq:ratio-ses}
\end{equation}
In particular, if $S_{N}(V)$ and $S_{D}(V)$ are the efficient scores
for $N$ and $D$, then $S_{R}(V)$ is the efficient score for $R.$
\end{lemma}}

\noindent \textcolor{black}{Poof. Let $R_{\theta},$ $N_{\theta}$
and $D_{\theta}$ denote the quantities $R$, $N$, and $D$ evaluated
with respect to the parametric submodel $f_{\theta}(V)$. By the chain
rule, we have 
\begin{eqnarray*}
\left.\dot{R}_{\theta}\right\rvert _{\theta=0} & = & \left.\frac{\dot{N}_{\theta}}{D}\right\rvert _{\theta=0}-R_{\theta}\left.\frac{\dot{D}_{\theta}}{D}\right\rvert _{\theta=0}\\
 & = & \frac{1}{D}\E[S_{N}(V)s(V)]-\frac{R}{D}\E[S_{D}(V)s(V)]\\
 & = & \E\left[\left\{ \frac{1}{D}S_{N}(V)-\frac{R}{D}S_{D}(V)\right\} s(V)\right],
\end{eqnarray*}
which leads to \eqref{eq:ratio-ses}.}

\subsection{Proof of Theorem 4 }

\textcolor{black}{We derive the efficient score for $\tau_{0}$. Given
(\ref{eq:att}), we can write $\tau_{0}=N_{0}/\pi_{0}$, where $N_{0}=\E[(1-S)\tau_{\varphi_{0}}(X)]$.
By Lemma \ref{lem:ratio}, the key is to derive the efficient scores
for $N_{0}$ and $\pi_{0}$. By chain rule, we have 
\begin{eqnarray*}
\dot{N}_{\theta}\mid_{\theta=0} & = & \E[(1-S)\tau_{\varphi_{0}}(X)s(X,S)]+\E[(1-S)\dot{\tau}_{\varphi_{0,\theta}}(X)\mid_{\theta=0}]\\
 & = & \E[\{(1-S)\tau_{\varphi_{0}}(X)-N_{0}\}s(V)]+\E\left[(1-S)\frac{\partial\tau_{\varphi_{0}}(X)}{\partial\varphi^{\T}}\right]\dot{\varphi}_{0,\theta}\mid_{\theta=0}\\
 & = & \E[\{(1-S)\tau_{\varphi_{0}}(X)-N_{0}\}s(V)]+\E\left[(1-S)\frac{\partial\tau_{\varphi_{0}}(X)}{\partial\varphi^{\T}}\right]\E[S_{\varphi_{0}}(V)s(V)]\\
 & = & \E\left[\left\{ (1-S)\tau_{\varphi_{0}}(X)-N_{0}+\E\left[(1-S)\frac{\partial\tau_{\varphi_{0}}(X)}{\partial\varphi^{\T}}\right]S_{\varphi_{0}}(V)\right\} s(V)\right].
\end{eqnarray*}
It is easy to verify that $(1-S)\tau_{\varphi_{0}}(X)-N_{0}\in\Lambda^{(3)}$.
Also, by construction, we have $S_{\varphi_{0}}(V)\in\Lambda$. Therefore,
we derive the efficient score for $N_{0}$ as 
\[
S_{N_{0}}(V)=(1-S)\tau_{\varphi_{0}}(X)-N_{0}+\E\left[(1-S)\frac{\partial\tau_{\varphi_{0}}(X)}{\partial\varphi^{\T}}\right]S_{\varphi_{0}}(V).
\]
By chain rule, we have 
\[
\dot{\pi}_{0,\theta}\mid_{\theta=0}=\E[(1-S)s(S)]=\E[(1-S-\pi_{0})s(V)].
\]
It is easy to verify that $1-S-\pi_{0}\in\Lambda^{(3)}$. Therefore,
we derive the efficient score for $\pi_{0}$ as 
\[
S_{\pi_{0}}(V)=1-S-\pi_{0}.
\]
Applying Lemma \ref{lem:ratio} yields the efficient score for $\tau_{0}$
as 
\begin{eqnarray*}
S_{\tau_{0}}(V) & = & \frac{1}{\pi_{0}}S_{N_{0}}(V)-\frac{\tau_{0}}{\pi_{0}}S_{\pi_{0}}(V)\\
 & = & \frac{(1-S)\tau_{\varphi_{0}}(X)}{\pi_{0}}-\tau_{0}+\E\left[\frac{1-S}{\pi_{0}}\frac{\partial\tau_{\varphi_{0}}(X)}{\partial\varphi^{\T}}\right]S_{\varphi_{0}}(V)\\
 &  & -\frac{\tau_{0}}{\pi_{0}}(1-S-\pi_{0})\\
 & = & \frac{1-S}{\pi_{0}}\{\tau_{\varphi_{0}}(X)-\tau_{0}\}+\E\left[\left.\frac{\partial\tau_{\varphi_{0}}(X)}{\partial\varphi^{\T}}\right\vert S=0\right]S_{\varphi_{0}}(V).
\end{eqnarray*}
}

\subsection{Proof of Theorem 5 }

\textcolor{black}{By the Taylor expansion, we have 
\begin{eqnarray}
N^{1/2}\left(\widehat{\tau}-\tau_{0}\right) & = & \frac{N^{1/2}}{m}\sum_{i=1}^{N}(1-S_{i})\left\{ \tau_{\widehat{\varphi}}(X_{i})-\tau_{0}\right\} \nonumber \\
 & = & \frac{N^{1/2}}{m}\sum_{i=1}^{N}(1-S_{i})\left\{ \tau_{\varphi_{0}}(X_{i})-\tau_{0}\right\} +\left\{ \frac{1}{m}\sum_{i=1}^{N}(1-S_{i})\frac{\partial\tau_{\varphi_{0}}(X_{i})}{\partial\varphi^{\T}}\right\} N^{1/2}\left(\widehat{\varphi}-\varphi_{0}\right)+o_{\bbP}(1)\nonumber \\
 & = & \frac{N^{1/2}}{m}\sum_{i=1}^{N}(1-S_{i})\left\{ \tau_{\varphi_{0}}(X_{i})-\tau_{0}\right\} +\Psi_{0}N^{1/2}\left(\widehat{\varphi}-\varphi_{0}\right)+o_{\bbP}(1).\label{eq:hat-tau}
\end{eqnarray}
Because the two terms in (\ref{eq:hat-tau}) are uncorrelated, we
have 
\begin{eqnarray*}
\V\left[N^{1/2}\left(\widehat{\tau}-\tau_{0}\right)\right] & = & \V\left[\frac{N^{1/2}}{m}\sum_{i=1}^{N}(1-S_{i})\left\{ \tau_{\varphi_{0}}(X_{i})-\tau_{0}\right\} \right]+\Psi_{0}^{\T}\V\left[N^{1/2}\left(\widehat{\varphi}-\varphi_{0}\right)\right]\Psi_{0}\\
 & = & \frac{1}{\pi_{0}}\V\left[\tau_{\varphi_{0}}(X)\mid S=0\right]+\Psi_{0}^{\T}\V\left[N^{1/2}\left(\widehat{\varphi}-\varphi_{0}\right)\right]\Psi_{0}.
\end{eqnarray*}
}

\section{Additional results from the simulation study\label{sec:AdditionalSim} }

\subsection{Visualization of the simulation results in Section 6}

Figure \ref{fig:point} displays the simulation results for Section
6 for point estimation.

\begin{figure}
\centering

\includegraphics[width=0.49\textwidth]{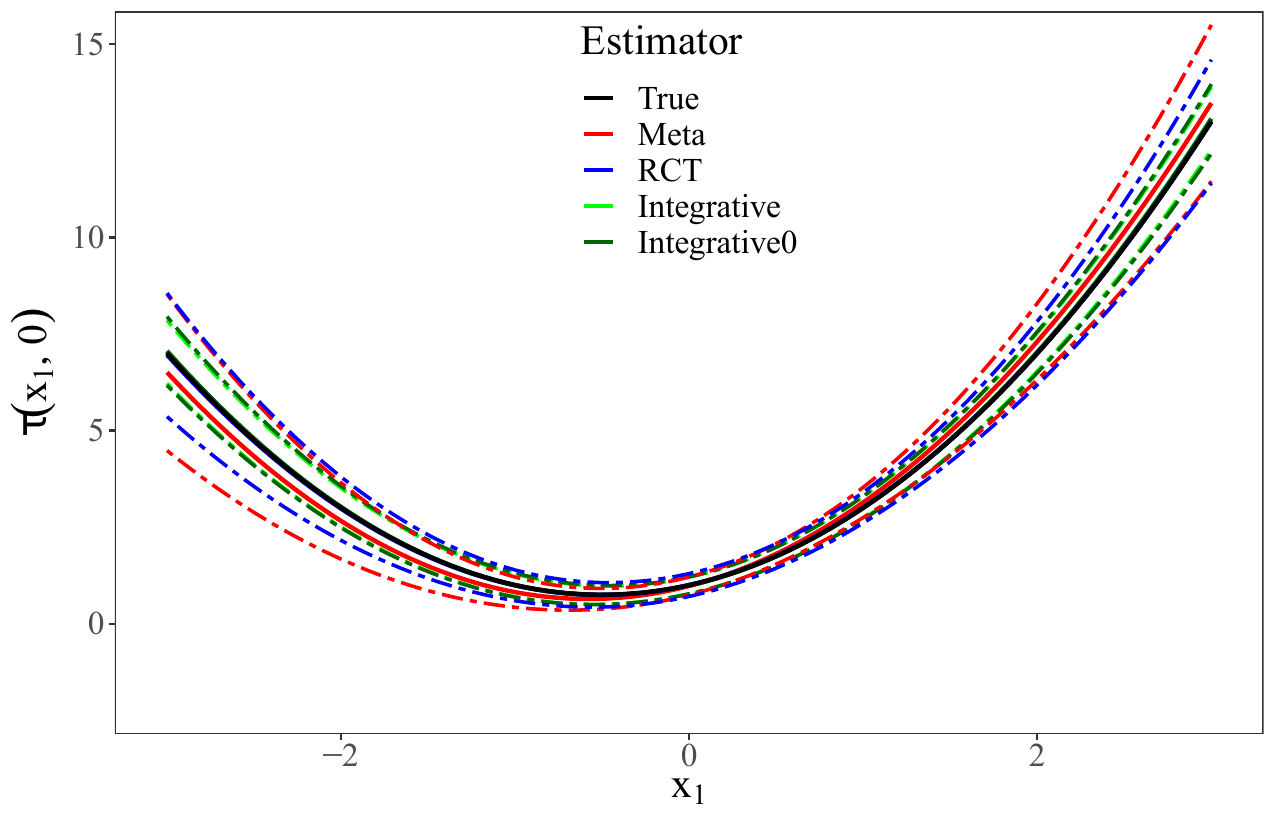}
\includegraphics[width=0.49\textwidth]{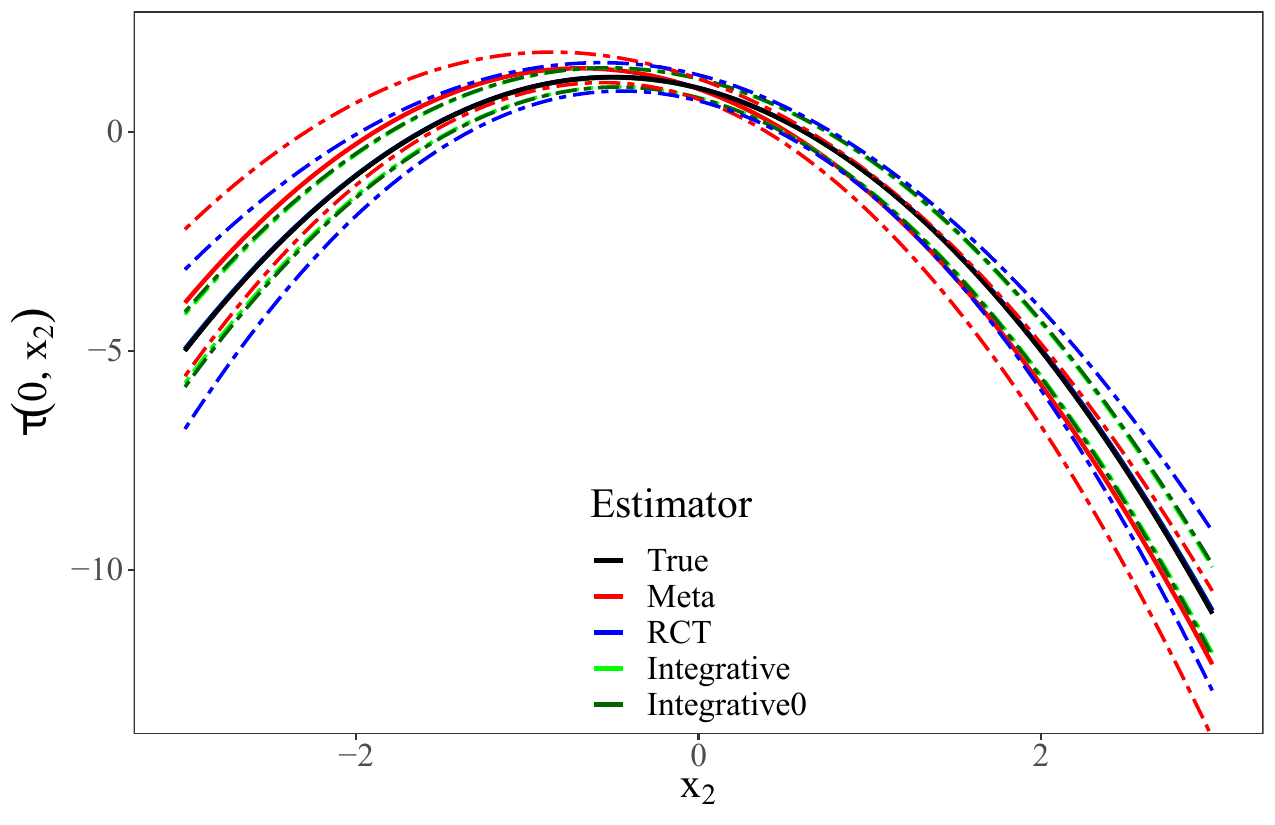}

\textbf{Setting 1} (without unmeasured confounding in the observational
study)

\vspace{0.5cm}

\includegraphics[width=0.49\textwidth]{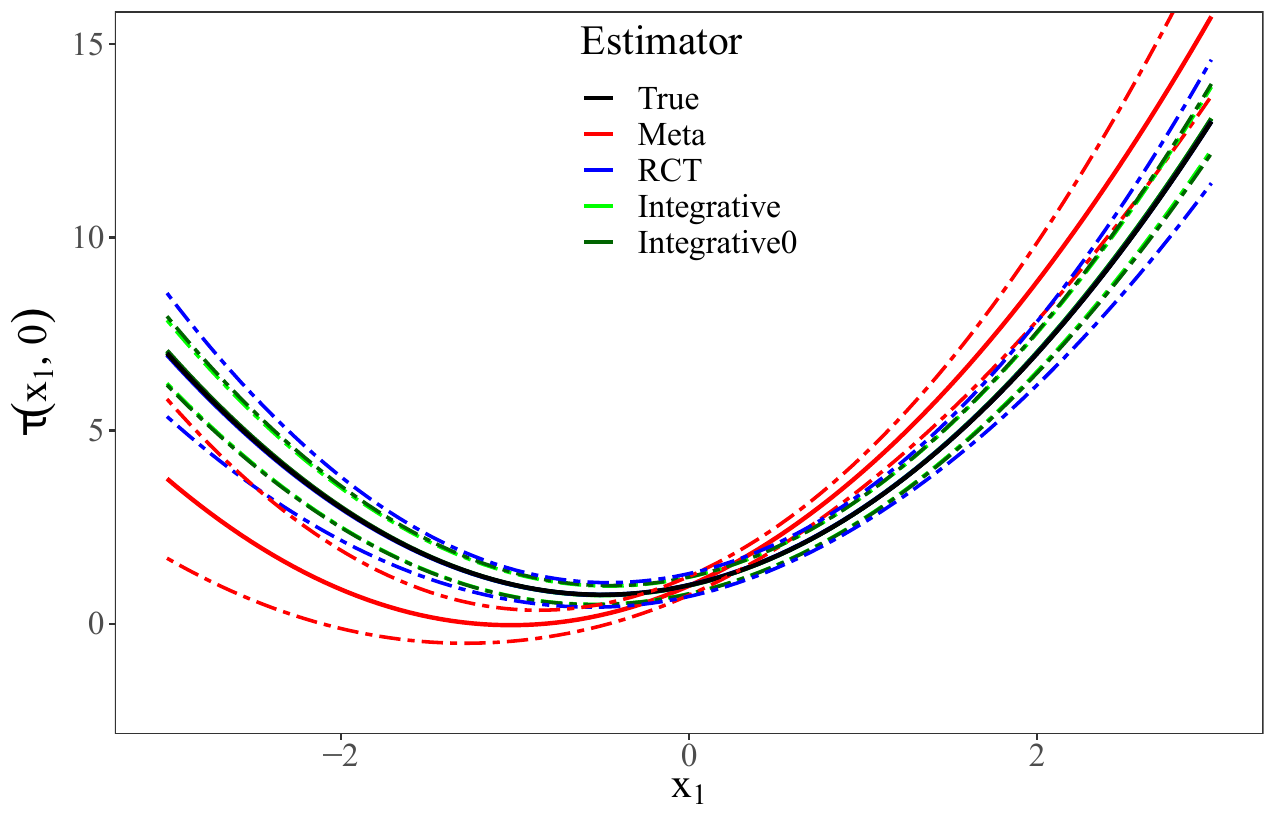}
\includegraphics[width=0.49\textwidth]{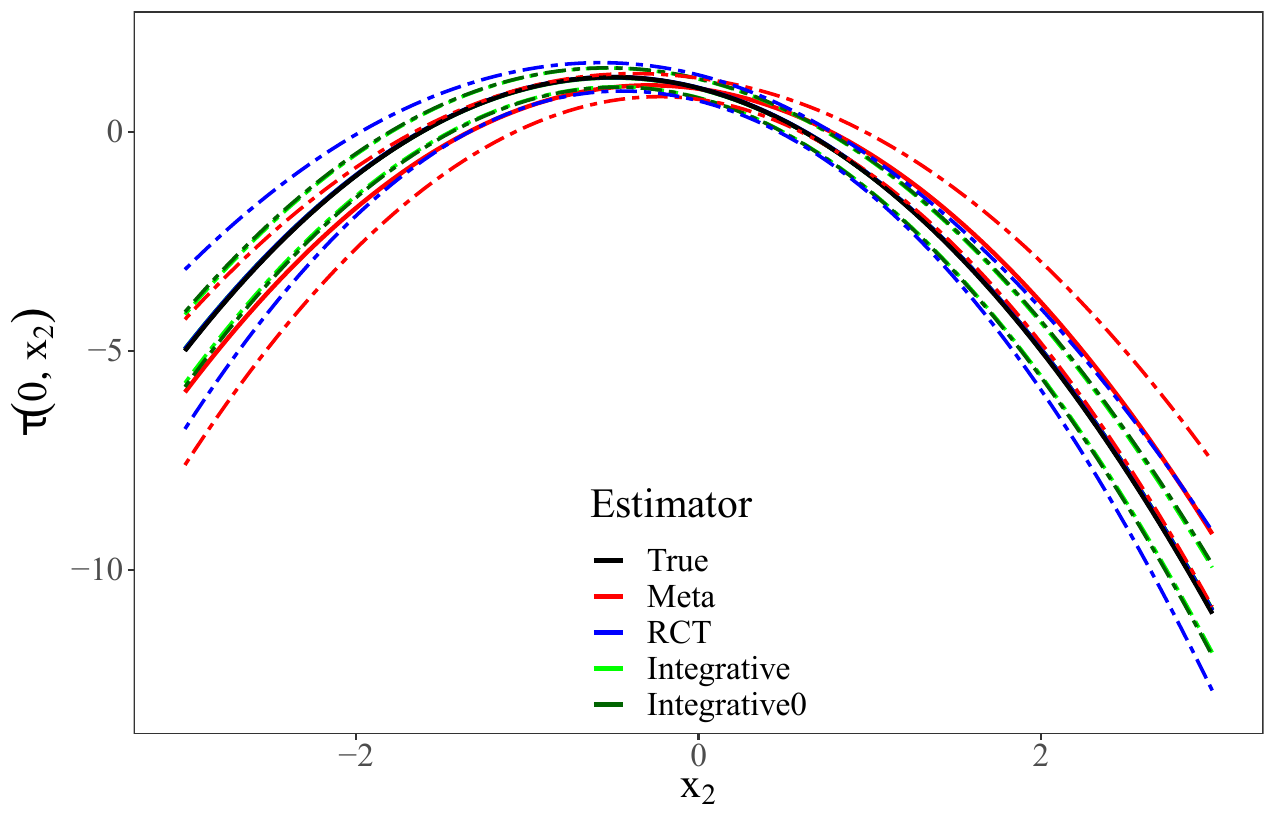}

\textbf{Setting 2} (with unmeasured confounding in the observational
study)

\vspace{0.5cm}

There are two panels: the left for $\tau(x_{1},0)$, where $x_{1}\in[-2,2]$
and $x_{2}=0$, and the right for $\tau(0,x_{2})$, where $x_{1}=0$
and $x_{2}\in[-2,2]$. The solid lines represent the Monte Carlo means
of the estimators, and the dashed bands represent the solid lines
$\pm1.96$ Monte \textcolor{black}{Carlo} standard deviations of the
estimators $\tau_{\widehat{\varphi}_{\meta}}(x)$ and $\tau_{\widehat{\varphi}_{\rct}}(x)$,
and the parametric-t wild bootstrap CIs of the integrative estimators
$\tau_{\widehat{\varphi}}(x)$ and $\tau_{\widehat{\varphi}_{\text{v1}}}(x)$.
The narrower the band, the more efficient the estimator.

\caption{\label{fig:point}Simulation results of various estimators.}
\end{figure}

{{}As mentioned in Section 4.1 of the main paper,
the Donsker requirement can be relaxed via cross-fitting in practice
\citep{chernozhukov2018double}. Under the same simulation settings
as in the main paper, we further adopt cross-fitting to estimate the
nuisance functions and reduce the Donsker assumption. Thus, various
machine learning algorithms can be adopted to fit the nuisance functions
$e(X,S),$ $\mu(X,S)$, and $\sigma_{a}^{2}(X,S)$. }{\par}

{{}The Donsker condition is frequently assumed
in practice to obtain the estimator, yet it requires a bounded condition
for the class of functions that contain the estimators of the nuisance
functions, which may not be satisfied by machine learning algorithms
where estimators are constructed in complex spaces \citep{chernozhukov2018double}.
Cross-fitting uses the idea of sample-splitting by switching the roles
of the auxiliary and main samples and averaging multiple estimators
to obtain the final estimator, which prevents overfitting and relaxes
the Donsker condition.}{\par}

{{}In the simulation study, we adopt the super
learner \citep{van2007super}, with the candidate learners including
generalized linear models, random forests, and boosting trees, which
can be carried out using off-the-shelf software, e.g., the ``SuperLearner''
function with specified algorithms in R. To employ cross-fitting,
we split the data into two folds of equal size, use one fold to estimate
the nuisance functions $e(X,S)$, $\mu(X,S)$, and $\sigma_{a}^{2}(X,S)$,
and obtain the estimator $\hat{\psi}^{(k)}$ by solving the score
equations based on the data from another fold for $k=1,\cdots,K$
and $K=2$. We repeat this estimation procedure twice by interchanging
the roles of each fold and derive the final estimator $\hat{\psi}=K^{-1}\sum_{k=1}^{K}\hat{\psi}^{(k)}$.
Parametric-t wild bootstrap is used to construct the CI, where we
follow \citet{chernozhukov2018double} to obtain the cross-fitting
variance estimator as $\hat{\mathbb{V}}(\hat{\psi})=\hat{J_{0}}^{-1}K^{-1}\sum_{k=1}^{K}\pr_{N_{K},K}\big\{ S_{\hat{\psi}^{(k)}}(V_{i})S_{\hat{\psi}^{(k)}}(V_{i})^{\T}\big\}(\hat{J_{0}}^{-1})^{\T}/N$.
Here, we denote $\hat{J_{0}}=K^{-1}\sum_{k=1}^{K}\pr_{N_{K},k}\big\{\hat{\Psi}^{(k)}(V_{i})\big\}$,
$\hat{\Psi}^{(k)}(V_{i})=\partial S_{\hat{\psi}^{(k)}}(V_{i})/\partial\psi$,
$N_{K}=N/K$, and $\pr_{N_{K},k}$ is the empirical average for the
$k$th fold. The simulation results presented in Tables \ref{tab:sim-cf}
and \ref{tab:simcvg-cf} are similar to the ones in Section 6 of the
main paper, validating the use of cross-fitting to reduce the Donsker
assumption. }{\par}

{{}}
\begin{table}[!ht]
{{}\centering\caption{\label{tab:sim-cf}Simulation results for point estimation under two
settings with cross-fitting, where the biases are scaled by $10^{-2}$
and the variances are scaled by $10^{-3}$.}
\resizebox{\textwidth}{!}{%%}%
\begin{tabular}{ccccccccccccccccccc}
\hline 
 & \multicolumn{2}{c}{{{}Meta}} & \multicolumn{2}{c}{{{}RCT}} & \multicolumn{2}{c}{{{}Integrative}} & \multicolumn{2}{c}{{{}Integrative0}} &  &  & \multicolumn{2}{c}{{{}Meta}} & \multicolumn{2}{c}{{{}RCT}} & \multicolumn{2}{c}{{{}Integrative}} & \multicolumn{2}{c}{{{}Integrative0}}\tabularnewline
 & {{}Bias} & {{}Var} & {{}Bias} & {{}Var} & {{}Bias} & {{}Var} & {{}Bias} & {{}Var} &  &  & {{}Bias} & {{}Var} & {{}Bias} & {{}Var} & {{}Bias} & {{}Var} & {{}Bias} & {{}Var}\tabularnewline
\hline 
\multicolumn{19}{c}{{{}Setting 1 (without unmeasured confounding
in the observational study)}}\tabularnewline
{{}$\tau(-3,0)$} & -42  & 975  & -7  & 1111  & 11  & 314  & 10  & 337  &  & {{}$\tau(0,-3)$} & 102  & 435  & 5  & 1288  & 2  & 343  & 5  & 334 \tabularnewline
{{}$\tau(-1.5,0)$} & -24  & 87  & -2  & 119  & 1  & 61  & 2  & 63  &  & {{}$\tau(0,-1.5)$} & 53  & 50  & 1  & 139  & -1  & 66  & -1  & 66 \tabularnewline
{{}$\tau(1.5,0)$} & 27  & 87  & -0  & 119  & 3  & 61  & 1  & 63  &  & {{}$\tau(0,1.5)$} & -60  & 50  & 4  & 139  & 1  & 66  & 3  & 66 \tabularnewline
{{}$\tau(3,0)$} & 59  & 975  & -4  & 1111  & 15  & 314  & 7  & 337  &  & {{}$\tau(0,3)$} & -123  & 435  & 11  & 1288  & 7  & 343  & 13  & 334 \tabularnewline
{{}$\tau(0,0)$} & -1  & 14  & 1  & 26  & -1  & 16  & -1  & 15  &  & {{}$\tau$} & -1  & 4  & 60  & 18  & 1  & 15  & 1  & 15 \tabularnewline
\multicolumn{19}{c}{{{}Setting 2 (with unmeasured confounding in
the observational study)}}\tabularnewline
{{}$\tau(-3,0)$} & -317  & 1005  & -7  & 1110  & 10  & 311  & 9  & 335  &  & {{}$\tau(0,-3)$} & -103  & 459  & 5  & 1288  & 1  & 341  & 5  & 334 \tabularnewline
{{}$\tau(-1.5,0)$} & -155  & 91  & -2  & 119  & 1  & 61  & 2  & 63  &  & {{}$\tau(0,-1.5)$} & -61  & 54  & 1  & 139  & -1  & 66  & -1  & 66 \tabularnewline
{{}$\tau(1.5,0)$} & 146  & 91  & -0  & 119  & 3  & 61  & 1  & 63  &  & {{}$\tau(0,1.5)$} & 78  & 54  & 4  & 139  & 1  & 66  & 3  & 66 \tabularnewline
{{}$\tau(3,0)$} & 284  & 1005  & -4  & 1110  & 13  & 311  & 7  & 335  &  & {{}$\tau(0,3)$} & 175  & 459  & 11  & 1288  & 6  & 341  & 12  & 334 \tabularnewline
{{}$\tau(0,0)$} & -1  & 15  & 1  & 26  & -1  & 16  & -1  & 15  &  & {{}$\tau$} & 2  & 5  & 60  & 18  & 1  & 15  & 1  & 15 \tabularnewline
\hline 
\end{tabular}{{}}}}
\end{table}
{\par}

{{}}
\begin{table}[!ht]
{{}\centering \caption{\label{tab:simcvg-cf}Simulation results for variance estimation and
coverage rate under two settings with cross-fitting, where the variances
are scaled by $10^{-3}$ and the coverage rates are scaled by $10^{-2}$.}
\resizebox{\textwidth}{!}{%%}%
\begin{tabular}{ccccccccccc}
\hline 
 & \multicolumn{2}{c}{{{}Integrative}} & \multicolumn{2}{c}{{{}Integrative0}} &  &  & \multicolumn{2}{c}{{{}Integrative}} & \multicolumn{2}{c}{{{}Integrative0}}\tabularnewline
 & {{}Var} & {{}CVG} & {{}Var} & {{}CVG} &  &  & {{}Var} & {{}CVG} & {{}Var} & {{}CVG}\tabularnewline
\hline 
\multicolumn{11}{c}{{{}Setting 1 (without unmeasured confounding
in the observational study)}}\tabularnewline
{{}$\tau(-3,0)$} & 314  & 93.3  & 337  & 93.8  &  & {{}$\tau(0,-3)$} & 343  & 93.6  & 334  & 93.2 \tabularnewline
{{}$\tau(-1.5,0)$} & 61  & 94.0  & 63  & 93.1  &  & {{}$\tau(0,-1.5)$} & 66  & 93.9  & 66  & 93.1 \tabularnewline
{{}$\tau(1.5,0)$} & 61  & 93.8  & 63  & 93.5  &  & {{}$\tau(0,1.5)$} & 66  & 93.7  & 66  & 93.3 \tabularnewline
{{}$\tau(3,0)$} & 314  & 93.5  & 337  & 93.2  &  & {{}$\tau(0,3)$} & 343  & 94.0  & 334  & 94.2 \tabularnewline
{{}$\tau(0,0)$} & 16  & 93.2  & 15  & 93.0  &  & {{}$\tau$} & 15  & 92.1  & 15  & 91.5 \tabularnewline
\multicolumn{11}{c}{{{}Setting 2 (with unmeasured confounding in
the observational study)}}\tabularnewline
{{}$\tau(-3,0)$} & 311  & 93.7  & 335  & 94.1  &  & {{}$\tau(0,-3)$} & 341  & 94.0  & 334  & 93.3 \tabularnewline
{{}$\tau(-1.5,0)$} & 61  & 94.1  & 63  & 93.2  &  & {{}$\tau(0,-1.5)$} & 66  & 93.8  & 66  & 92.9 \tabularnewline
{{}$\tau(1.5,0)$} & 61  & 93.9  & 63  & 93.0  &  & {{}$\tau(0,1.5)$} & 66  & 93.6  & 66  & 93.4 \tabularnewline
{{}$\tau(3,0)$} & 311  & 93.7  & 335  & 93.4  &  & {{}$\tau(0,3)$} & 341  & 94.3  & 334  & 94.3 \tabularnewline
{{}$\tau(0,0)$} & 16  & 93.5  & 15  & 93.0  &  & {{}$\tau$} & 15  & 92.0  & 15  & 91.4 \tabularnewline
\hline 
\end{tabular}{{}}}}
\end{table}
{\par}

\subsection{{Simulation results under model misspecification}}

{{}The simulations in the main paper assume
the correct parametric structural models of the heterogeneity of treatment
effect function $\tau(X)$ and the confounding function $\lambda(X)$.
We now incorporate possible model misspecification of $\tau(X)$ and
$\lambda(X)$ in the simulation to assess our proposed estimators.
After sampling the covariates $X_{j}\sim N(0,1)$, we create another
covariate $Z_{j}$ by performing a nonlinear transformation of $X_{1}$
as $Z_{1}=\log(|X_{1}|)-X_{1}^{2}/3$ and $Z_{j}=X_{j}$ for $j=2,\cdots,5$.
The treatment $A$ and the latent variable $U$ are generated in the
same way as in Section 6 of the main paper, and the potential outcomes
$Y(a)$ is generated by $Y(a)\mid(X,S=s)=a\tau(X)+(X_{1}+X_{2}X_{3}/4-X_{4}X_{5}/4)+(X^{\T}\beta)U+\varepsilon(a)$,
where $\varepsilon(a)\sim N(0,1)$. Different formations of $\tau(X)$
and $\lambda(X)$ are considered as: Case 1 with a misspecified $\tau(X)$
and a correct $\lambda(X)$, where we set the true heterogeneity of
treatment effect function as $\tau(Z)=1+Z_{1}-Z_{1}^{2}+Z_{2}-Z_{2}^{2}$
and the true confounding function as $\lambda(X)=X^{\T}\beta$ while
using the covariates $X$ to estimate both functions; Case 2 with
a correct $\tau(X)$ and a misspecified $\lambda(X)$, where we set
the true heterogeneity of treatment effect function as $\tau(X)=1+X_{1}-X_{1}^{2}+X_{2}-X_{2}^{2}$
and the true confounding function as $\lambda(Z)=Z^{\T}\beta$ while
using $X$ to estimate both functions. Here, we again consider different
incorporations of unmeasured confounders by setting $\beta=(0,\cdots,0)^{\T}$
under Setting 1 to depict a scenario without unmeasured confounders
in the observational study or setting $\beta=(1,\cdots,1)^{\T}$ under
Setting 2 with unmeasured confounders. }{\par}

{{}When $\tau(X)$ and $\lambda(X)$ are misspecified,
Remark 2 in the main paper implies that the parametric structural
models $\tau_{\varphi_{0}}(X)$ and $\lambda_{\phi_{0}}(X)$ remain
to be the best approximations in the sense that $(\varphi_{0},\phi_{0})$
minimizes the loss function $L(\varphi,\phi)$ defined in Section
\ref{sec:Proof-of-Remark1}. Under Case 1 when $\tau(X)$ is misspecified,
Table \ref{tab:sim-mis-hte} entails that the estimators $\tau_{\hat{\varphi}_{\rct}}(x)$
and $\tau_{\hat{\varphi}}(x)$ are consistent with the pseudo-truth,
and the integrative estimators have satisfying coverage rates. }{\par}

{{}}
\begin{table}[!ht]
{{}\centering\caption{\label{tab:sim-mis-hte}Simulation results for point estimation under
two settings with the model misspecification of $\lambda(X)$, where
the biases are scaled by $10^{-2}$ and the variances are scaled by
$10^{-3}$.}
\resizebox{\textwidth}{!}{%%}%
\begin{tabular}{cccccccccccccc}
\hline 
 & \multicolumn{3}{c}{{{}RCT}} & \multicolumn{3}{c}{{{}Integrative}} &  & \multicolumn{2}{c}{{{}RCT}} &  & \multicolumn{2}{c}{{{}Integrative}} & \tabularnewline
 & {{}Bias} & {{}Var} & CVG & {{}Bias} & {{}Var} & CVG &  & {{}Bias} & {{}Var} & CVG & {{}Bias} & {{}Var} & CVG\tabularnewline
\hline 
\multicolumn{14}{c}{{{}Setting 1 (without unmeasured confounding
in the observational study)}}\tabularnewline
{{}$\tau_{\varphi_{0}}(-3,0)$} & 4  & 561  & 94.1  & -3  & 176  & 93.4  & {{}$\tau_{\varphi_{0}}(0,-3)$} & -1  & 1163  & 91.7  & -1  & 310  & 93.6 \tabularnewline
{{}$\tau_{\varphi_{0}}(-1.5,0)$} & -1  & 76  & 93.3  & -2  & 40  & 93.5  & {{}$\tau_{\varphi_{0}}(0,-1.5)$} & -2  & 141  & 91.7  & -2  & 66  & 92.4 \tabularnewline
{{}$\tau_{\varphi_{0}}(1.5,0)$} & -1  & 76  & 92.9  & -1  & 40  & 92.3  & {{}$\tau_{\varphi_{0}}(0,1.5)$} & 0  & 141  & 90.8  & -0  & 66  & 93.1 \tabularnewline
{{}$\tau_{\varphi_{0}}(3,0)$} & 4  & 561  & 93.4  & -1  & 176  & 94.0  & {{}$\tau_{\varphi_{0}}(0,3)$} & 4  & 1163  & 92.5  & 2  & 310  & 93.2 \tabularnewline
{{}$\tau_{\varphi_{0}}(0,0)$} & -2  & 33  & 92.3  & -1  & 19  & 93.0  & {{}$\tau_{\varphi_{0}}$} & 38  & 15  & 12.2  & -1  & 19  & 96.1 \tabularnewline
\multicolumn{14}{c}{{{}Setting 2 (with unmeasured confounding in
the observational study): using the pseudo truth}}\tabularnewline
{{}$\tau_{\varphi_{0}}(-3,0)$} & 4  & 561  & 94.1  & -2  & 177  & 93.5  & {{}$\tau_{\varphi_{0}}(0,-3)$} & -1  & 1163  & 91.8  & -1  & 331  & 93.7 \tabularnewline
{{}$\tau_{\varphi_{0}}(-1.5,0)$} & -1  & 76  & 93.4  & -2  & 40  & 93.2  & {{}$\tau_{\varphi_{0}}(0,-1.5)$} & -2  & 141  & 91.7  & -2  & 67  & 92.5 \tabularnewline
{{}$\tau_{\varphi_{0}}(1.5,0)$} & -1  & 76  & 92.9  & -1  & 40  & 92.1  & {{}$\tau_{\varphi_{0}}(0,1.5)$} & 0  & 141  & 90.7  & -0  & 67  & 93.2 \tabularnewline
{{}$\tau_{\varphi_{0}}(3,0)$} & 4  & 561  & 93.4  & -0  & 177  & 94.6  & {{}$\tau_{\varphi_{0}}(0,3)$} & 4  & 1163  & 92.5  & 2  & 331  & 92.6 \tabularnewline
{{}$\tau_{\varphi_{0}}(0,0)$} & -2  & 33  & 92.3  & -1  & 19  & 93.1  & {{}$\tau_{\varphi_{0}}$} & 38  & 15  & 12.2  & -1  & 19  & 95.7 \tabularnewline
\hline 
\end{tabular}{{}}}}
\end{table}
{\par}

{{}}{\par}

{{}Under Case 2 when $\tau(X)$ is correctly
specified and $\lambda(X)$ is misspecified, the RCT-only estimator
$\tau_{\hat{\varphi}_{\rct}}(x)$ remains intact since it does not
suffer from the confounding issue; while the integrative estimators
$\tau_{\hat{\varphi}}(x)$ and $\tau_{\hat{\varphi}_{\text{v1}}}(x)$
converge to the pseudo heterogeneity of treatment effect function
$\tau_{\varphi_{0}}(x)$, where $(\varphi_{0},\phi_{0})$ is the minimizer
of the loss function $L(\varphi,\phi)$ defined in Section \ref{sec:Proof-of-Remark1}.
Table \ref{tab:sim-mis-cf} presents the simulation results under
Setting 2. The integrative estimators $\tau_{\hat{\varphi}}(x)$ deviate
from $\tau_{\hat{\varphi}_{\rct}}(x)$ yet are consistent with the
pseudo heterogeneity of treatment effect function $\tau_{\varphi_{0}}(x)$
with good coverage rates that are close to the nominal value.}{\par}

{{}}
\begin{table}[!ht]
{{}\centering\caption{\label{tab:sim-mis-cf}Simulation results for point estimation under
two settings with the model misspecification of $\lambda(X)$ under
Setting 2, where the biases are scaled by $10^{-2}$ and the variances
are scaled by $10^{-3}$.}
\resizebox{\textwidth}{!}{%%}%
\begin{tabular}{cccccccccccccc}
\hline 
 & \multicolumn{3}{c}{{{}RCT}} & \multicolumn{3}{c}{{{}Integrative}} &  & \multicolumn{2}{c}{{{}RCT}} &  & \multicolumn{2}{c}{{{}Integrative}} & \tabularnewline
 & {{}Bias} & {{}Var} & CVG & {{}Bias} & {{}Var} & CVG &  & {{}Bias} & {{}Var} & CVG & {{}Bias} & {{}Var} & CVG\tabularnewline
\hline 
{{}$\tau_{\varphi_{0}}(-3,0)$} & -161  & 317  & 13.8  & -9  & 139  & 91.5  & {{}$\tau_{\varphi_{0}}(0,-3)$} & 22  & 423  & 88.8  & 4  & 112  & 93.7 \tabularnewline
{{}$\tau_{\varphi_{0}}(-1.5,0)$} & -34  & 37  & 46.1  & -2  & 24  & 93.7  & {{}$\tau_{\varphi_{0}}(0,-1.5)$} & 16  & 49  & 68.8  & 1  & 23  & 94.0 \tabularnewline
{{}$\tau_{\varphi_{0}}(1.5,0)$} & -13  & 37  & 79.8  & -1  & 24  & 92.3  & {{}$\tau_{\varphi_{0}}(0,1.5)$} & 19  & 49  & 82.8  & 3  & 23  & 94.0 \tabularnewline
{{}$\tau_{\varphi_{0}}(3,0)$} & -118  & 317  & 18.7  & -7  & 139  & 90.8  & {{}$\tau_{\varphi_{0}}(0,3)$} & 28  & 423  & 89.8  & 7  & 112  & 93.4 \tabularnewline
{{}$\tau_{\varphi_{0}}(0,0)$} & 15  & 9  & 61.4  & 1  & 7  & 93.8  & {{}$\tau_{\varphi_{0}}$} & 59  & 10  & 11.6  & 1  & 7  & 95.2 \tabularnewline
\hline 
\end{tabular}{{}}}}
\end{table}
{\par}

{{}}{\par}

\section{Additional results from the application\label{sec:Application}}

Table \ref{tab4} reports the results of the model parameters in the
heterogeneity of treatment effect $\tau_{\varphi}(X)$. The trial
estimator is not significant due to the small sample size. The Meta
and integrative estimator gain efficiency over the trial estimator
by leveraging the observational study. However, the Meta estimator
may be biased due to the unmeasured confounding in the observational
study. Table \ref{tab4} reports the results of the model parameters
in the linear confounding function 
\begin{eqnarray*}
\lambda_{\phi}(X) & = & \phi_{1}+\phi_{2}{\rm Age}+\phi_{3}{\rm Tumor}\ {\rm size}+\phi_{4}{\rm Gender}+\phi_{5}{\rm Histology}\\
 &  & +\phi_{6}{\rm Race}+\phi_{7}{\rm Charlson}+\phi_{8}\text{Income}+\phi_{9}{\rm Insurance}+\phi_{10}{\rm Travel}.
\end{eqnarray*}
Because $\widehat{\phi}_{2}$, $\widehat{\phi}_{4}$ and $\widehat{\phi}_{5}$
are significant, the no unmeasured confounding assumption may not
hold for the observational sample.

\begin{table}
\centering\caption{\label{tab:res}Point estimate, standard error, and 95\% Wald confidence
interval of the model parameters in the heterogeneity of treatment
effect $\tau_{\varphi}(X)$ between adjuvant chemotherapy and observation:
tumor size{*}$=(\text{tumor size}-4.8)/1.7$, and all numbers are
multiplied by $10^{3}$.}

\begin{tabular}{cccccccccccccccc}%{cccc||c||c||c|ccc||c||c|ccc||c}
\hline
 & \multicolumn{6}{c}{RCT} & \multicolumn{5}{c}{Meta} & \multicolumn{4}{c}{Integrative}\tabularnewline
 & Est  & SE  & \multicolumn{4}{c}{95\% CI} & Est  & SE  & \multicolumn{3}{c}{95\% CI} & Est  & SE  & \multicolumn{2}{c}{95\% CI}\tabularnewline
 \hline
1 $\varphi_{1}$  & -104  & 64  & \multicolumn{4}{c}{(-229, 22)} & -74  & 9  & \multicolumn{3}{c}{(-91, -57)} & -96  & 37  & \multicolumn{2}{c}{(-170, -23)}\tabularnewline
tumor size{*} $\varphi_{2}$  & -14  & 47  & \multicolumn{4}{c}{(-106,79)} & 98  & 12  & \multicolumn{3}{c}{(75, 121)} & -8  & 29  & \multicolumn{2}{c}{(-65, 49)}\tabularnewline
(tumor size{*})$^{2}$ $\varphi_{3}$  & 11  & 27  & \multicolumn{4}{c}{(-43, 64)} & -15  & 2  & \multicolumn{3}{c}{(-19, -10)} & 5  & 2  & \multicolumn{2}{c}{(1, 10)}\tabularnewline
\hline
\end{tabular}
\end{table}

\begin{table}[htbp]
\centering \caption{Point estimate, standard error and 95\% Wald CI of model parameters
in $\lambda_{\phi}(X)$: : tumor size{*}$=(\text{tumor size}-4.8)/1.7$,
and all numbers are multiplied by $10^{3}$.}
\begin{tabular}{lccccc}
\hline
 & Est  & SE  & \multicolumn{2}{c}{95\% CI} & \tabularnewline
\hline
1 $\phi_{1}$  & 13  & 54  & (-93,  & 120)  & \tabularnewline
Age $\phi_{2}$  & -33  & 7  & (-47,  & -20)  & {*}\tabularnewline
Tumor size $\phi_{3}$  & -21  & 30  & (-80,  & 38)  & \tabularnewline
Gender $\phi_{4}$  & -26  & 12  & (-49,  & -2)  & {*}\tabularnewline
Histology $\phi_{5}$  & -64  & 11  & (-86,  & -43)  & {*}\tabularnewline
Race $\phi_{6}$  & 14  & 19  & (-24,  & 52)  & \tabularnewline
Charlson $\phi_{7}$  & -15  & 8  & (-31,  & 2)  & \tabularnewline
Income $\phi_{8}$  & 2  & 6  & (-9,  & 13)  & \tabularnewline
Insurance $\phi_{9}$  & 8  & 12  & (-15,  & 30)  & \tabularnewline
Travel $\phi_{10}$  & 5  & 10  & (-15,  & 24)  & \tabularnewline
\hline
\end{tabular}\label{tab4} 
\end{table}

%% if your bibliography is in bibtex format, uncomment commands:
% \bibliographystyle{Chicago} % Style BST file (imsart-number.bst or imsart-nameyear.bst)
% \bibliography{ci,ci7,ciwDuplicate}       % Bibliography file (usually '*.bib')

%% or include bibliography directly:
% \begin{thebibliography}{}
% \bibitem[\protect\citeauthoryear{???}{???}]{b1}
% \end{thebibliography}

\end{document}